# Stability Results for Steady, Spatially–Periodic Planforms

Benoit Dionne
Dept. of Mathematics
University of Ottawa
585 King Edward
Ottawa, Ontario K1N 6N5
Canada

Mary Silber
Dept. of Engineering Sciences
and Applied Mathematics
Northwestern University
Evanston, IL 60208
U.S.A.

Anne C. Skeldon
Dept. of Mathematics
City University
Northampton Square
London, EC1V 0HB
United Kingdom

## Abstract

We consider the symmetry-breaking steady state bifurcation of a spatially-uniform equilibrium solution of $E(2)$-equivariant partial differential equations (PDEs). We restrict the space of solutions to those that are doubly-periodic with respect to a square or hexagonal lattice, and consider the bifurcation problem restricted to a finite-dimensional center manifold. For the square lattice we assume that the kernel of the linear operator, at the bifurcation point, consists of four complex Fourier modes, with wave vectors $\mathbf{K}_1 = (\alpha, \beta)$, $\mathbf{K}_2 = (-\beta, \alpha)$, $\mathbf{K}_3 = (\beta, \alpha)$, and $\mathbf{K}_4 = (-\alpha, \beta)$, where $\alpha > \beta > 0$ are integers. For the hexagonal lattice, we assume that the kernel of the linear operator at the bifurcation point consists of six complex Fourier modes, also parameterized by an integer pair $(\alpha, \beta)$. We derive normal forms for the bifurcation problems, which we use to compute the linear, orbital stability of those solution branches guaranteed to exist by the equivariant branching lemma. These solutions consist of rolls, squares, hexagons, and a countable set of rhombs, and also a countable set of planforms that are superpositions of all of the Fourier modes in the kernel. Since rolls and squares (hexagons) are common to all of the bifurcation problems posed on square (hexagonal) lattices, this framework can be used to determine their stability relative to a countable set of perturbations by varying $\alpha$ and $\beta$. For the square lattice an $\mathcal{O}(2(\alpha + \beta) - 1)$ truncation of the normal form is required to completely determine the stability of the planforms, although many of the stability results are established at cubic order. For the hexagonal lattice, all of the solution branches guaranteed by the equivariant branching lemma are, generically, unstable due to the presence of a quadratic term in the normal form. We analyze the degenerate bifurcation problem that is obtained by setting the coefficient of the quadratic term to zero. We must retain terms through $\mathcal{O}(2\alpha - 1)$ in the normal form of the bifurcation problem. The unfolding of the degenerate bifurcation problem reveals a new class of secondary bifurcations on the hexagons and rhombs solution branches. We also analyze the bifurcation problems for $E(2) + \mathbf{Z}_2$-equivariant PDEs, which leads to new results in the hexagonal lattice case, only.

# 1 Introduction.

Equivariant bifurcation theory [12] is a powerful tool for investigating pattern-forming instabilities in physical and chemical systems. This approach distinguishes between those aspects of the bifurcation problem that are a consequence of symmetry and those aspects that depend on the specifics of the mathematical model. For example, the normal form of the bifurcation problem is derived using symmetry considerations alone, with details of the mathematical model appearing only in the numerical values of the normal form coefficients. Consequently, disparate physical and chemical systems, that nonetheless share the same symmetries, can exhibit strikingly similar behavior. This paper uses equivariant bifurcation theory to investigate the evolution of symmetry-breaking, steady state bifurcations in parameterized families of $E(2)$-equivariant partial differential equations (PDEs), where $E(2)$ is the Euclidean group of rotations, reflections and translations in a plane. Our results, which are based solely on the symmetries of the PDEs and certain features of the linear instability, apply to a wide variety of pattern forming systems, *e.g.*, Rayleigh-Bénard convection [3], models of steady cellular patterns in combustion [25] and solidification [5] and reaction-diffusion systems in the Turing instability regime [26].

Our analysis is pertinent to PDEs posed with periodic boundary conditions, and also to spatially-periodic states of PDEs posed on unbounded domains. The features of the linear instability that we assume are: (1) there is a time-independent, spatially-uniform trivial solution that loses stability in a steady state bifurcation as a parameter $\lambda$ is increased through the bifurcation point $\lambda = \lambda_c$, and (2) at $\lambda = \lambda_c$, the trivial solution is neutrally stable to perturbations in the form of Fourier modes $e^{2\pi i \mathbf{k} \cdot \mathbf{x}}$ ($\mathbf{k}, \mathbf{x} \in \mathbf{R}^2$) with finite critical wavenumber, $|\mathbf{k}| = k_c \neq 0$. A consequence of the rotational symmetry of the PDEs is that the kernel of the linearized problem at $\lambda = \lambda_c$ is infinite dimensional (*i.e.*, the length, but not the direction, of the critical wave vectors $\mathbf{k}$ is determined). One of the fundamental nonlinear problems in pattern formation is to determine which superpositions of Fourier modes lead to stable, steady state solutions, and to provide a simple characterization of these solutions, *e.g.* in terms of their symmetries. Our work is motivated by this problem.

This paper develops a bifurcation theoretic framework for determining aspects of the stability of a class of spatially-periodic equilibrium solution branches that bifurcate from the trivial solution at $\lambda = \lambda_c$. The solutions that we investigate are periodic in two directions such that they tile either a square or hexagonal lattice. For example, on the square lattice, with fundamental domain $\mathbf{x} \in [0,1) \times [0,1)$, the steady solutions are of the form

$$z_1 e^{2\pi i \mathbf{K}_1 \cdot \mathbf{x}} + z_2 e^{2\pi i \mathbf{K}_2 \cdot \mathbf{x}} + z_3 e^{2\pi i \mathbf{K}_3 \cdot \mathbf{x}} + z_4 e^{2\pi i \mathbf{K}_4 \cdot \mathbf{x}} + \text{c.c.} + \text{harmonics}, \quad (1.1)$$

where $\mathbf{z} = (z_1, z_2, z_3, z_4) \in \mathbf{C}^4$, and

$$\mathbf{K}_1 = (\alpha, \beta), \qquad \mathbf{K}_2 = (-\beta, \alpha), \qquad \mathbf{K}_3 = (\beta, \alpha), \qquad \mathbf{K}_4 = (-\alpha, \beta) . \quad (1.2)$$

Here $\alpha > \beta > 0$ are integers, and we assume that lengths are scaled such that $|\mathbf{K}_j|^2 = k_c^2 = \alpha^2 + \beta^2$, $j = 1, ..., 4$. Note that the wave vectors $\mathbf{K}_1$ and $\mathbf{K}_2$ are orthogonal, as are the wave vectors $\mathbf{K}_3$ and $\mathbf{K}_4$, but that the angle between $\mathbf{K}_1$ and $\mathbf{K}_3$ depends on $\alpha$ and $\beta$. Thus by setting two of

the amplitudes $z_j$ in (1.1) to zero, we recover the bifurcation problems that address the relative stability of rolls and squares, or rolls and rhombs. The specific solutions that we investigate here are of the form

1. Rolls (*a.k.a.* stripes): $\mathbf{z} = (x, 0, 0, 0)$, $x \in \mathbf{R}$.

2. Simple squares: $\mathbf{z} = (x, x, 0, 0)$, $x \in \mathbf{R}$.

3. Two different rhombs (*a.k.a.* rectangles): $\mathbf{z} = (x, 0, x, 0)$ and $\mathbf{z} = (x, 0, 0, x)$, $x \in \mathbf{R}$.

4. Super squares: $\mathbf{z} = (x, x, x, x)$, $x \in \mathbf{R}$.

5. Anti-squares: $\mathbf{z} = (x, x, -x, -x)$, $x \in \mathbf{R}$.

Throughout we identify all solutions that are symmetry-related (*e.g.*, we do not distinguish between rolls that are related by a rotation or translation). Note that the rolls and simple square states are the same, up to an overall length scaling factor, for every value of $\alpha$ and $\beta$. In contrast, we obtain a countable set of distinct rhombs, super squares and anti-squares by varying $\alpha$ and $\beta$. Dionne and Golubitsky [7] used the equivariant branching lemma [12, 27] to prove existence of the above steady solution branches for the generic $E(2)$-equivariant steady state bifurcation problem. The equivariant branching lemma provides an algebraic criterion for existence of steady solution branches in steady state bifurcation problems with symmetry. Roughly, the solution branches that are guaranteed by the equivariant branching lemma are those that are completely characterized by their symmetry and by a *single* real amplitude; these solution branches are referred to as *axial*.

Our bifurcation analysis proceeds by first restricting the space of solutions of the PDEs to those that are periodic with respect to some square or hexagonal lattice. Then, within this subspace of solutions, we invoke the center manifold theorem to reduce the bifurcation problem to a finite-dimensional one

$$\dot{\mathbf{z}} = \mathbf{g}(\mathbf{z}, \lambda), \quad \mathbf{g} : \mathbf{C}^s \times \mathbf{R} \to \mathbf{C}^s. \quad (1.3)$$

For example, in the case of the square lattice problem (1.1) with $k_c^2 = \alpha^2 + \beta^2$, $\alpha > \beta > 0$, we have $\mathbf{z} \in \mathbf{C}^4$ and $\mathbf{g} : \mathbf{C}^4 \times \mathbf{R} \to \mathbf{C}^4$. The vector field $\mathbf{g}$ inherits certain symmetries from the PDEs. Specifically, the group of symmetries is $\mathbf{H}\dotplus\mathbf{T}^2$, where $\mathbf{H}$ characterizes the discrete symmetries of a fundamental domain of the lattice and $\mathbf{T}^2$ is the torus of translation symmetries associated with doubly-periodic solutions. For the square lattice $\mathbf{H} = \mathbf{D}_4$, and for the hexagonal lattice $\mathbf{H} = \mathbf{D}_6$. In the case that the PDEs are posed on a bounded domain with periodic boundary conditions, the periodicity of the solutions is prescribed, rather than imposed, and the $\mathbf{T}^2$ symmetry is a consequence of the boundary conditions.

The translation symmetry on the square lattice ensures that there are no even terms in the Taylor expansion of $\mathbf{g}$ in (1.3). However, this is not the case for the analogous bifurcation problem on the hexagonal lattice. The significance of this observation is that the quadratic terms in the Taylor expansion of $\mathbf{g}$ (generically) force *all* of the axial solution branches to bifurcate unstably [15]. Our approach to investigating stable solution branches for the hexagonal lattice bifurcation problem is similar to that of Golubitsky, Swift, and Knobloch [13]. Specifically, we consider the following two problems: (1) the degenerate bifurcation problem in which the coefficient of the quadratic term in the normal form is zero, and (2) the bifurcation problem for PDEs that are $E(2) + \mathbf{Z}_2$-equivariant, where the extra $\mathbf{Z}_2$ reflection symmetry kills the even terms in the Taylor expansion of $\mathbf{g}$. Both of these problems arise naturally in Rayleigh-Bénard convection; the degenerate bifurcation problem arises when the linearized operator is self-adjoint [22]; and the $\mathbf{Z}_2$ symmetry corresponds to a reflection in the midplane of the fluid layer that is present in the Boussinesq approximation [13].

For the hexagonal lattice, we consider solutions of the form

$$z_1 e^{2\pi i \mathbf{K}_1 \cdot \mathbf{x}} + z_2 e^{2\pi i \mathbf{K}_2 \cdot \mathbf{x}} + z_3 e^{2\pi i \mathbf{K}_3 \cdot \mathbf{x}} + z_4 e^{2\pi i \mathbf{K}_4 \cdot \mathbf{x}} + z_5 e^{2\pi i \mathbf{K}_5 \cdot \mathbf{x}} + z_6 e^{2\pi i \mathbf{K}_6 \cdot \mathbf{x}} + \text{c.c.} + \text{harmonics} , \quad (1.4)$$

where $\mathbf{z} = (z_1, z_2, z_3, z_4, z_5, z_6) \in \mathbf{C}^6$. The angle between $\mathbf{K}_1$ and $\mathbf{K}_4$ is determined by an integer pair $\alpha > \beta > \alpha/2 > 0$. In particular, we assume that lengths have been scaled so that $|\mathbf{K}_j|^2 = k_c^2 = \alpha^2 + \beta^2 - \alpha\beta$, $j = 1, ..., 6$, where

$$\mathbf{K}_1 = \alpha\Big(0, 1\Big) + \beta\Big(\frac{\sqrt{3}}{2}, -\frac{1}{2}\Big), \qquad \mathbf{K}_4 = \alpha\Big(0, 1\Big) + (\alpha - \beta)\Big(\frac{\sqrt{3}}{2}, -\frac{1}{2}\Big) . \quad (1.5)$$

The wave vectors $\mathbf{K}_2, \mathbf{K}_3$ are obtained by rotating $\mathbf{K}_1$ by $\pm\frac{2\pi}{3}$, with $\mathbf{K}_5$ and $\mathbf{K}_6$ obtained from $\mathbf{K}_4$ in the same way. In the absence of the extra $\mathbf{Z}_2$ symmetry, the solution branches guaranteed by the equivariant branching lemma are [7]

1. Rolls: $\mathbf{z} = (x, 0, 0, 0, 0, 0)$, $x \in \mathbf{R}$.

2. Simple hexagons: $\mathbf{z} = (x, x, x, 0, 0, 0)$, $x \in \mathbf{R}$.

3. Three different rhombs: $\mathbf{z} = (x, 0, 0, x, 0, 0)$, $\mathbf{z} = (x, 0, 0, 0, x, 0)$, and $\mathbf{z} = (x, 0, 0, 0, 0, x)$, $x \in \mathbf{R}$.

4. Super hexagons: $\mathbf{z} = (x, x, x, x, x, x)$, $x \in \mathbf{R}$.

Moreover, the simple and super hexagons bifurcate transcritically, so we distinguish between the branch with $z_1 > 0$ and the branch with $z_1 < 0$. The rolls and simple hexagons are the same state for every $(\alpha, \beta)$, and there is a countable set of rhombs and super hexagons. The branching of super hexagons in Rayleigh-Bénard convection has been investigated by Kirschgässner [16].

In the case that the PDEs are $E(2) + \mathbf{Z}_2$-equivariant, we use the equivariant branching lemma to show that there are five additional axial solution branches to those enumerated 1-4 above. These are

1. Simple triangles: $\mathbf{z} = (ix, ix, ix, 0, 0, 0)$, $x \in \mathbf{R}$.

2. Rhombs (called the "patchwork quilt" in [13]): $\mathbf{z} = (x, x, 0, 0, 0, 0)$, $x \in \mathbf{R}$.

3. Anti-hexagons: $\mathbf{z} = (x, x, x, -x, -x, -x)$, $x \in \mathbf{R}$.

4. Super triangles: $\mathbf{z} = (ix, ix, ix, ix, ix, ix)$, $x \in \mathbf{R}$.

5. Anti-triangles: $\mathbf{z} = (ix, ix, ix, -ix, -ix, -ix)$, $x \in \mathbf{R}$.

The countable set of anti-hexagons, super triangles and anti-triangles solution branches is new. The simple triangles and patchwork quilt rhombs are investigated in [13]; they show that, generically, this branch of rhombs, composed of rolls rotated by $\pi/3$ relative to each other, bifurcates unstably.

The goal of our analysis is to derive the normal form of the equivariant bifurcation problem and then to use it to determine the branching and linear (orbital) stability of the solutions enumerated above. We treat the square and hexagonal lattice bifurcation problems separately. We proceed by first characterizing the symmetry $\Sigma_{\mathbf{z}_\lambda}$ of each of the axial solution branches $\mathbf{z}_\lambda$. These symmetries put restrictions on the form of the Jacobian matrix $\mathbf{Dg}$ evaluated on $\mathbf{z}_\lambda$. Specifically, $\mathbf{Dg}(\mathbf{z}_\lambda)$ commutes with the symmetry group $\Sigma_{\mathbf{z}_\lambda}$. We exploit this observation to determine the eigenvalues of $\mathbf{Dg}$ and their multiplicities for each axial solution branch. We then Taylor expand the equivariant bifurcation problem to sufficiently high order so that the signs of the eigenvalues are determined. Provided certain nondegeneracy conditions are satisfied, a cubic truncation of the normal form $\mathbf{g}$ is sufficient for determining the stability of rolls, rhombs and simple squares, only. The stability of simple hexagons and simple triangles depend on quartic order terms in the Taylor expansion (or quintic order terms in the case that there is an extra $\mathbf{Z}_2$ symmetry). The stability of super squares/hexagons/triangles and anti-squares/hexagons/triangles all depend on higher order terms, where the order is determined by the integers $\alpha$ and $\beta$ in (1.2) and (1.5). For example, we find that an eigenvalue (of multiplicity 2) of $\mathbf{Dg}$ evaluated on the super squares or anti-squares branches is zero unless we retain terms through $\mathcal{O}(2(\alpha + \beta) - 1)$ in the Taylor expansion of $\mathbf{g}$. However, we also find that if these two solutions are neutrally stable at cubic order, then, generically, one and only one of them is stable when the $\mathcal{O}(2(\alpha + \beta) - 1)$ terms are taken into account.

The bifurcation framework developed in this paper is a natural one for PDEs posed on a square or hexagonal domain with periodic boundary conditions. In particular, it applies when the size of the domain is larger than the wavelength of the instability, $1/k_c$. For example, in the case of a square box of sidelength $\ell$, our bifurcation analysis applies when $\ell$ is best approximated by $\sqrt{\alpha^2 + \beta^2}/k_c$ for integers $\alpha, \beta \in \mathbf{Z}$ that satisfy $\alpha > \beta > 0$. In the case that the PDEs are posed on an unbounded domain, our bifurcation analysis allows us to compute the stability of the periodic solutions in the form of rolls, simple squares, simple hexagons and simple triangles to an infinite number of perturbations by varying $\alpha$ and $\beta$. This stability computation for the simple hexagons in the Bénard problem is presented in [17].

Our paper is organized as follows. In section 2 we present a mathematical formulation of the bifurcation problem. We give the action of the symmetry group on the space of spatially-periodic solutions on the square and hexagonal lattices. We also define some terminology and outline our computations of the linear stability of the axial solution branches. In section 3 we characterize the solutions guaranteed by the equivariant branching lemma in terms of their symmetries; the necessary group theoretic computations that generate these results are banished to the appendix. The role of "hidden" Euclidean symmetries in the hexagonal lattice bifurcation problems is described. We also present some examples of the planforms associated with the axial solutions. Section 4 contains our analysis of the square lattice bifurcation problem. We compute the eigenvalues of the axial solution branches in terms of the coefficients of the normal form of the bifurcation problem.

From this information we draw a number of conclusions about the branching and (bi)stability of the solutions. In section 5 we consider two bifurcation problems associated with the hexagonal lattices. We compute stability of the axial solutions for the degenerate bifurcation problem in which the coefficient of the quadratic term is zero. We also briefly discuss the unfolding of this bifurcation problem and present an example bifurcation diagram that indicates the secondary bifurcation points on the axial solution branches. We then consider the bifurcation problem in the case that there is an extra $\mathbf{Z}_2$ symmetry. Section 6 contains our conclusions.

## 2 Problem Formulation.

We consider parameterized families of partial differential equations which we write in evolutionary form,

$$\frac{\partial}{\partial t}\mathbf{u}(\mathbf{x}, t) = \mathbf{F}(\mathbf{u}(\mathbf{x}, t), \lambda) , \qquad (2.1)$$

where $\mathbf{F} : \mathcal{X} \times \mathbf{R} \to \mathcal{Y}$ is a nonlinear operator between suitably chosen function spaces, $\mathcal{X}$ and $\mathcal{Y}$, and $\lambda \in \mathbf{R}$ is the bifurcation parameter. Here $\mathbf{u} : \mathbf{R}^2 \times \mathbf{R} \to \mathbf{R}^n$ is a function in $\mathcal{X}$ of a spatial variable $\mathbf{x} \in \mathbf{R}^2$ and time $t$. For simplicity we have suppressed any possible dependence of $\mathbf{u}$ on a third bounded spatial variable $y$. The $y$-dependence is important to our analysis only in so far as it can introduce additional symmetry to the problem. In this paper, we consider only the case where the additional symmetry is a reflection.

### 2.1 Symmetries of the PDEs.

We assume that (2.1) has Euclidean symmetry. The Euclidean group $E(2)$ is the group of motions in $\mathbf{R}^2$ that preserve distances, *i.e.* rotations, reflections and translations. We denote elements of $E(2)$ by $(h, \mathbf{d})$ where $h \in O(2)$ is an orthogonal transformation (a reflection or rotation) and $\mathbf{d} \in \mathbf{R}^2$ is a translation. The action of $(h, \mathbf{d})$ on $\mathbf{x} \in \mathbf{R}^2$ is defined by

$$(h, \mathbf{d})\mathbf{x} = h\mathbf{x} + \mathbf{d} . \qquad (2.2)$$

This action of $E(2)$ on $\mathbf{R}^2$ forces the product of $(h_1, \mathbf{d_1})$ and $(h_2, \mathbf{d_2})$ to be defined by

$$(h_1, \mathbf{d_1})(h_2, \mathbf{d_2}) = (h_1 h_2, \mathbf{d_1} + h_1 \mathbf{d_2}) . \qquad (2.3)$$

Hence $E(2)$ is the semi-direct product (denoted by $\dotplus$) of the groups of orthogonal transformations and translations; specifically, $E(2) = O(2) \dotplus \mathbf{R}^2$, where $\mathbf{R}^2$ is a normal subgroup of $E(2)$.

The Euclidean group acts on the vector-valued function $\mathbf{u} : \mathbf{R}^2 \times \mathbf{R} \to \mathbf{R}^n$ as follows

$$\gamma.\mathbf{u}(\mathbf{x}, t) = \mathbf{A}_\gamma \mathbf{u}(\gamma^{-1}\mathbf{x}, t) \qquad (2.4)$$

for all $\gamma \in E(2)$, where $\mathbf{A}_\gamma$ is an orthogonal $n \times n$ matrix. For example, when (2.1) is a system of $n$ reaction-diffusion equations, $\mathbf{A}_\gamma = \mathbf{I}_n$ for all $\gamma \in E(2)$, where $\mathbf{I}_n$ is the $n \times n$ identity matrix. For the Navier-Stokes equations in the plane, $\mathbf{A}_\gamma = \mathbf{I}_2$ when $\gamma$ is a translation, and $\mathbf{A}_\gamma$ is a $2 \times 2$ orthogonal matrix when $\gamma = (h, \mathbf{0})$, $h \in O(2)$. Our assumption that (2.1) has Euclidean symmetry means that $\mathbf{F}$ is $E(2)$-equivariant, i.e.,

$$\gamma.\mathbf{F}(\mathbf{u}(\mathbf{x}, t), \lambda) = \mathbf{F}(\gamma.\mathbf{u}(\mathbf{x}, t), \lambda) \tag{2.5}$$

for all $\gamma \in E(2)$, where the action of $\gamma$ on the vector-valued function $\mathbf{F}$ is given by (2.4).

The symmetry of the problem is enlarged from $E(2)$ to $E(2) + \mathbf{Z}_2$ for some of the motivating applications. For example, in certain Rayleigh-Bénard convection problems $\mathbf{Z}_2$ is a reflection in the mid-plane of the fluid layer [13]. Reaction-diffusion systems,

$$\frac{\partial \mathbf{u}}{\partial t} = \mathcal{D} \nabla^2 \mathbf{u} + \mathbf{g}(\mathbf{u}), \quad \mathbf{g}(-\mathbf{u}) = -\mathbf{g}(\mathbf{u}), \tag{2.6}$$

where $\mathcal{D}$ is a (constant) matrix of diffusion constants, also possess a reflection symmetry $\mathbf{u} \to -\mathbf{u}$.

## 2.2 Linear analysis and the symmetry-breaking bifurcation.

We assume that there is a Euclidean-invariant time-independent solution of (2.1) for all values of $\lambda$. This corresponds to a spatially uniform equilibrium, which, without loss of generality, we take to be $\mathbf{u} = \mathbf{0}$. We assume that this trivial solution is linearly stable for $\lambda < \lambda_c$, and linearly unstable for $\lambda > \lambda_c$. Moreover, we assume that there is a symmetry-breaking steady state bifurcation at $\lambda = \lambda_c$. At this bifurcation point the zero solution is neutrally stable to perturbations in the form of spatial Fourier modes $e^{2\pi i \mathbf{k} \cdot \mathbf{x}}$ with $\mathbf{k} \in \mathbf{R}^2$, $|\mathbf{k}| = k_c$. The *neutral stability curve*, $\lambda = \lambda(|\mathbf{k}|)$, is determined by seeking equilibrium solutions $\mathbf{u} = \mathbf{u}_{\mathbf{k}} e^{2\pi i \mathbf{k} \cdot \mathbf{x}}$ of the linearization $\frac{\partial \mathbf{u}}{\partial t} = \mathbf{L}_\lambda \mathbf{u}$ of (2.1) at $\mathbf{u} = \mathbf{0}$. (Here $\mathbf{u}_{\mathbf{k}}$ is a constant $n$-dimensional vector.) We refer to the equilibrium solutions $\mathbf{u}_{\mathbf{k}} e^{2\pi i \mathbf{k} \cdot \mathbf{x}}$, $|\mathbf{k}| = k_c$, of the linearized problem at $\lambda = \lambda_c$ as the *critical* or *neutral* modes. A typical neutral stability curve is depicted in Figure 1a. Note that the minimum of the neutral curve occurs at $(|\mathbf{k}|, \lambda) = (k_c, \lambda_c)$, where $k_c$ is nonzero and finite. A consequence of the $O(2) \subset E(2)$ symmetry is that the minimum corresponds to a circle of radius $k_c$ in the two-dimensional $\mathbf{k}$-space (see Figure 1b). Without loss of generality, we set $\lambda_c = 0$ in the remainder of the paper.

## 2.3 Spatially doubly-periodic solutions.

We restrict our bifurcation analysis to solutions $\mathbf{u}(\mathbf{x}, t)$ of (2.1) that are doubly-periodic with respect to some square or hexagonal lattice $\mathcal{L}$. Specifically, the planar lattice $\mathcal{L}$ is generated by two linearly independent vectors $\ell_1, \ell_2 \in \mathbf{R}^2$, i.e.,

$$\mathcal{L} = \{n_1 \ell_1 + n_2 \ell_2 \in \mathbf{R}^2 : n_1, n_2 \in \mathbf{Z}\}. \tag{2.7}$$

In this paper we consider two cases that satisfy $|\ell_1| = |\ell_2|$: (1) the square lattice with

$$\ell_1 = (1, 0), \qquad \ell_2 = (0, 1), \tag{2.8}$$

and (2) the hexagonal lattice with

$$\ell_1 = \left(\frac{1}{\sqrt{3}}, 1\right), \qquad \ell_2 = \left(\frac{2}{\sqrt{3}}, 0\right). \tag{2.9}$$

We say that a function $\mathbf{u}(\mathbf{x}, t)$ is $\mathcal{L} - periodic$ if

$$\mathbf{u}(\mathbf{x} + \ell, t) = \mathbf{u}(\mathbf{x}, t) \quad \text{for all} \quad \ell \in \mathcal{L}. \tag{2.10}$$

Moreover, we assume that $\mathcal{L}$-periodic solutions of (2.1) can be expressed in a Fourier series

$$u_j(\mathbf{x}, t) = \sum_{\mathbf{k} \in \mathcal{L}^*} \left( \hat{u}_{j,\mathbf{k}}(t) e^{2\pi i \mathbf{k} \cdot \mathbf{x}} + c.c. \right), \quad j = 1, \ldots, n, \tag{2.11}$$

where $\hat{u}_{j,\mathbf{k}} \in \mathbf{C}$ is the time-dependent amplitude of the $\mathbf{k}^{th}$ Fourier mode. The wave vectors $\mathbf{k}$ lie in the dual lattice to $\mathcal{L}$, denoted $\mathcal{L}^*$. Specifically, $\mathcal{L}^*$ is generated by two linearly independent vectors $\mathbf{k}_1, \mathbf{k}_2 \in \mathbf{R}^2$, where $\mathbf{k}_i \cdot \ell_j = \delta_{i,j}$ (the Kronecker delta):

$$\mathcal{L}^* = \{ n_1 \mathbf{k}_1 + n_2 \mathbf{k}_2 \in \mathbf{R}^2 : n_1, n_2 \in \mathbf{Z} \}. \tag{2.12}$$

An important consequence of restricting the solution space of (2.1) to $\mathcal{L}$-periodic functions is that the spectrum of the linear operator $\mathbf{L}_\lambda$ is rendered discrete. Hence, we expect the center manifold theorem [19] to apply at the bifurcation point. Specifically, for the problems of interest, this restriction ensures that there are only a finite number of zero eigenvalues at the bifurcation point, with all other eigenvalues bounded away from the imaginary axis. The dimension of the bifurcation problem depends on the number of points $\mathbf{k} \in \mathcal{L}^*$ that lie on the critical circle of radius $k_c$. For the square lattice, we consider the case where the critical circle intersects 8 points in $\mathcal{L}^*$, and, for the hexagonal lattice, we consider the case where the center manifold is 12-dimensional (see Figure 2). We note that the simpler cases where 4 points intersect the critical circle for the square lattice, and 6 points for the hexagonal lattice have already been analyzed (see [12], and references therein); we recover many of the results of these earlier studies in the course of our analysis.

In what follows we identify the kernel of the linear operator $\mathbf{L}_0$,

$$\ker(\mathbf{L}_0) = \{ \mathbf{u} = \sum_{j=1}^{s} z_j e^{2\pi i \mathbf{K}_j \cdot \mathbf{x}} \mathbf{u}_j + c.c. : z_j \in \mathbf{C}, |\mathbf{K}_j| = k_c \}, \tag{2.13}$$

with the vector space

$$V = \{ v = \sum_{j=1}^{s} z_j e^{2\pi i \mathbf{K}_j \cdot \mathbf{x}} + c.c. : z_j \in \mathbf{C}, |\mathbf{K}_j| = k_c \} \cong \mathbf{C}^s. \tag{2.14}$$

In (2.13) $\mathbf{u}_j$ is a constant $n$-dimensional vector associated with the $\mathbf{K}_j$ Fourier mode. The isomorphism between $V$ and $\mathbf{C}^s$ is defined by

$$v \mapsto \mathbf{z} = (z_1, z_1, \ldots, z_s). \tag{2.15}$$

As a vector space over the reals, $\dim(V) = 2s$. As mentioned above, this paper focuses on the case $s = 4$ for the square lattice and $s = 6$ for the hexagonal lattice.

The PDEs, restricted to the center manifold, lead to a system of ordinary differential equations

$$\dot{\mathbf{z}} = \mathbf{g}(\mathbf{z}, \lambda), \quad \mathbf{g} : \mathbf{C}^s \times \mathbf{R} \to \mathbf{C}^s . \tag{2.16}$$

Here $g(\mathbf{0}, \lambda) = \mathbf{0}$ and the Jacobian matrix at the bifurcation point, $D\mathbf{g}(\mathbf{0}, 0)$, is the zero matrix. In the next section we describe the symmetries inherited by the bifurcation problem from the PDEs. In particular, if $\Gamma$ is the symmetry group of the bifurcation problem (2.16), then $\mathbf{g}(\mathbf{z}, \lambda)$ satisfies the usual equivariance condition

$$\gamma \mathbf{g}(\mathbf{z}, \lambda) = \mathbf{g}(\gamma \mathbf{z}, \lambda), \quad \text{for all } \gamma \in \Gamma . \tag{2.17}$$

## 2.4 Symmetry of the restricted bifurcation problem.

The symmetry of the PDEs (2.1), reformulated in the space $\mathcal{X}_\mathcal{L}$ of $\mathcal{L}$-periodic functions, is a compact group $\Gamma$. Specifically, $\Gamma$ is the largest group, constructed from $E(2)$, that preserves $\mathcal{X}_\mathcal{L}$, i.e., $\gamma . \mathcal{X}_\mathcal{L} \subset \mathcal{X}_\mathcal{L}$ for all $\gamma \in \Gamma$. As with $E(2)$, $\Gamma$ has a semi-direct product structure, namely $\Gamma = \mathbf{H} \dotplus \mathbf{T}^2$, where $\mathbf{H} \subset O(2)$ is the finite group of rotations and reflections that preserve the lattice and $\mathbf{T}^2 \simeq \mathbf{R}^2/\mathcal{L}$ is the torus of translations. The discrete group $\mathbf{H}$ is called the *holohedry* of the lattice; in the case of the square lattice, $\mathbf{H} = \mathbf{D}_4$, while $\mathbf{H} = \mathbf{D}_6$ for the hexagonal lattice. (Recall that $\mathbf{D}_n$, the dihedral group of order $2n$, is the group of symmetries of a regular $n$-gon.) In this paper we also consider the case where $\Gamma$ is enlarged to $\Gamma + \mathbf{Z}_2$. In the remainder of the paper, let $\Gamma_s \equiv \mathbf{D}_4 \dotplus \mathbf{T}^2$ and $\Gamma_h \equiv \mathbf{D}_6 \dotplus \mathbf{T}^2$, while $\Gamma$, without a subscript, refers to $\Gamma_s(+\mathbf{Z}_2)$ and/or $\Gamma_h(+\mathbf{Z}_2)$.

We note that if the PDEs (2.1) are posed with periodic boundary conditions in two linearly independent directions, then the symmetry of the full problem is exactly $\Gamma = \mathbf{H} \dotplus \mathbf{T}^2$. In this situation, the $\mathbf{T}^2$ symmetry is a consequence of the periodic boundary conditions, and $\mathbf{H}$ characterizes the symmetries of the spatial domain.

In this paper, we assume that the action (2.4) of $\gamma \in \Gamma$ on vector-valued functions $\mathbf{u}(\mathbf{x}, t)$ leads to an action of $\Gamma$ on real-valued scalar functions $v(\mathbf{x}, t)$ in (2.14) given by

$$\gamma . v(\mathbf{x}, t) = v(\gamma^{-1} \mathbf{x}, t) . \tag{2.18}$$

This is the case for all of the applications mentioned in the introduction. (See [1] for examples of "pseudo-scalar" PDEs that do not meet this criterion.)

**Square lattice case.**

For doubly-periodic solutions on a square lattice we take the generators of the dual lattice $\mathcal{L}^*$ to be

$$\mathbf{k}_1 = (1, 0) \quad \text{and} \quad \mathbf{k}_2 = (0, 1) . \tag{2.19}$$

Thus the wave vectors $\mathbf{k} \in \mathcal{L}^*$ in (2.11) have the form $(n_1, n_2)$, where $n_1$ and $n_2$ are integers. Moreover, we assume that lengths in the original PDEs have been scaled so that $k_c = \sqrt{\alpha^2 + \beta^2} \neq 0$ for some integers $\alpha$ and $\beta$. Alternatively, we could hold $k_c$ fixed and scale the lattice $\mathcal{L}$.

Table 1: Translation Free (absolutely) Irreducible Representations For the Square Lattice.

| dim($V$) | $\mathbf{K}'s$ |
|---|---|
| 4 (s=2) | $\mathbf{K}_1 = \mathbf{k}_1 = (1,0)$ |
|  | $\mathbf{K}_2 = \mathbf{k}_2 = (0,1)$ |
| 8 (s=4) | $\mathbf{K}_1 = \alpha \mathbf{k}_1 + \beta \mathbf{k}_2 = (\alpha, \beta)$ |
|  | $\mathbf{K}_2 = -\beta \mathbf{k}_1 + \alpha \mathbf{k}_2 = (-\beta, \alpha)$ |
|  | $\mathbf{K}_3 = \beta \mathbf{k}_1 + \alpha \mathbf{k}_2 = (\beta, \alpha)$ |
|  | $\mathbf{K}_4 = -\alpha \mathbf{k}_1 + \beta \mathbf{k}_2 = (-\alpha, \beta)$ |
|  | $\alpha, \beta \in \mathbf{Z},\ \alpha > \beta > 0,\ (\alpha, \beta) = 1$ *, |
|  | $\alpha$ and $\beta$ are not both odd. |

* $(\alpha, \beta) = 1$ means that $\alpha$ and $\beta$ are relatively prime.

The relevant representation of the symmetry group $\Gamma_s = \mathbf{D}_4 \dotplus \mathbf{T}^2$ is determined by considering its action on the complex amplitudes $z_j$ of the critical Fourier modes in (2.14). The irreducible representations of $\Gamma_s$ are either 4-dimensional or 8-dimensional, in which case there are two or four complex Fourier amplitudes, respectively (*i.e.*, $s = 2$ or $s = 4$). Examples of these two different cases are depicted in Figure 2a for $k_c = 1$ and $k_c = \sqrt{5}$, *i.e.*, for $(\alpha, \beta) = (1, 0)$ and $(\alpha, \beta) = (2, 1)$. Note that it is also possible for the critical circle to intersect more than eight points in the dual lattice, *e.g.*, if $k_c = 5$ then there are four (real) Fourier modes associated with $(\alpha, \beta) = (5, 0)$ and eight associated with $(\alpha, \beta) = (4, 3)$. We do not consider these special cases here. (See Crawford [6] for an application of these higher-dimensional reducible representations.)

Following Dionne and Golubitsky [7] we require the representation of $\Gamma_s$ to be not only irreducible, but also *translation free*. A representation is translation free if there are no (non-trivial) translations in $\Gamma_s$ that act trivially on (2.14). This requirement ensures that we have found the finest lattice $\mathcal{L}$ that supports the neutral modes (2.14) [7]. Table 1 gives the values of the critical wave vectors for the translation free (absolutely) irreducible representations, henceforth simply called representations. Note that there is just one four-dimensional representation. It is the one that applies when the periodicity of functions in $\mathcal{X}_\mathcal{L}$ coincides with the wavelength of the instability, *i.e.* $k_c = |\mathbf{k}_1| = |\mathbf{k}_2|$. The bifurcation problem associated with this representation of $\Gamma_s$ has been studied extensively. In this paper we focus on the eight-dimensional representations associated with the integer pairs $(\alpha, \beta)$ where $\alpha > \beta > 0$ (see Figure 3). The additional requirements in Table 1, namely that $\alpha$ and $\beta$ are relatively prime and not both odd, ensure that the representation is translation free, and hence that the set of all critical modes (2.14) cannot be supported by a finer lattice $\mathcal{L}$ (see [7]).

$\mathbf{D}_4 \subset \Gamma_s$ is generated by a counterclockwise rotation $R_{\pi/2}$ by $\pi/2$ about the origin and a reflection $\tau_{x_1}$ through the $x_1$-axis. The elements of $\mathbf{T}^2 \subset \Gamma_s$ are denoted by $\Theta = (\theta_1, \theta_2)$, where $\theta_1, \theta_2 \in [0, 1)$. The action of $\Gamma_s$ on $V$ given by (2.14) with $s = 2$ in Table 1 induces an action of $\Gamma_s$ on $\mathbf{C}^2$ generated by

$$R_{\pi/2}(\mathbf{z}) = (\overline{z}_2, z_1),  \qquad (2.20)$$

$$\tau_{x_1}(\mathbf{z}) = (z_1, \overline{z}_2) \qquad (2.21)$$

and

$$\begin{aligned} \Theta(\mathbf{z}) &= (e^{-2\pi i \mathbf{k}_1 \cdot \Theta} z_1, e^{-2\pi i \mathbf{k}_2 \cdot \Theta} z_2) \\ &= (e^{-2\pi i \theta_1} z_1, e^{-2\pi i \theta_2} z_2) \,. \end{aligned} \tag{2.22}$$

The action of $\Gamma_s$ on $V$ for $s = 4$ in Table 1 induces an action of $\Gamma_s$ on $\mathbf{C}^4$ generated by (*cf.* Figure 3)

$$R_{\pi/2}(\mathbf{z}) = (\overline{z}_2, z_1, \overline{z}_4, z_3) \,, \tag{2.23}$$

$$\tau_{x_1}(\mathbf{z}) = (\overline{z}_4, \overline{z}_3, \overline{z}_2, \overline{z}_1) \,, \tag{2.24}$$

and

$$\begin{aligned} \Theta(\mathbf{z}) &= (e^{-2\pi i \mathbf{K}_1 \cdot \Theta} z_1, e^{-2\pi i \mathbf{K}_2 \cdot \Theta} z_2, e^{-2\pi i \mathbf{K}_3 \cdot \Theta} z_3, e^{-2\pi i \mathbf{K}_4 \cdot \Theta} z_4) \\ &= (e^{-2\pi i (\alpha\theta_1 + \beta\theta_2)} z_1, e^{-2\pi i (-\beta\theta_1 + \alpha\theta_2)} z_2, e^{-2\pi i (\beta\theta_1 + \alpha\theta_2)} z_3, e^{-2\pi i (-\alpha\theta_1 + \beta\theta_2)} z_4) \,. \end{aligned} \tag{2.25}$$

**Hexagonal lattice case.**

For doubly-periodic solutions on a hexagonal lattice the generators of the dual lattice $\mathcal{L}^*$ are

$$\mathbf{k}_1 = (0, 1) \quad \text{and} \quad \mathbf{k}_2 = (\sqrt{3}/2, -1/2) \,. \tag{2.26}$$

We assume that lengths in the original PDEs have been scaled so that $k_c = \sqrt{\alpha^2 + \beta^2 - \alpha\beta} \neq 0$ for some integers $\alpha$ and $\beta$.

The relevant representation of the symmetry group $\Gamma_h = \mathbf{D}_6 \dot{+} \mathbf{T}^2$ is determined by considering its action on the complex amplitudes of the critical Fourier modes at the bifurcation point. As for the square case, the neutrally stable modes at $\lambda = 0$ are given by (2.14). In this case the irreducible representations of $\Gamma_h$ are either 6-dimensional or 12-dimensional, in which case there are three or six complex amplitudes, respectively (*i.e.*, $s = 3$ or $s = 6$). Examples of these two different cases are depicted in Figure 2b for $k_c = 1$ and $k_c = \sqrt{7}$, *i.e.*, for $(\alpha, \beta) = (1, 0)$ and $(\alpha, \beta) = (3, 2)$. The values of the critical wave vectors for the translation free (absolutely) irreducible representations are summarized in Table 2. Note that there is just one six-dimensional representation which is associated with the case where the periodicity of functions in $\mathcal{X}_\mathcal{L}$ coincides with the wavelength of the instability, *i.e.* $k_c = |\mathbf{k}_1| = |\mathbf{k}_2|$. The bifurcation problem associated with this representation of $\Gamma_h$ has been studied extensively [4, 13]. In this paper we focus on the twelve-dimensional representations associated with the integer pairs $(\alpha, \beta)$, $\alpha > \beta > \alpha/2 > 0$ (see Figure 4). The restriction $(\alpha, \beta) = (3, \alpha + \beta) = 1$ ensures that the representations are translation free.

$\mathbf{D}_6 \subset \Gamma_h$ is generated by a counterclockwise rotation $R_{\pi/3}$ by $\pi/3$ about the origin and a reflection $\tau_{x_1}$ through the $x_1$-axis. The elements of $\mathbf{T}^2 \subset \Gamma_h$ are denoted by $\Theta = \theta_1 \ell_1 + \theta_2 \ell_2$, where $\ell_1 = (1/\sqrt{3}, 1)$, $\ell_2 = (2/\sqrt{3}, 0)$, and $\theta_1, \theta_2 \in [0, 1)$. The action of $\Gamma_h$ on $V$ given by (2.14) with $s = 3$ in Table 2 induces an action of $\Gamma_h$ on $\mathbf{C}^3$ generated by

$$R_{\pi/3}(\mathbf{z}) = (\overline{z}_3, \overline{z}_1, \overline{z}_2) \,, \tag{2.27}$$

Table 2: Translation Free (absolutely) Irreducible Representations For the Hexagonal Lattice.

| dim($V$) | $\mathbf{K}'s$ |
|---|---|
| 6 | $\mathbf{K}_1 = \mathbf{k}_1 = (0,1)$ |
| s=3 | $\mathbf{K}_2 = \mathbf{k}_2 = (\sqrt{3}/2, -1/2)$ |
|  | $\mathbf{K}_3 = -\mathbf{k}_1 - \mathbf{k}_2 = (-\sqrt{3}/2, -1/2)$ |
| 12 | $\mathbf{K}_1 = \alpha \mathbf{k}_1 + \beta \mathbf{k}_2$ |
| s=6 | $\mathbf{K}_2 = (-\alpha + \beta)\mathbf{k}_1 - \alpha \mathbf{k}_2$ |
|  | $\mathbf{K}_3 = -\beta \mathbf{k}_1 + (\alpha - \beta)\mathbf{k}_2$ |
|  | $\mathbf{K}_4 = \alpha \mathbf{k}_1 + (\alpha - \beta)\mathbf{k}_2$ |
|  | $\mathbf{K}_5 = -\beta \mathbf{k}_1 - \alpha \mathbf{k}_2$ |
|  | $\mathbf{K}_6 = (-\alpha + \beta)\mathbf{k}_1 + \beta \mathbf{k}_2$ |
|  | $\alpha, \beta \in \mathbf{Z}, \alpha > \beta > \alpha/2 > 0$, |
|  | $(\alpha, \beta) = 1$ and $(3, \alpha + \beta) = 1$ *. |

* $(3, \alpha + \beta) = 1$ means that $\alpha + \beta$ is not a multiple of 3.

$$\tau_{x_1}(\mathbf{z}) = (\overline{z}_1, \overline{z}_3, \overline{z}_2) \tag{2.28}$$

and

$$\begin{aligned}\Theta(\mathbf{z}) &= \left(e^{-2\pi i \mathbf{k}_1 \cdot \Theta} z_1, e^{-2\pi i \mathbf{k}_2 \cdot \Theta} z_2, e^{-2\pi \mathbf{k}_3 \cdot \Theta} z_3\right) \\ &= \left(e^{-2\pi i \theta_1} z_1, e^{-2\pi i \theta_2} z_2, e^{2\pi i (\theta_1 + \theta_2)} z_3\right).\end{aligned} \tag{2.29}$$

The action of $\Gamma_h$ on $V$ for $s = 6$ in Table 2 induces an action of $\Gamma_h$ on $\mathbf{C}^6$ generated by (*cf.* Figure 4)

$$R_{\pi/3}(\mathbf{z}) = (\overline{z}_2, \overline{z}_3, \overline{z}_1, \overline{z}_5, \overline{z}_6, \overline{z}_4), \tag{2.30}$$

$$\tau_{x_1}(\mathbf{z}) = (z_6, z_5, z_4, z_3, z_2, z_1) \tag{2.31}$$

and

$$\begin{aligned}\Theta(\mathbf{z}) &= \left(e^{-2\pi i \mathbf{K}_1 \cdot \Theta} z_1, e^{-2\pi i \mathbf{K}_2 \cdot \Theta} z_2, e^{-2\pi i \mathbf{K}_3 \cdot \Theta} z_3, e^{-2\pi i \mathbf{K}_4 \cdot \Theta} z_4, e^{-2\pi i \mathbf{K}_5 \cdot \Theta} z_5, e^{-2\pi i \mathbf{K}_6 \cdot \Theta} z_6\right) \\ &= \left(e^{-2\pi i(\alpha \theta_1 + \beta \theta_2)} z_1, e^{-2\pi i((-\alpha + \beta)\theta_1 - \alpha \theta_2)} z_2, e^{-2\pi i(-\beta \theta_1 + (\alpha - \beta)\theta_2)} z_3,\right.\\ &\quad \left.e^{-2\pi i(\alpha \theta_1 + (\alpha - \beta)\theta_2)} z_4, e^{-2\pi i(-\beta \theta_1 - \alpha \theta_2)} z_5, e^{-2\pi i((-\alpha + \beta)\theta_1 + \beta \theta_2)} z_6\right).\end{aligned} \tag{2.32}$$

**Additional $\mathbf{Z}_2$ symmetry.**

In this paper we consider the possibility that there is an additional $\mathbf{Z}_2$ symmetry so that the bifurcation problems are equivariant with respect to $(\mathbf{H} \dot{+} \mathbf{T}^2) + \mathbf{Z}_2$, where $\mathbf{H} = \mathbf{D}_4$ or $\mathbf{H} = \mathbf{D}_6$. We assume that $\kappa \in \mathbf{Z}_2$ takes $v$ to $-v$, where $v \in V$ is given by (2.14). This induces the following action on $\mathbf{z} \in \mathbf{C}^s$:

$$\kappa(\mathbf{z}) = -\mathbf{z}. \tag{2.33}$$

The additional reflection symmetry has no effect on the bifurcation problems associated with the square lattice. This observation, for the four-dimensional representation of $(\mathbf{D}_4 \dot{+} \mathbf{T}^2) + \mathbf{Z}_2$, is made in [24]. The case of the eight-dimensional representations in Table 1 is the same. Specifically, we note that the translation $(\frac{1}{2}, \frac{1}{2}) \in \mathbf{T}^2$ in (2.25) acts on $\mathbf{z}$ in the same way as the reflection $\kappa$ in (2.33). Hence, we need consider the effect of the additional reflection for bifurcation problems associated with hexagonal lattices only.

## 2.5 Overview of calculations.

The goal of this work is to compute stability of particular solution branches of the bifurcation problem (2.16). In particular, we compute the stability, at bifurcation, of those solutions that are guaranteed to exist by the equivariant branching lemma [12]. This lemma provides an algebraic criterion for existence of solution branches associated with particular subgroups of $\Gamma$. Specifically, we specify the symmetry of an equilibrium solution $\mathbf{z} \in \mathbf{C}^s$ by the isotropy subgroup $\Sigma_{\mathbf{z}} \subset \Gamma$, where

$$\Sigma_{\mathbf{z}} = \{\sigma \in \Gamma : \sigma \mathbf{z} = \mathbf{z}\} . \tag{2.34}$$

A subgroup $\Sigma \subset \Gamma$ is an *isotropy* subgroup if there exists a $\mathbf{z} \in \mathbf{C}^s$ for which $\Sigma_{\mathbf{z}} = \Sigma$. Associated with each isotropy subgroup $\Sigma \subset \Gamma$ is a vector subspace of $\mathbf{C}^s$, called the fixed point subspace and denoted Fix($\Sigma$), where

$$\text{Fix}(\Sigma) = \{\mathbf{z} \in \mathbf{C}^s : \sigma \mathbf{z} = \mathbf{z}, \text{ for all } \sigma \in \Sigma\} . \tag{2.35}$$

The equivariant branching lemma states that provided certain (generic) conditions are satisfied by the bifurcation, there exists a branch of equilibrium solutions, bifurcating from the origin at $\lambda = 0$, with symmetry $\Sigma$ for each isotropy subgroup $\Sigma \subset \Gamma$ that satisfies dim(Fix($\Sigma$))=1 (see [12]). In the next section we determine, up to conjugacy, all isotropy subgroups for the 8-dimensional representations of $\mathbf{D}_4 \dot{+} \mathbf{T}^2$ and the 12-dimensional representations of $\mathbf{D}_6 \dot{+} \mathbf{T}^2$ and $(\mathbf{D}_6 \dot{+} \mathbf{T}^2) + \mathbf{Z}_2$. We then depict examples of those planforms guaranteed to exist by the equivariant branching lemma. Following [11], we refer to isotropy subgroups with 1-dimensional fixed point spaces as *axial* and the associated spatially doubly-periodic solutions as *axial planforms*.

To determine the linear stability of an equilibrium solution branch $\mathbf{z}_\lambda$ of the $\Gamma$-equivariant bifurcation problem $\dot{\mathbf{z}} = \mathbf{g}(\mathbf{z}, \lambda)$, we compute the eigenvalues of the Jacobian matrix $D\mathbf{g}(\mathbf{z}_\lambda, \lambda)$. A simple consequence of the equivariance (2.17) of $\mathbf{g} : \mathbf{C}^s \times \mathbf{R} \to \mathbf{C}^s$ is that the Jacobian matrix evaluated at $\mathbf{z} = \mathbf{z}_\lambda$ commutes with all $\sigma \in \Sigma_{\mathbf{z}_\lambda}$. This condition puts restrictions on the form of $D\mathbf{g}(\mathbf{z}_\lambda, \lambda)$. We exploit this observation to determine the eigenvalues of the Jacobian matrix for each axial subgroup. We then determine which of the eigenvalues are forced, by the continuous translation symmetry $\mathbf{T}^2$, to be zero. Finally we determine expressions for the nonzero eigenvalues in terms of the leading order terms in the Taylor expansion, at $\mathbf{z} = \mathbf{0}$, of the general smooth $\Gamma$-equivariant vector field $\mathbf{g}(\mathbf{z}, \lambda)$. These results determine conditions, for stability of the axial planforms, that must be satisfied by the coefficients of certain terms in the Taylor expansion of $\mathbf{g}$.

In the case of the hexagonal lattice with $\mathbf{D}_6 \dot{+} \mathbf{T}^2$ symmetry, the Taylor expansion of the general equivariant vector field possesses a term that is quadratic in the $z_j$. It then follows, from a theorem due to Ihrig and Golubitsky [15], that generically the axial planforms all bifurcate unstably.

Therefore, in order to investigate stable solutions within the setting of a local bifurcation problem, we study the degenerate bifurcation problem obtained by setting the coefficient of the quadratic term to zero. We also briefly indicate some of the interesting possibilities for secondary bifurcations within an unfolding of this degenerate bifurcation problem.

# 3 Group Theoretic Results.

## 3.1 Isotropy subgroups.

In this paper we follow the convention of identifying all solution branches that are on the group orbit $\Gamma \mathbf{z}_\lambda$ of a particular branch $\mathbf{z}_\lambda$. Thus we classify isotropy subgroups of $\Gamma$ by conjugacy class since the isotropy of a point $\mathbf{z}_\lambda \in \mathbf{C}^s$ is conjugate to the isotropy of a point on its group orbit. Specifically, $\Sigma_{\gamma \mathbf{z}_\lambda} = \gamma \Sigma_{\mathbf{z}_\lambda} \gamma^{-1}$. (Recall that two subgroups $\Sigma_1, \Sigma_2 \subset \Gamma$ are conjugate if $\Sigma_2 = \gamma \Sigma_1 \gamma^{-1}$ for some $\gamma \in \Gamma$.)

In this section, we determine, up to conjugacy, all isotropy subgroups and their fixed point spaces for the 8-dimensional representations of $\Gamma_s$, and for the 12-dimensional representations of $\Gamma_h$ and $\Gamma_h + \mathbf{Z}_2$. In each case we summarize the results in a table; the computations that are necessary to generate the tables are relegated to the appendix.

### Square lattice case.

We list in Table 3 the isotropy subgroups, up to conjugacy, of $\Gamma_s$ acting on $\mathbf{C}^4$ together with their generators, their fixed point subspaces and the dimensions of the fixed point subspaces. Note that the pure translation subgroups, denoted $\mathbf{S}^1$, $\mathbf{S}_{1,2}$, $\mathbf{S}_{1,3}$ and $\mathbf{S}_{1,4}$, depend on the values $\alpha$ and $\beta$ and hence are not the same for all 8-dimensional representations. Associated with these fixed point subspaces are planforms in $V$ that are periodic with respect to a finer lattice than $\mathcal{L}$. There are six isotropy subgroups for which dim(Fix($\Sigma$))=1; these are the axial subgroups. The equivariant branching lemma guarantees the (generic) existence of a branch of equilibria associated with each of these subgroups $\Sigma$.

### Hexagonal lattice cases.

Table 4 contains the isotropy subgroups, up to conjugacy, of $\Gamma_h$ acting on $\mathbf{C}^6$ together with their generators, their fixed point subspaces and the dimensions of the fixed point subspaces. The isotropy subgroups associated with $\Gamma_h + \mathbf{Z}_2$ are listed in Tables 4 and 6. The equivariant branching lemma applies to six of the isotropy subgroups in the case that $\Gamma = \Gamma_h$ and ten in the case that there is an extra $\mathbf{Z}_2$ symmetry. Note that some of the isotropy subgroups in Table 4 are modified by the extra $\mathbf{Z}_2$ symmetry as indicated in Table 5. In this table we denote elements of $\Gamma_h = \mathbf{D}_6 \dotplus \mathbf{T}^2$ by $(h, \Theta)$ and elements of $\Gamma_h + \mathbf{Z}_2$ by $((h, \Theta), Id)$ and $((h, \Theta), \kappa)$. Here $h \in \mathbf{D}_6$, $\Theta \in \mathbf{T}^2$ and $\mathbf{Z}_2 = \{Id, \kappa\}$, where $Id$ specifies the identity element of a group.

Table 3: Isotropy Subgroups $\Sigma$ (up to conjugacy) of $\Gamma_s$.

| $\Sigma$ | Generators of $\Sigma$ | Fix($\Sigma$) | Dim(Fix($\Sigma$)) |
|---|---|---|---|
| $\Gamma_s$ | $R_{\pi/2}, \tau_{x_1}, \mathbf{T}^2$ [a] | $\mathbf{z} = (z_1, z_2, z_3, z_4) = \mathbf{0}$ | 0 |
| $\mathbf{D}_4$ | $R_{\pi/2}, \tau_{x_1}$ | $z_1 = z_2 = z_3 = z_4 \in \mathbf{R}$ | 1 |
| $\widetilde{\mathbf{D}}_4$ | $R_{\pi/2}, (\tau_{x_1}, (\frac{1}{2}, \frac{1}{2}))$ | $z_1 = z_2 = -z_3 = -z_4 \in \mathbf{R}$ | 1 |
| $\mathbf{D}_2^d$ | $R_\pi, \tau_d$ [b] | $z_1 = z_3, z_2 = z_4 \in \mathbf{R}$ | 2 |
| $\mathbf{D}_2^x$ | $R_\pi, \tau_{x_1}$ | $z_1 = z_4, z_2 = z_3 \in \mathbf{R}$ | 2 |
| $\widehat{\mathbf{D}}_2^x$ | $R_\pi, (\tau_{x_1}, (\frac{1}{2}, 0))$ | $z_1 = -z_4, z_2 = z_3 \in \mathbf{R}$ if $\alpha$ is odd. | 2 |
|  |  | $z_1 = z_4, z_2 = -z_3 \in \mathbf{R}$ if $\beta$ is odd. | 2 |
| $\widetilde{\mathbf{D}}_2^x$ | $R_\pi, (\tau_{x_1}, (\frac{1}{2}, \frac{1}{2}))$ | $z_1 = -z_4, z_2 = -z_3 \in \mathbf{R}$ | 2 |
| $\mathbf{Z}_4$ | $R_{\pi/2}$ | $z_1 = z_2, z_3 = z_4 \in \mathbf{R}$ | 2 |
| $\mathbf{Z}_2^d$ | $\tau_d$ | $z_1 = z_3, z_2 = \overline{z}_4 \in \mathbf{C}$ | 4 |
| $\mathbf{Z}_2^x$ | $\tau_{x_1}$ | $z_1 = \overline{z}_4, z_2 = \overline{z}_3 \in \mathbf{C}$ | 4 |
| $\widehat{\mathbf{Z}}_2^x$ | $(\tau_{x_1}, (\frac{1}{2}, 0))$ | $z_1 = -\overline{z}_4, z_2 = \overline{z}_3 \in \mathbf{C}$ if $\alpha$ is odd. | 4 |
|  |  | $z_1 = \overline{z}_4, z_2 = -\overline{z}_3 \in \mathbf{C}$ if $\beta$ is odd. | 4 |
| $\mathbf{Z}_2^c$ | $R_\pi$ | $z_1, z_2, z_3, z_4 \in \mathbf{R}$ | 4 |
| $\mathbf{1}$ | $Id$ | $z_1, z_2, z_3, z_4 \in \mathbf{C}$ | 8 |
| $\mathbf{Z}_2^c \dot{+} \mathbf{S}^1$ | $R_\pi, \mathbf{S}^1$ [c] | $z_1 \in \mathbf{R}, z_2 = z_3 = z_4 = 0$ | 1 |
| $\mathbf{Z}_4 \dot{+} \mathbf{S}_{1,2}$ | $R_{\pi/2}, \mathbf{S}_{1,2}$ [d] | $z_1 = z_2 \in \mathbf{R}, z_3 = z_4 = 0$ | 1 |
| $\mathbf{Z}_2^c \dot{+} \mathbf{S}_{1,2}$ | $R_\pi, \mathbf{S}_{1,2}$ | $z_1, z_2 \in \mathbf{R}, z_3 = z_4 = 0$ | 2 |
| $\mathbf{D}_2^d \dot{+} \mathbf{S}_{1,3}$ | $R_\pi, \tau_d, \mathbf{S}_{1,3}$ [e] | $z_1 = z_3 \in \mathbf{R}, z_2 = z_4 = 0$ | 1 |
| $\mathbf{Z}_2^c \dot{+} \mathbf{S}_{1,3}$ | $R_\pi, \mathbf{S}_{1,3}$ | $z_1, z_3 \in \mathbf{R}, z_2 = z_4 = 0$ | 2 |
| $\mathbf{D}_2^x \dot{+} \mathbf{S}_{1,4}$ | $R_\pi, \tau_{x_1}, \mathbf{S}_{1,4}$ [f] | $z_1 = z_4 \in \mathbf{R}, z_2 = z_3 = 0$ | 1 |
| $\mathbf{Z}_2^c \dot{+} \mathbf{S}_{1,4}$ | $R_\pi, \mathbf{S}_{1,4}$ | $z_1, z_4 \in \mathbf{R}, z_2 = z_3 = 0$ | 2 |

[a] the generators of $\Gamma_s$ are given in equations 2.30-2.32.
[b] $R_\pi \equiv R_{\pi/2}^2$; $\tau_d \equiv \tau_{x_1} R_{\pi/2}^3$ is a reflection through the line containing $\ell_1 + \ell_2$ (cf. Fig. 3).
[c] $\mathbf{S}^1 = \{(\beta s, -\alpha s) \in \mathbf{T}^2 : s \in \mathbf{R}\}$.
[d] $\mathbf{S}_{1,2}$ is generated by $(\frac{\alpha}{\alpha^2+\beta^2}, \frac{\beta}{\alpha^2+\beta^2}), (\frac{-\beta}{\alpha^2+\beta^2}, \frac{\alpha}{\alpha^2+\beta^2}) \in \mathbf{T}^2$.
[e] $\mathbf{S}_{1,3}$ is generated by $(\frac{\alpha}{\alpha^2-\beta^2}, \frac{-\beta}{\alpha^2-\beta^2}), (\frac{-\beta}{\alpha^2-\beta^2}, \frac{\alpha}{\alpha^2-\beta^2}) \in \mathbf{T}^2$.
[f] $\mathbf{S}_{1,4}$ is generated by $(\frac{1}{2\alpha}, \frac{1}{2\beta}), (\frac{-1}{2\alpha}, \frac{1}{2\beta}) \in \mathbf{T}^2$.

Table 4: Isotropy Subgroups $\Sigma$ (up to conjugacy) of $\Gamma_h$ and $\Gamma_h + \mathbf{Z}_2$.

| $\Sigma$ | Generators of $\Sigma$ | Fix($\Sigma$) | Dim(Fix($\Sigma$)) |
|---|---|---|---|
| $\Gamma_h(+\mathbf{Z}_2)$ | $R_{\pi/3}, \tau_{x_1}, \mathbf{T}^2, (\kappa)$ [a] | $\mathbf{z} = (z_1, z_2, z_3, z_4, z_6) = \mathbf{0}$ | 0 |
| $\mathbf{D}_6$ | $R_{\pi/3}, \tau_{x_1}$ | $z_1 = z_2 = z_3 = z_4 = z_5 = z_6 \in \mathbf{R}$ | 1 |
| $\mathbf{D}_3^n$ | $R_{\pi/3}^2, \tau_n$ [b] | $z_1 = z_2 = z_3 = \overline{z}_4 = \overline{z}_5 = \overline{z}_6 \in \mathbf{C}$ | 2 |
| $\mathbf{D}_3$ | $R_{\pi/3}^2, \tau_{x_1}$ | $z_1 = z_2 = z_3 = z_4 = z_5 = z_6 \in \mathbf{C}$ | 2 |
| $\mathbf{Z}_6$ | $R_{\pi/3}$ | $z_1 = z_2 = z_3, z_4 = z_5 = z_6 \in \mathbf{R}$ | 2 |
| $\mathbf{D}_2^x$ | $R_\pi, \tau_{x_1}$ | $z_1 = z_6, z_2 = z_5, z_3 = z_4 \in \mathbf{R}$ | 3 |
| $\mathbf{Z}_3$ | $R_{\pi/3}^2$ | $z_1 = z_2 = z_3, z_4 = z_5 = z_6 \in \mathbf{C}$ | 4 |
| $\mathbf{Z}_2^n$ | $\tau_n$ | $z_1 = \overline{z}_4, z_2 = \overline{z}_6, z_3 = \overline{z}_5 \in \mathbf{C}$ | 6 |
| $\mathbf{Z}_2^x$ | $\tau_{x_1}$ | $z_1 = z_6, z_2 = z_5, z_3 = z_4 \in \mathbf{C}$ | 6 |
| $\mathbf{Z}_2^c$ | $R_\pi$ | $z_1, z_2, z_3, z_4, z_5, z_6 \in \mathbf{R}$ | 6 |
| $\mathbf{1}$ | $Id$ | $z_1, z_2, z_3, z_4, z_5, z_6 \in \mathbf{C}$ | 12 |
| $\mathbf{Z}_2^c \dotplus \mathbf{S}^1$ | $R_\pi, \mathbf{S}^1$ [c] | $z_1 \in \mathbf{R}, z_2 = z_3 = z_4 = z_5 = z_6 = 0$ | 1 |
| $\mathbf{D}_2^n \dotplus \mathbf{S}_{1,4}$ | $R_\pi, \tau_n, \mathbf{S}_{1,4}$ | $z_1 = z_4 \in \mathbf{R}, z_2 = z_3 = z_5 = z_6 = 0$ | 1 |
| $\mathbf{Z}_2^c \dotplus \mathbf{S}_{1,4}$ | $R_\pi, \mathbf{S}_{1,4}$ | $z_1, z_4 \in \mathbf{R}, z_2 = z_3 = z_5 = z_6 = 0$ | 2 |
| $\mathbf{D}_2^m \dotplus \mathbf{S}_{1,5}$ | $R_\pi, \tau_m$ [d], $\mathbf{S}_{1,5}$ | $z_1 = z_5 \in \mathbf{R}, z_2 = z_3 = z_4 = z_6 = 0$ | 1 |
| $\mathbf{Z}_2^c \dotplus \mathbf{S}_{1,5}$ | $R_\pi, \mathbf{S}_{1,5}$ | $z_1, z_5 \in \mathbf{R}, z_2 = z_3 = z_4 = z_6 = 0$ | 2 |
| $\mathbf{D}_2^x \dotplus \mathbf{S}_{1,6}$ | $R_\pi, \tau_{x_1}, \mathbf{S}_{1,6}$ | $z_1 = z_6 \in \mathbf{R}, z_2 = z_3 = z_4 = z_5 = 0$ | 1 |
| $\mathbf{Z}_2^c \dotplus \mathbf{S}_{1,6}$ | $R_\pi, \mathbf{S}_{1,6}$ | $z_1, z_6 \in \mathbf{R}, z_2 = z_3 = z_4 = z_5 = 0$ | 2 |
| $\mathbf{Z}_6 \dotplus \mathbf{S}_{1,2,3}$ | $R_{\pi/3}, \mathbf{S}_{1,2,3}$ | $z_1 = z_2 = z_3 \in \mathbf{R}, z_4 = z_5 = z_6 = 0$ | 1 |
| $\mathbf{Z}_3 \dotplus \mathbf{S}_{1,2,3}$ | $R_{\pi/3}^2, \mathbf{S}_{1,2,3}$ | $z_1 = z_2 = z_3 \in \mathbf{C}, z_4 = z_5 = z_6 = 0$ | 2 |
| $\mathbf{Z}_2^c \dotplus \mathbf{S}_{1,2,3}$ | $R_\pi, \mathbf{S}_{1,2,3}$ | $z_1, z_2, z_3 \in \mathbf{R}, z_4 = z_5 = z_6 = 0$ | 3 |
| $\mathbf{S}_{1,2,3}$ | $\mathbf{S}_{1,2,3}$ | $z_1, z_2, z_3 \in \mathbf{C}, z_4 = z_5 = z_6 = 0$ | 6 |

[a] the generators of $\Gamma_h$ are given in equations 2.23-2.25, and $\kappa \in \mathbf{Z}_2$ is given in (2.33).
[b] $\tau_n \equiv R_{\pi/3}\tau_{x_1}$ is a reflection through the line containing the vector $\ell_1 - 2\ell_2$ (cf. Fig. 4).
[c] The generators of $\mathbf{S}^1$, $\mathbf{S}_{1,4}$, $\mathbf{S}_{1,5}$, $\mathbf{S}_{1,6}$, and $\mathbf{S}_{1,2,3}$ are given in Table 5.
[d] $\tau_m \equiv R_{\pi/3}^5 \tau_{x_1}$ is a reflection through the line containing the vector $\ell_1 + \ell_2$ (cf. Fig. 4).

Table 5: Generators of Subgroups $\mathbf{S}^1$ and $\mathbf{S}_*$ of $\Gamma_h$ and $\Gamma_h \dotplus \mathbf{Z}_2$.

| $\mathbf{S}^1/\mathbf{S}_*$ | Generators of $\mathbf{S}^1, \mathbf{S}_* \subset \Gamma_h$ | Generators of $\mathbf{S}^1, \mathbf{S}_* \subset \Gamma_h \dotplus \mathbf{Z}_2$ |
|---|---|---|
| $\mathbf{S}^1$ | $(Id, \beta s\,\ell_1 - \alpha s\,\ell_2)$, where $s \in \mathbf{R}$ | $((Id, \beta s\,\ell_1 + (\frac{1}{2\beta} - \alpha s)\,\ell_2), \kappa)$, where $s \in \mathbf{R}$ |
| $\mathbf{S}_{1,4}$ | $(Id, \frac{\alpha-\beta}{\alpha^2-2\alpha\beta}\,\ell_1 - \frac{1}{\alpha-2\beta}\,\ell_2)$, $(Id, \frac{-\beta}{\alpha^2-2\alpha\beta}\,\ell_1 + \frac{1}{\alpha-2\beta}\,\ell_2)$ | $((Id, \frac{\alpha-2\beta}{2(\alpha^2-2\alpha\beta)}\,\ell_1), \kappa)$, $((Id, \frac{1}{2(\alpha-2\beta)}\,(\ell_1 - 2\ell_2)), \kappa)$ |
| $\mathbf{S}_{1,5}$ | $(Id, \frac{\alpha}{\alpha^2-\beta^2}\,\ell_1 - \frac{\beta}{\alpha^2-\beta^2}\,\ell_2)$, $(Id, \frac{\beta}{\alpha^2-\beta^2}\,\ell_1 - \frac{\alpha}{\alpha^2-\beta^2}\,\ell_2)$ | $((Id, \frac{1}{2(\alpha-\beta)}\,(\ell_1 - \ell_2)), \kappa)$, $((Id, \frac{1}{2(\alpha+\beta)}\,(\ell_1 + \ell_2)), \kappa)$ |
| $\mathbf{S}_{1,6}$ | $(Id, \frac{1}{2\alpha-\beta}\,\ell_1 + \frac{\alpha-\beta}{2\alpha\beta-\beta^2}\,\ell_2)$, $(Id, \frac{-1}{2\alpha-\beta}\,\ell_1 + \frac{\alpha}{2\alpha\beta-\beta^2}\,\ell_2)$ | $((Id, \frac{1}{2\beta}\,\ell_2), \kappa)$, $((Id, \frac{1}{2(2\alpha-\beta)}\,(2\ell_1 - \ell_2)), \kappa)$. |
| $\mathbf{S}_{1,2,3}$ | $(Id, \frac{\alpha}{\alpha^2-\alpha\beta+\beta^2}\,\ell_1 - \frac{\alpha-\beta}{\alpha^2-\alpha\beta+\beta^2}\,\ell_2)$, $(Id, \frac{\beta}{\alpha^2-\alpha\beta+\beta^2}\,\ell_1 - \frac{\alpha}{\alpha^2-\alpha\beta+\beta^2}\,\ell_2)$ | $((Id, \frac{\alpha}{\alpha^2-\alpha\beta+\beta^2}\,\ell_1 - \frac{\alpha-\beta}{\alpha^2-\alpha\beta+\beta^2}\,\ell_2), Id)$, $((Id, \frac{\beta}{\alpha^2-\alpha\beta+\beta^2}\,\ell_1 - \frac{\alpha}{\alpha^2-\alpha\beta+\beta^2}\,\ell_2), Id)$ |
| $\mathbf{S}_{1,2}$ | not applicable | $((Id, \frac{\alpha+\beta}{2(\alpha^2-\alpha\beta+\beta^2)}\,\ell_1 - \frac{2\alpha-\beta}{2(\alpha^2-\alpha\beta+\beta^2)}\,\ell_2), \kappa)$, $((Id, \frac{\alpha-\beta}{2(\alpha^2-\alpha\beta+\beta^2)}\,\ell_1 + \frac{\beta}{2(\alpha^2-\alpha\beta+\beta^2)}\,\ell_2), \kappa)$ |

Table 6: Isotropy Subgroups $\Sigma$ (up to conjugacy) of $\Gamma_h \dotplus \mathbf{Z}_2$. Also see Table 4.

| $\Sigma$ | Generators of $\Sigma$ | Fix($\Sigma$) | Dim(Fix($\Sigma$)) |
|---|---|---|---|
| $\mathbf{D}_6$ | $((R_{\pi/3}, 0), \kappa), ((\tau_{x_1}, 0), \kappa)$ | $z_1 = z_2 = z_3 = -z_4 = -z_5 = -z_6 \in \mathbf{R}i$ | 1 |
| $\mathbf{D}_6$ | $((R_{\pi/3}, 0), \kappa), ((\tau_{x_1}, 0), Id)$ | $z_1 = z_2 = z_3 = z_4 = z_5 = z_6 \in \mathbf{R}i$ | 1 |
| $\mathbf{D}_6$ | $((R_{\pi/3}, 0), Id), ((\tau_{x_1}, 0), \kappa)$ | $z_1 = z_2 = z_3 = -z_4 = -z_5 = -z_6 \in \mathbf{R}$ | 1 |
| $\mathbf{D}_3$ | $((R^2_{\pi/3}, 0), Id), ((\tau_{x_1}, 0), \kappa)$ | $z_1 = z_2 = z_3 = -z_4 = -z_5 = -z_6 \in \mathbf{C}$ | 2 |
| $\mathbf{D}_3$ | $((R^2_{\pi/3}, 0), Id), ((\tau_n, 0), \kappa)$ | $z_1 = z_2 = z_3 = -\overline{z}_4 = -\overline{z}_5 = -\overline{z}_6 \in \mathbf{C}$ | 2 |
| $\mathbf{Z}_6$ | $((R_{\pi/3}, 0), \kappa)$ | $z_1 = z_2 = z_3, z_4 = z_5 = z_6 \in \mathbf{R}i$ | 2 |
| $\mathbf{D}_2$ | $((\tau_{x_1}, 0), \kappa), ((R_\pi, 0), \kappa)$ | $z_1 = -z_6, z_2 = -z_5, z_3 = -z_4 \in \mathbf{R}i$ | 3 |
| $\mathbf{D}_2$ | $((\tau_{x_1}, 0), \kappa), ((R_\pi, 0), Id)$ | $z_1 = -z_6, z_2 = -z_5, z_3 = -z_4 \in \mathbf{R}$ | 3 |
| $\mathbf{D}_2$ | $((\tau_{x_1}, 0), Id), ((R_\pi, 0), \kappa)$ | $z_1 = z_6, z_2 = z_5, z_3 = z_4 \in \mathbf{R}i$ | 3 |
| $\mathbf{Z}_2$ | $((\tau_{x_1}, 0), \kappa)$ | $z_1 = -\overline{z}_6, z_2 = -\overline{z}_5, z_3 = \overline{z}_4 \in \mathbf{C}$ | 6 |
| $\mathbf{Z}_2$ | $((\tau_n, 0), \kappa)$ | $z_1 = -\overline{z}_4, z_2 = -\overline{z}_6, z_3 = \overline{z}_5 \in \mathbf{C}$ | 6 |
| $\mathbf{Z}_2$ | $((R_\pi, 0), \kappa)$ | $z_1, z_2, z_3, z_4, z_5, z_6 \in \mathbf{R}i$ | 6 |
| $\mathbf{Z}_6 \dotplus \mathbf{S}_{1,2,3}$ | $((R_{\pi/3}, 0), \kappa), \mathbf{S}_{1,2,3}$ [a] | $z_1 = z_2 = z_3 \in \mathbf{R}i, z_4 = z_5 = z_6 = 0$ | 1 |
| $\mathbf{Z}_2 \dotplus \mathbf{S}_{1,2,3}$ | $((R_\pi, 0), \kappa), \mathbf{S}_{1,2,3}$ | $z_1, z_2, z_3 \in \mathbf{R}i, z_4 = z_5 = z_6 = 0$ | 3 |
| $\mathbf{Z}_2^c \dotplus \mathbf{S}_{1,2}$ | $((R_\pi, 0), Id), \mathbf{S}_{1,2}$ [a] | $z_1, z_2 \in \mathbf{R}, z_3 = z_4 = z_5 = z_6 = 0$ | 2 |

[a] The generators of $\mathbf{S}_{1,2}$ and $\mathbf{S}_{1,2,3}$ are given in Table 5.

### 3.2 Hidden symmetries.

In the case of the hexagonal lattice, we must take certain *hidden symmetries* into account in our analysis. The hidden symmetries are elements of the full Euclidean group that are not present when the PDE is restricted to a periodic domain. Nonetheless, these hidden symmetries are manifest in certain fixed point subspaces of $\mathbf{C}^6$ and put restrictions on the normal form of the bifurcation problem (2.16). We refer the reader to Crawford [6] for a detailed treatment of hidden Euclidean symmetries in $\Gamma_s$ mode interaction problems.

For the 12-dimensional representation of $\Gamma_h$, the hidden Euclidean symmetries are generated by a reflection through the line containing the vector $\beta \ell_1 - \alpha \ell_2$, denoted by $\tilde{\tau}_{x_1} \in E(2)$. This reflection acts on $\mathbf{z} \in \text{Fix}(\mathbf{S}_{1,2,3}) \cong \{(z_1, z_2, z_3, 0, 0, 0) : z_i \in \mathbf{C}\}$ in the same manner that $\tau_{x_1} \in \mathbf{D}_6$ acts on the six-dimensional representation of $\Gamma_h$ (*cf.* equation 2.28). Specifically,

$$\tilde{\tau}_{x_1}(z_1, z_2, z_3, 0, 0, 0) = (\overline{z}_1, \overline{z}_3, \overline{z}_2, 0, 0, 0). \tag{3.1}$$

Once the hidden reflection $\tilde{\tau}_{x_1} \in E(2)$ is included we can reformulate the $\Gamma_h$-equivariant bifurcation problem (2.16) restricted to the subspace $\{(z_1, z_2, z_3, 0, 0, 0) : z_i \in \mathbf{C}\}$ as a bifurcation problem for the six-dimensional representation of $\Gamma_h$ in Table 2. This is consistent with the observation that when the three amplitudes $z_4, z_5, z_6$ are zero, the solutions are all periodic with respect to a finer lattice; the basis vectors for the dual to this finer lattice are $\mathbf{K}_1$ and $\mathbf{K}_2$.

Only in the case of $\Gamma_h + \mathbf{Z}_2$ does the inclusion of the hidden reflection ensure the existence of an additional axial planform. Specifically, the line containing $\mathbf{z} = (1, 1, 0, 0, 0, 0)$ in the two-dimensional fixed point space $\text{Fix}(\mathbf{Z}_2^c \dot{+} \mathbf{S}_{1,2}) \cong \{(x_1, x_2, 0, 0, 0, 0) : x_i \in \mathbf{R}\}$ is fixed by $\tilde{\tau}_{x_1} R_{\pi/3}$.

For the 8-dimensional representation of $\Gamma_s$, there is a hidden reflection that acts on $\mathbf{z} \in \{(z_1, z_2, 0, 0) : z_i \in \mathbf{C}\}$ as follows:

$$\tilde{\tau}_{x_1}(z_1, z_2, 0, 0) = (z_1, \overline{z}_2, 0, 0). \tag{3.2}$$

However, including this hidden symmetry does not change the normal form of the bifurcation problem nor does it lead to new axial planforms. (Note that every point in the subspace $\{(z_1, z_2, 0, 0) : z_i \in \mathbf{C}\}$ is on the group orbit of one in $\text{Fix}(\mathbf{Z}_4 \dot{+} \mathbf{S}_{1,2}) \cong \{(x_1, x_2, 0, 0) : x_i \in \mathbf{R}\}$, and that the hidden reflection $\tilde{\tau}_{x_1}$ acts trivially on points in $\text{Fix}(\mathbf{Z}_4 \dot{+} \mathbf{S}_{1,2})$.)

### 3.3 Axial planforms.

In this section we present examples of the planforms associated with the axial subgroups. Specifically, we present grey scale plots of the function $v(\mathbf{x})$ in (2.14) for a representative point $\mathbf{z} \in \text{Fix}(\Sigma)$. We do this for each conjugacy class of subgroups $\Sigma \subset \Gamma$ that fix one-dimensional subspaces.

In the case of the square lattice, there are six conjugacy classes of isotropy subgroups that fix one-dimensional subspaces of $\mathbf{C}^4$. These are listed in Table 7. Figure 5 presents the associated axial planforms $v$ in the case that $(\alpha, \beta) = (2, 1)$. The rhombic, super square and anti-square states depend on $(\alpha, \beta)$. The rolls and simple squares are, up to scaling of the spatial variable $\mathbf{x}$, the same for each 8-dimensional representation of $\Gamma_s$.

Table 7: Axial Planforms for 8-dimensional representations of $\Gamma_s$.

| Name | Nomenclature | representative point $\mathbf{z} \in \mathrm{Fix}(\Sigma) \subset \mathbf{C}^4$ |
|---|---|---|
| Rolls [a] | R | $\mathbf{z} = (1, 0, 0, 0)$ |
| Simple Squares | SiS | $\mathbf{z} = (1, 1, 0, 0)$ |
| Rhombs [b] | $\mathrm{Rh}_{s1,\alpha,\beta}$ | $\mathbf{z} = (1, 0, 1, 0)$ |
| Rhombs | $\mathrm{Rh}_{s2,\alpha,\beta}$ | $\mathbf{z} = (1, 0, 0, 1)$ |
| Super Squares [c] | $\mathrm{SuS}_{\alpha,\beta}$ | $\mathbf{z} = (1, 1, 1, 1)$ |
| Anti-squares | $\mathrm{AS}_{\alpha,\beta}$ | $\mathbf{z} = (1, 1, -1, -1)$ |

[a] This state is also called "stripes".
[b] Rhombs are also called "rectangles".
[c] Super squares are called simply "squares" in Dionne and Golubitsky [7].

In the hexagonal lattice case there are six different axial planforms for the twelve-dimensional representations of $\Gamma = \Gamma_h$ and eleven for $\Gamma = \Gamma_h \dot{+} \mathbf{Z}_2$, including the state $\mathrm{Rh}_{h0}$ determined by taking into account the hidden reflection symmetry. These are listed in Table 8 and depicted in Figures 6 and 7. In the case of $\Gamma_h \dot{+} \mathbf{Z}_2$, the SiH$^-$ (SuH$^-$) branch of simple (super) hexagons is on the group orbit of the SiH$^+$ (SuH$^+$) branch since $\kappa(\mathbf{z}) = -\mathbf{z}$. Note that the only states that are the same (after rescaling $\mathbf{x}$) for every value of $\alpha$ and $\beta$ are the rolls, the rhombs $\mathrm{Rh}_{h0}$, the simple hexagons, and the simple triangles. Rolls are the only state that is common to both the square and the hexagonal lattices.

Only the super- and the anti-states of Tables 7 and 8 are characterized by translation free isotropy subgroups. Hence these are the only axial planforms with (smallest) periodicity determined by $\ell_1$ and $\ell_2$. All of the other axial planforms are periodic on a finer square, hexagonal, or rhombic lattice. In particular, the wavelength of their periodicity is $1/k_c$, where $k_c = \sqrt{\alpha^2 + \beta^2}$ for square lattice states and $k_c = \sqrt{\alpha^2 + \beta^2 - \alpha\beta}$ for hexagonal lattice states. Note that while the periodicity of super- and anti-states is given by $|\ell_1| = |\ell_2| \gg 1/k_c$, the lengthscale $1/k_c$ is also evident in the patterns. This lengthscale shows up as small scale structure in the patterns; compare, for example, simple hexagons with super hexagons in Figure 6.

The axial planforms, rhombs $\mathrm{Rh}_{h0}$, simple squares, and simple triangles/hexagons, are listed in Tables 3, 4 and 6 as possessing $\mathbf{Z}_n$ symmetry, where $n = 2, 4, 6$, respectively. However, on the appropriate (finer) square or hexagonal lattice that supports these planforms, they are $\mathbf{D}_n$-symmetric. Once we include the hidden reflection symmetry $\widetilde{\tau}_{x_1}$ we recover the full $\mathbf{D}_n$ symmetry that is manifest in Figures 5-7.

Finally, we note that there is a countable set of rhombs that are periodic on square or hexagonal lattices. In Table 9 we characterize the rhombs on the square and hexagonal lattices in two different ways. We give the angle between the wave vectors associated with the critical modes, e.g. the angle between $\mathbf{K}_1$ and $\mathbf{K}_3$ for $\mathrm{Rh}_{s1,\alpha,\beta}$. We also give the aspect ratio of the rectangles evident in the rhomb patterns in Figures 5-7.

Table 8: Axial Planforms for 12-dimensional representations of $\Gamma_h$ and $\Gamma_h + \mathbf{Z}_2$.

| Name | Nomenclature | representative point $\mathbf{z} \in \text{Fix}(\Sigma) \subset \mathbf{C}^6$ | $\Gamma$ |
|---|---|---|---|
| Rolls | R | $\mathbf{z} = (1,0,0,0,0,0)$ | $\Gamma_h$ and $\Gamma_h + \mathbf{Z}_2$ |
| Simple Hexagons [a] | SiH$^\pm$ | $\mathbf{z} = \pm(1,1,1,0,0,0)$ | $\Gamma_h$ and $\Gamma_h + \mathbf{Z}_2$ |
| Simple Triangles [b] | SiT | $\mathbf{z} = (i,i,i,0,0,0)$ | $\Gamma_h + \mathbf{Z}_2$ only |
| Rhombs [c] | Rh$_{h0}$ | $\mathbf{z} = (1,1,0,0,0,0)$ | $\Gamma_h + \mathbf{Z}_2$ only |
| Rhombs | Rh$_{h1,\alpha,\beta}$ | $\mathbf{z} = (1,0,0,1,0,0)$ | $\Gamma_h$ and $\Gamma_h + \mathbf{Z}_2$ |
| Rhombs | Rh$_{h2,\alpha,\beta}$ | $\mathbf{z} = (1,0,0,0,1,0)$ | $\Gamma_h$ and $\Gamma_h + \mathbf{Z}_2$ |
| Rhombs | Rh$_{h3,\alpha,\beta}$ | $\mathbf{z} = (1,0,0,0,0,1)$ | $\Gamma_h$ and $\Gamma_h + \mathbf{Z}_2$ |
| Super Hexagons [d] | SuH$^\pm_{\alpha,\beta}$ | $\mathbf{z} = \pm(1,1,1,1,1,1)$ | $\Gamma_h$ and $\Gamma_h + \mathbf{Z}_2$ |
| Anti-hexagons | AH$_{\alpha,\beta}$ | $\mathbf{z} = (1,1,1,-1,-1,-1)$ | $\Gamma_h + \mathbf{Z}_2$ only |
| Super Triangles | SuT$_{\alpha,\beta}$ | $\mathbf{z} = (i,i,i,i,i,i)$ | $\Gamma_h + \mathbf{Z}_2$ only |
| Anti-triangles | AT$_{\alpha,\beta}$ | $\mathbf{z} = (i,i,i,-i,-i,-i)$ | $\Gamma_h + \mathbf{Z}_2$ only |

[a] Simple hexagons are simply "hexagons" in [13].
[b] Simple triangles are called "right triangles" in [13].
[c] Golubitsky *et al.* call this rhombs state the "patchwork quilt"[13].
[d] Super hexagons are called "hexagons" in Dionne and Golubitsky[7].

Table 9: Characterization of the rhombs.

| Lattice | Rhombs | aspect ratio | angle |
|---|---|---|---|
| Square | Rh$_{s1,\alpha,\beta}$ | $\frac{\alpha-\beta}{\alpha+\beta}$ | $\cos^{-1}\left(\frac{2\alpha\beta}{\alpha^2+\beta^2}\right)$ |
| Square | Rh$_{s2,\alpha,\beta}$ | $\frac{\beta}{\alpha}$ | $\cos^{-1}\left(\frac{\alpha^2-\beta^2}{\alpha^2+\beta^2}\right)$ |
| Hexagonal | Rh$_{h1,\alpha,\beta}$ | $\frac{2\beta-\alpha}{\sqrt{3}\,\alpha}$ | $\cos^{-1}\left(\frac{\alpha^2+2\alpha\beta-2\beta^2}{2(\alpha^2-\alpha\beta+\beta^2)}\right)$ |
| Hexagonal | Rh$_{h2,\alpha,\beta}$ | $\frac{\sqrt{3}\,(\alpha-\beta)}{\alpha+\beta}$ | $\cos^{-1}\left(-\frac{\alpha^2-4\alpha\beta+\beta^2}{2(\alpha^2-\alpha\beta+\beta^2)}\right)$ |
| Hexagonal | Rh$_{h3,\alpha,\beta}$ | $\frac{2\alpha-\beta}{\sqrt{3}\,\beta}$ | $\cos^{-1}\left(\frac{2\alpha^2-2\alpha\beta-\beta^2}{2(\alpha^2-\alpha\beta+\beta^2)}\right)$ |
| Hexagonal | Rh$_{h0}$ | $\frac{1}{\sqrt{3}}$ | $\frac{\pi}{3}$ |

# 4 Stability Results: Square Lattice.

In this section we compute the linear stability, at bifurcation, of the six axial planforms listed in Table 7. We do this within the center manifold framework of a general $\Gamma_s$-equivariant bifurcation problem $\dot{\mathbf{z}} = \mathbf{g}(\mathbf{z}, \lambda)$, $\mathbf{g} : \mathbf{C}^4 \times \mathbf{R} \to \mathbf{C}^4$. An equilibrium solution branch $\mathbf{z}_\lambda$ of the bifurcation problem is linearly (orbitally) stable if all eigenvalues of the Jacobian matrix $\mathbf{Dg}(\mathbf{z}_\lambda, \lambda)$, not forced by symmetry to be zero, have negative real part for $\lambda$ sufficiently close to zero. If any eigenvalue has positive real part then the planform is unstable.

We begin by determining the restrictions that symmetry imposes on the $8 \times 8$ (real) Jacobian matrix $\mathbf{Dg}$ evaluated on each of the axial solution branches. This allows us to determine the eigenvalues of the Jacobian matrix in terms of certain entries. We then determine the Taylor expansion, about $\mathbf{z} = \mathbf{0}$, of a smooth $\Gamma_s$-equivariant vector field to high enough order to determine the sign of the real part of each eigenvalue.

## 4.1 Commuting linear maps.

Let $g_j = g_j^r + i g_j^i$, $j = 1, ..., 4$, where the $r$ and $i$ superscripts specify the real and imaginary parts, and let $z_j = x_j + i y_j$, where $x_j$ and $y_j$ are the real and imaginary parts of $z_j$, respectively. Thus $\dot{x}_j = g_j^r$ and $\dot{y}_j = g_j^i$, $j = 1, ..., 4$, are the eight components of the equivariant vector field over the reals. Table 10 gives the eigenvalues of the Jacobian matrix evaluated on each of the axial solution branches. We use two different approaches to determining the eigenvalues.

The first approach exploits the observation that the Jacobian matrix evaluated on a solution branch $\mathbf{z}_\lambda$ commutes with each element $\sigma \in \Sigma_{\mathbf{z}_\lambda}$. For example, the Jacobian matrix $\mathbf{Dg}$ evaluated on the rolls solution branch commutes with the linear transformations associated with the generators of $\Sigma = \mathbf{Z}_2^c \dot{+} \mathbf{S}^1$ in Table 3, namely

$$\mathbf{z} \to \overline{\mathbf{z}} \tag{4.1}$$

and

$$\mathbf{z} \to (z_1, e^{2\pi i(\alpha^2+\beta^2)s} z_2, e^{2\pi i(\alpha^2-\beta^2)s} z_3, e^{4\pi i \alpha \beta s} z_4), \quad s \in \mathbf{R}. \tag{4.2}$$

We choose an ordering of the coordinates of $\mathbf{R}^8$ to be $(x_1, x_2, x_3, x_4, y_1, y_2, y_3, y_4)$. It follows from the observation that $\mathbf{Dg}$, evaluated on the rolls solution branch, must commute with the above transformations that $\mathbf{Dg}$ is diagonal and three of the eigenvalues have multiplicity two. Moreover, the group orbit of rolls is one-dimensional so there is a zero eigenvalue associated with translation along the group orbit. The null direction is determined by computing the tangent vector to the group orbit, e.g.,

$$\left. \frac{\partial}{\partial \theta_1} \right|_{\Theta=\mathbf{0}} \Theta(x_1, 0, 0, 0) = (-2\pi i \alpha x_1, 0, 0, 0), \ x_1 \in \mathbf{R}, \tag{4.3}$$

where the action of $\Theta$ on $\mathbf{C}^4$ is given by (2.25). It follows that $(0, 0, 0, 0, 1, 0, 0, 0)^\top$ is a null eigenvector of $\mathbf{Dg}$ and that the eigenvalue $\frac{\partial g_1^i}{\partial y_1}$ is zero.

The second approach to computing the eigenvalues relies on forming the *isotypic decomposition* of $\mathbf{C}^4$ for the isotropy subgroup $\Sigma_{\mathbf{z}_\lambda}$ of a solution $\mathbf{z}_\lambda$. This decomposition determines coordinates that block-diagonalize $D\mathbf{g}$ [12]. The isotypic decomposition proceeds by first decomposing $\mathbf{C}^4$ into $\Sigma$-irreducible subspaces $V_j$ so that $\mathbf{C}^4 = V_0 \oplus V_1 \oplus \cdots \oplus V_\ell$. (Recall that a representation is $\Sigma$-irreducible if the only $\Sigma$-invariant subspace of $V_j$, other than $\{\mathbf{0}\}$, is $V_j$ itself.) The isotypic components $W_j$ are then formed by combining the irreducible subspaces that are $\Sigma$-*isomorphic*. (Two $\Sigma$-irreducible subspaces are $\Sigma$-isomorphic if there exists a linear isomorphic mapping between them which commutes with the action of $\Sigma$.) The isotypic decomposition is $\mathbf{C}^4 = W_0 \oplus W_1 \oplus \cdots \oplus W_k$, $k \leq \ell$, where the $W_j$ are uniquely determined.

As an example of the second approach, which block-diagonalizes $D\mathbf{g}$, we determine the isotypic components of $\mathbf{C}^4$ for $\Sigma = \mathbf{D}_4[R_{\pi/2}, \tau_{x_1}]$ that applies to the super squares state. A decomposition of $\mathbf{C}^4$ into irreducible subspaces is

$$\begin{aligned}\mathbf{C}^4 &= \mathbf{R}\{(1,1,1,1)\} \oplus \mathbf{R}\{(1,1,-1,-1)\} \oplus \mathbf{R}\{(1,-1,1,-1)\} \oplus \mathbf{R}\{(1,-1,-1,1)\} \\ &\oplus \mathbf{R}\{(i,i,i,i),(-i,i,-i,i)\} \oplus \mathbf{R}\{(i,0,0,i),(0,i,-i,0)\}\,.\end{aligned} \quad (4.4)$$

The one-dimensional subspaces are non-isomorphic representations of $\mathbf{D}_4$, whereas the two-dimensional subspaces are $\mathbf{D}_4$-isomorphic and hence are in the same isotypic component (identify $(i,i,i,i)$ with $(i,0,0,i)$ and $(-i,i,-i,i)$ with $(0,i,-i,0)$). From the one-dimensional isotypic components we can immediately determine four of the eigenvalues of $D\mathbf{g}$ evaluated on the super squares solution branch; these are

$$\begin{aligned}&\frac{\partial g_1^r}{\partial x_1} + \frac{\partial g_1^r}{\partial x_2} + \frac{\partial g_1^r}{\partial x_3} + \frac{\partial g_1^r}{\partial x_4}\,, \quad \frac{\partial g_1^r}{\partial x_1} + \frac{\partial g_1^r}{\partial x_2} - \frac{\partial g_1^r}{\partial x_3} - \frac{\partial g_1^r}{\partial x_4}\,, \\ &\frac{\partial g_1^r}{\partial x_1} - \frac{\partial g_1^r}{\partial x_2} + \frac{\partial g_1^r}{\partial x_3} - \frac{\partial g_1^r}{\partial x_4}\,, \quad \frac{\partial g_1^r}{\partial x_1} - \frac{\partial g_1^r}{\partial x_2} - \frac{\partial g_1^r}{\partial x_3} + \frac{\partial g_1^r}{\partial x_4}\,.\end{aligned} \quad (4.5)$$

Symmetry places additional restrictions on the matrix obtained by restricting $D\mathbf{g}$ to the four-dimensional isotypic component $W_4$. Specifically, $D\mathbf{g}|_{W_4}$ commutes with the action of $\mathbf{D}_4$ on $W_4 \cong \mathbf{R}^4$, where $W_4$ is the direct sum of two isomorphic $\mathbf{D}_4$-absolutely irreducible subspaces. (Recall that a representation of a group $\Gamma$ acts *absolutely irreducibly* on a space $V$ if the only linear maps on $V$ commuting with $\Gamma$ are multiples of the identity.) It follows that $\mathbf{A} = D\mathbf{g}|_{W_4}$ has the form

$$\mathbf{A} = \begin{pmatrix} a\mathbf{I}_2 & b\mathbf{I}_2 \\ c\mathbf{I}_2 & d\mathbf{I}_2 \end{pmatrix}, \quad (4.6)$$

where $\mathbf{I}_2$ is the $2 \times 2$ identity matrix and $a,b,c,d \in \mathbf{R}$. Each eigenvalue in this matrix has multiplicity two. Moreover, two of the eigenvalues must be zero because the group orbit of super squares is two-dimensional. Thus the eigenvalues are determined by simply computing $\text{Tr}(\mathbf{A})$, which, in terms of real coordinates, is

$$\text{Tr}(\mathbf{A}) = \frac{\partial g_1^i}{\partial y_1} + \frac{\partial g_2^i}{\partial y_2} + \frac{\partial g_3^i}{\partial y_3} + \frac{\partial g_4^i}{\partial y_4}\,. \quad (4.7)$$

This can be further simplified by noting that since $D\mathbf{g}$ commutes with the transformations $R_{\pi/2}$ and $\tau_{x_1}$

$$\frac{\partial g_1^i}{\partial y_1} = \frac{\partial g_2^i}{\partial y_2} = \frac{\partial g_3^i}{\partial y_3} = \frac{\partial g_4^i}{\partial y_4} \quad (4.8)$$

on the super squares solution branch. Thus the repeated eigenvalue, $\frac{1}{2}\text{Tr}(\mathbf{A})$, is simply $2\frac{\partial g^i}{\partial y_1}$.

The isotypic decomposition of $\mathbf{C}^4$ for $\Sigma = \widetilde{\mathbf{D}}_4$ is the same as the $\mathbf{D}_4$-isotypic decomposition presented above. The details of the computations of the eigenvalues for the remaining axial planforms are omitted. Note that symmetry considerations alone determine that the eigenvalues of $\mathbf{Dg}$ are real for all of the axial planforms.

## 4.2  Normal form for $\mathbf{D}_4 \dotplus \mathbf{T}^2$.

In this section we determine the Taylor expansion of the equivariant bifurcation problem (2.16) to sufficient order to determine the signs of the eigenvalues of $\mathbf{Dg}$ given in Table 10.

The equivariance condition (2.17) for $\Gamma = \Gamma_s$ is satisfied if (see, for example, Appendix A.3 in [6])

$$\begin{aligned}
\dot{z}_1 &= g_1(z_1, z_2, z_3, z_4) \\
\dot{z}_2 &= g_1(z_2, \overline{z}_1, z_4, \overline{z}_3) \\
\dot{z}_3 &= g_1(z_3, \overline{z}_4, z_1, \overline{z}_2) \\
\dot{z}_4 &= g_1(z_4, z_3, z_2, z_1) ,
\end{aligned} \quad (4.9)$$

where

$$\overline{g_1(\mathbf{z})} = g_1(\overline{\mathbf{z}}) , \quad (4.10)$$

and

$$\Theta\left(\overline{z}_1 g_1(\mathbf{z})\right) = \overline{z}_1 g_1(\mathbf{z}) , \text{ for all } \Theta \in \mathbf{T}^2 . \quad (4.11)$$

Equivariance with respect to $\mathbf{D}_4 \subset \Gamma_s$ is guaranteed by conditions (4.9) and (4.10), while equivariance with respect to $\mathbf{T}^2 \subset \Gamma_s$ is equivalent to condition (4.11), *i.e.*, to $\overline{z}_1 g_1(\mathbf{z})$ being an invariant function of the $\mathbf{T}^2$-action. The problem of determining the general form of the $\Gamma_s$-equivariant vector field then reduces to one of finding the most general function $g_1(\mathbf{z})$ that satisfies (4.10) and (4.11).

We assume that local to the bifurcation point $\mathbf{z} = \mathbf{0}$, $\lambda = 0$ the equivariant normal form can be expanded in a Taylor series about $\mathbf{z} = \mathbf{0}$. We proceed by determining the general form of a $\mathbf{T}^2$-invariant function $h = \overline{z}_1 g_1(\mathbf{z})$. Let

$$h(\mathbf{z}) = \sum_{k_i \geq 0} a_{\mathbf{k}} \, z_1^{k_1} \overline{z}_1^{k_2} z_2^{k_3} \overline{z}_2^{k_4} z_3^{k_5} \overline{z}_3^{k_6} z_4^{k_7} \overline{z}_4^{k_8} , \quad (4.12)$$

where $k_2 > 0$. The condition (4.10) determines that the coefficients $a_{\mathbf{k}}$ are real. Given the action (2.25) of $\mathbf{T}^2$ on $\mathbf{C}^4$, the $\mathbf{T}^2$-invariance of $h$ determines that $a_{\mathbf{k}} = 0$ unless

$$\begin{aligned}
(k_1 - k_2)(\alpha\theta_1 + \beta\theta_2) + (k_3 - k_4)(-\beta\theta_1 + \alpha\theta_2) & \\
+ (k_5 - k_6)(\beta\theta_1 + \alpha\theta_2) + (k_7 - k_8)(-\alpha\theta_1 + \beta\theta_2) &= 0
\end{aligned} \quad (4.13)$$

for all $(\theta_1, \theta_2) \in \mathbf{T}^2$. Clearly, $k_1 = k_2$, $k_3 = k_4$, $k_5 = k_6$, and $k_7 = k_8$ is a solution to equation (4.13), which yields the $\mathbf{T}^2$-invariants $|z_j|^2$, $j = 1, 2, 3, 4$. In the following we factor out all powers of $|z_j|^2$ from the translation invariant monomials $P_{\mathbf{k}}(z) = z_1^{k_1} \bar{z}_1^{k_2} z_2^{k_3} \bar{z}_2^{k_4} z_3^{k_5} \bar{z}_3^{k_6} z_4^{k_7} \bar{z}_4^{k_8}$, and only consider monomials of the form $z_1^m z_2^n z_3^p z_4^q$, where we adopt the convention that $z_j^m \equiv \bar{z}_j^{|m|}$ if $m < 0$. In this way we reduce the problem of finding all $\mathbf{T}^2$-invariant monomials to one of finding $m, n, p, q \in \mathbf{Z}$ such that

$$m(\alpha\theta_1 + \beta\theta_2) + n(-\beta\theta_1 + \alpha\theta_2) + p(\beta\theta_1 + \alpha\theta_2) + q(-\alpha\theta_1 + \beta\theta_2) = 0. \tag{4.14}$$

Table 10: Eigenvalues and their multiplicities for axial planforms associated with 8-dim. representations of $\Gamma_s$.

| Axial Planform | Eigenvalues |
|---|---|
| Rolls<br>$\mathbf{z} = x(1,0,0,0), x \in \mathbf{R}$<br>$\Sigma = \mathbf{Z}_2^c \dot{+} \mathbf{S}^1$ | $\frac{\partial g_1^r}{\partial x_1}$, $\frac{\partial g_2^r}{\partial x_2}$ (mult. 2), $\frac{\partial g_3^r}{\partial x_3}$ (mult. 2), $\frac{\partial g_4^r}{\partial x_4}$ (mult. 2), 0 |
| Simple Squares<br>$\mathbf{z} = x(1,1,0,0), x \in \mathbf{R}$<br>$\Sigma = \mathbf{Z}_4 \dot{+} \mathbf{S}_{1,2}$ | $\frac{\partial g_1^r}{\partial x_1} + \frac{\partial g_1^r}{\partial x_2}$, $\frac{\partial g_1^r}{\partial x_1} - \frac{\partial g_1^r}{\partial x_2}$, $\frac{\partial g_3^r}{\partial x_3}$ (mult. 4), 0 (mult. 2) |
| Rhombs $(\mathrm{Rh}_{s1,\alpha,\beta})$<br>$\mathbf{z} = x(1,0,1,0), x \in \mathbf{R}$<br>$\Sigma = \mathbf{D}_2^d \dot{+} \mathbf{S}_{1,3}$ | $\frac{\partial g_1^r}{\partial x_1} + \frac{\partial g_1^r}{\partial x_3}$, $\frac{\partial g_1^r}{\partial x_1} - \frac{\partial g_1^r}{\partial x_3}$, $\frac{\partial g_2^r}{\partial x_2}$ (mult. 4), 0 (mult. 2) |
| Rhombs $(\mathrm{Rh}_{s2,\alpha,\beta})$<br>$\mathbf{z} = x(1,0,0,1), x \in \mathbf{R}$<br>$\Sigma = \mathbf{D}_2^x \dot{+} \mathbf{S}_{1,4}$ | $\frac{\partial g_1^r}{\partial x_1} + \frac{\partial g_1^r}{\partial x_4}$, $\frac{\partial g_1^r}{\partial x_1} - \frac{\partial g_1^r}{\partial x_4}$, $\frac{\partial g_2^r}{\partial x_2}$ (mult. 4), 0 (mult. 2) |
| Super Squares<br>$\mathbf{z} = x(1,1,1,1), x \in \mathbf{R}$<br>$\Sigma = \mathbf{D}_4$ | $\frac{\partial g_1^r}{\partial x_1} + \frac{\partial g_1^r}{\partial x_2} + \frac{\partial g_1^r}{\partial x_3} + \frac{\partial g_1^r}{\partial x_4}$, $\frac{\partial g_1^r}{\partial x_1} + \frac{\partial g_1^r}{\partial x_2} - \frac{\partial g_1^r}{\partial x_3} - \frac{\partial g_1^r}{\partial x_4}$,<br>$\frac{\partial g_1^r}{\partial x_1} - \frac{\partial g_1^r}{\partial x_2} + \frac{\partial g_1^r}{\partial x_3} - \frac{\partial g_1^r}{\partial x_4}$, $\frac{\partial g_1^r}{\partial x_1} - \frac{\partial g_1^r}{\partial x_2} - \frac{\partial g_1^r}{\partial x_3} + \frac{\partial g_1^r}{\partial x_4}$,<br>$2\frac{\partial g_1^i}{\partial y_1}$ (mult. 2), 0 (mult. 2) |
| Anti-Squares<br>$\mathbf{z} = x(1,1,-1,-1), x \in \mathbf{R}$<br>$\Sigma = \widetilde{\mathbf{D}}_4$ | Same as Super Squares |

Since $\theta_1$ and $\theta_2$ are independent this requires

$$(m - q)\alpha - (n - p)\beta = 0, \qquad (n + p)\alpha + (m + q)\beta = 0. \tag{4.15}$$

Furthermore, since $(\alpha, \beta) = 1$, this implies

$$m - q = j\beta, \qquad n - p = j\alpha, \tag{4.16}$$
$$n + p = k\beta, \qquad m + q = -k\alpha,$$

where $j, k \in \mathbf{Z}$. Solving for $m, n, p$ and $q$ gives

$$m = -\frac{1}{2}(k\alpha - j\beta), \qquad n = \frac{1}{2}(j\alpha + k\beta), \tag{4.17}$$
$$p = -\frac{1}{2}(j\alpha - k\beta), \qquad q = -\frac{1}{2}(k\alpha + j\beta),$$

Since $(\alpha, \beta) = 1$ and $\alpha$ and $\beta$ are not both odd, the system of equations (4.17) has no (nontrivial) solution if any two of $m, n, p$ or $q$ are zero. It then follows that both $k$ and $j$ are even and equations (4.17) may be replaced by

$$m = -(k'\alpha - j'\beta), \qquad n = (j'\alpha + k'\beta), \tag{4.18}$$
$$p = -(j'\alpha - k'\beta), \qquad q = -(k'\alpha + j'\beta),$$

where $j', k' \in \mathbf{Z}$. If $j' = 0, k' \neq 0$ or $j' \neq 0, k' = 0$ then (4.18) yields the translation invariant monomials

$$\bar{z}_1^\alpha z_2^\beta z_3^\beta \bar{z}_4^\alpha, \quad z_1^\beta z_2^\alpha \bar{z}_3^\alpha \bar{z}_4^\beta, \tag{4.19}$$

and their complex conjugates.

The monomials (4.19) are order $2(\alpha + \beta)$. We now show that these are the lowest order (nontrivial) translation invariant monomials of the form $z_1^m z_2^n z_3^p z_4^q$. To do this we consider the remaining cases for which $j'k' \neq 0$ in 4.18:

1. $j' > 0, k' > 0$
2. $j' > 0, k' < 0$
3. $j' < 0, k' > 0$
4. $j' < 0, k' < 0$.

In case 1, since no solution to (4.18) exists if two of $m, n, p$ and $q$ are zero, the order of the invariant is

$$\begin{aligned} |m| + |n| + |p| + |q| &> |n| + |q| \\ &\geq (j' + k')(\alpha + \beta) \\ &\geq 2(\alpha + \beta) \quad \text{since} \quad j', k' > 0, \end{aligned}$$

Similarly, in case 2 the order of the invariant

$$\begin{aligned} |m| + |n| + |p| + |q| &> |m| + |p| \\ &\geq (j' + |k'|)(\alpha + \beta) \\ &\geq 2(\alpha + \beta). \end{aligned}$$

The argument that $|m| + |n| + |p| + |q| > 2(\alpha + \beta)$ is similar for cases 3 and 4. Hence no invariants of order less than or equal to $2(\alpha + \beta)$ occur for $j'k' \neq 0$.

We use (4.11) to compute the leading order terms in the Taylor expansion of $g_1(\mathbf{z})$ from the $\mathbf{T}^2$-invariants, $|z_j|^2$, (4.19), and their complex conjugates. In particular, we find

$$\dot{z}_1 = z_1 f(|z_1|^2, |z_2|^2, |z_3|^2, |z_4|^2) + b_1 \bar{z}_1^{\beta-1} \bar{z}_2^{\alpha} z_3^{\alpha} z_4^{\beta} + b_2 \bar{z}_1^{\alpha-1} z_2^{\beta} z_3^{\beta} \bar{z}_4^{\alpha} + \mathcal{O}(2(\alpha + \beta)), \qquad (4.20)$$

where it follows from (4.10) that $f$ is a real-valued function of its arguments and that $b_1, b_2 \in \mathbf{R}$. Condition (4.9) determines the remaining components of $\mathbf{g}$ from $g_1$.

### 4.3 Stability.

We use the leading order terms in the Taylor expansion of the normal form (4.20) and the expressions for the eigenvalues given in table 10 to compute the signs of the eigenvalues of $\mathbf{Dg}$ at bifurcation. The results for the axial planforms are summarized in Table 11. The eigenvalues for the rolls, simple squares and rhombs are determined using a cubic truncation of the normal form,

$$\begin{aligned} \dot{z}_1 &= \lambda z_1 + z_1(a_1|z_1|^2 + a_2|z_2|^2 + a_3|z_3|^2 + a_4|z_4|^2) + \mathcal{O}(|\mathbf{z}|^5) \\ \dot{z}_2 &= \lambda z_2 + z_2(a_1|z_2|^2 + a_2|z_1|^2 + a_3|z_4|^2 + a_4|z_3|^2) + \mathcal{O}(|\mathbf{z}|^5) \\ \dot{z}_3 &= \lambda z_3 + z_3(a_1|z_3|^2 + a_2|z_4|^2 + a_3|z_1|^2 + a_4|z_2|^2) + \mathcal{O}(|\mathbf{z}|^5) \\ \dot{z}_4 &= \lambda z_4 + z_4(a_1|z_4|^2 + a_2|z_3|^2 + a_3|z_2|^2 + a_4|z_1|^2) + \mathcal{O}(|\mathbf{z}|^5). \end{aligned} \qquad (4.21)$$

Here we assumed, without loss of generality, that time has been scaled so that the linear term in the normal form is simply $\lambda \mathbf{z}$. We note that the computation of the last (repeated) eigenvalue for the super squares and anti-squares solution branches is simplified by the observation that

$$\left.\frac{\partial g_1^i}{\partial y_j}\right|_{z=\bar{z}} = \left.\left(\frac{\partial g_1}{\partial z_j} - \frac{\partial g_1}{\partial \bar{z}_j}\right)\right|_{z=\bar{z}}. \qquad (4.22)$$

The sign of this eigenvalue is determined by keeping all terms through $\mathcal{O}(2(\alpha + \beta) - 1)$ in the Taylor expansion of the normal form (4.20).

Table 11: Stability results for the square lattice bifurcation problem, from Table 10, and equations 4.20, 4.21.

| Axial Planform | Signs of Nonzero Eigenvalues | Branching Equation |
|---|---|---|
| Rolls $\mathbf{z} = (x,0,0,0)$ | $sgn(a_1)$, $sgn(a_2 - a_1)$, $sgn(a_3 - a_1)$, $sgn(a_4 - a_1)$ | $0 = \lambda x + a_1 x^3 + \cdots$ |
| Simple Squares $\mathbf{z} = (x,x,0,0)$ | $sgn(a_1 + a_2)$, $sgn(a_1 - a_2)$, $sgn(a_3 + a_4 - a_1 - a_2)$ | $0 = \lambda x + (a_1 + a_2) x^3 + \cdots$ |
| Rhombs ($Rh_{s1,\alpha,\beta}$) $\mathbf{z} = (x,0,x,0)$ | $sgn(a_1 + a_3)$, $sgn(a_1 - a_3)$, $sgn(a_2 + a_4 - a_1 - a_3)$ | $0 = \lambda x + (a_1 + a_3) x^3 + \cdots$ |
| Rhombs ($Rh_{s2,\alpha,\beta}$) $\mathbf{z} = (x,0,0,x)$ | $sgn(a_1 + a_4)$, $sgn(a_1 - a_4)$, $sgn(a_2 + a_3 - a_1 - a_4)$ | $0 = \lambda x + (a_1 + a_4) x^3 + \cdots$ |
| Super Squares $\mathbf{z} = (x,x,x,x)$ | $sgn(a_1 + a_2 + a_3 + a_4)$, $sgn(a_1 + a_2 - a_3 - a_4)$, $sgn(a_1 + a_3 - a_2 - a_4)$, $sgn(a_1 + a_4 - a_2 - a_3)$, $sgn(-b_1\beta - b_2\alpha)$ | $0 = \lambda x + (a_1 + a_2 + a_3 + a_4)x^3 + \cdots$ $+(b_1 + b_2)x^{2(\alpha+\beta)-1} + \cdots$ |
| Anti-squares $\mathbf{z} = (x,x,-x,-x)$ | $sgn(a_1 + a_2 + a_3 + a_4)$, $sgn(a_1 + a_2 - a_3 - a_4)$, $sgn(a_1 + a_3 - a_2 - a_4)$, $sgn(a_1 + a_4 - a_2 - a_3)$, $sgn(b_1\beta + b_2\alpha)$ | $0 = \lambda x + (a_1 + a_2 + a_3 + a_4)x^3 + \cdots$ $-(b_1 + b_2)x^{2(\alpha+\beta)-1} + \cdots$ |

We assume that the following nondegeneracy conditions are satisfied:

$$a_1 \neq 0, \pm a_2, \pm a_3, \pm a_4 , \qquad b_1 \beta + b_2 \alpha \neq 0 ,$$
$$(a_1 + a_2) \neq \pm(a_3 + a_4) , \qquad (a_1 - a_2) \neq \pm(a_3 - a_4) . \qquad (4.23)$$

In this case we can draw a number of conclusions from Table 11.

1. Any one of the axial solution branches can bifurcate supercritically to produce a stable solution.

2. If the super squares and anti-squares are neutrally stable at cubic order, then one and only one of these two states bifurcates stably.

3. If all of the axial planforms bifurcate supercritically, then at least one of them must be stable.

4. If *any* axial solution branch bifurcates subcritically, then rolls, super squares and anti-squares are all unstable.

5. If rolls, super squares, or anti-squares bifurcate subcritically, then *all* axial planforms are unstable at bifurcation.

6. If simple squares is the only axial solution branch to bifurcate subcritically, then it is still possible that one, but not both, of the rhombs solutions is stable. Similarly, if one of the rhombs solutions bifurcates subcritically, then it is possible that simple squares or the other rhombs solution branch is stable, though they cannot both be stable in this case.

7. The only solution branches that can co-exist stably are simple squares SiS and the rhombs $\text{Rh}_{s1,\alpha,\beta}$, $\text{Rh}_{s2,\alpha,\beta}$. Any combination of two of these states can bifurcate stably, but not all three.

## 5  Stability Results: Hexagonal Lattice.

In this section we compute the linear stability, at bifurcation, of the axial planforms that are associated with the twelve-dimensional representations of $\Gamma_h$ and $\Gamma_h + \mathbf{Z}_2$ (see Table 8). As with the square lattice case, we do this within the framework of a general $\Gamma$-equivariant bifurcation problem $\dot{\mathbf{z}} = \mathbf{g}(\mathbf{z}, \lambda)$, where $\mathbf{g} : \mathbf{C}^6 \times \mathbf{R} \to \mathbf{C}^6$.

In the case of $\Gamma = \Gamma_h + \mathbf{Z}_2$ there are only odd terms in the Taylor expansion of $\mathbf{g}$ due to the $\mathbf{Z}_2$ symmetry. However, if the $\mathbf{Z}_2$ symmetry is absent, then even terms are admissible. In particular, we find that the coefficients of most, but not all, quadratic terms in the Taylor expansion of $\mathbf{g}_1$ are zero; the exception is $\epsilon \equiv \frac{1}{2} \frac{\partial^2 g_1}{\partial \bar{z}_2 \partial \bar{z}_3}$, *i.e.*, the following vector is $\Gamma_h$-equivariant

$$(\bar{z}_2 \bar{z}_3, \bar{z}_3 \bar{z}_1, \bar{z}_1 \bar{z}_2, \bar{z}_5 \bar{z}_6, \bar{z}_6 \bar{z}_4, \bar{z}_4 \bar{z}_5)^\top \qquad (5.1)$$

The presence of such a quadratic term ensures that generically all of the axial planforms bifurcate unstably [15]. In order to obtain stable axial solution branches we focus on the degenerate bifurcation problem defined by $\epsilon = 0$. We then discuss briefly the unfolding of this bifurcation problem (*i.e.*, the case $0 < |\epsilon| \ll 1$), before analyzing the generic $\Gamma_h + \mathbf{Z}_2$-equivariant bifurcation problem.

## 5.1 Commuting linear maps.

We begin by determining the restrictions that symmetry places on the eigenvalues of the $12 \times 12$ real Jacobian matrix $\mathbf{Dg}$ when it is evaluated on an axial solution branch. The results are summarized in Table 12. We denote the real and imaginary parts of $g_j$ by $g_j^r$ and $g_j^i$, respectively, and the real and imaginary parts of $z_j$ by $x_j$ and $y_j$, respectively, $j = 1,...,6$.

We exploit the observation that $\mathbf{Dg}$, evaluated at $\mathbf{z}$, commutes with the generators of the isotropy subgroup $\Sigma_\mathbf{z}$ to determine the general form of the Jacobian matrix. For example, in the case of rhombs, $\text{Rh}_{h1,\alpha,\beta}$, we find that the Jacobian matrix has the form

$$\mathbf{Dg} = \begin{pmatrix} \mathbf{A} & \mathbf{0} \\ \mathbf{0} & \mathbf{B} \end{pmatrix}, \tag{5.2}$$

where the $6 \times 6$ real matrices $\mathbf{A}$ and $\mathbf{B}$ have the form

$$\mathbf{A} = \begin{pmatrix} a_{11} & 0 & 0 & a_{14} & 0 & 0 \\ 0 & a_{22} & a_{23} & 0 & 0 & 0 \\ 0 & a_{32} & a_{33} & 0 & 0 & 0 \\ a_{14} & 0 & 0 & a_{11} & 0 & 0 \\ 0 & 0 & 0 & 0 & a_{33} & a_{32} \\ 0 & 0 & 0 & 0 & a_{23} & a_{22} \end{pmatrix},$$

$$\mathbf{B} = \begin{pmatrix} b_{11} & 0 & 0 & b_{14} & 0 & 0 \\ 0 & a_{22} & -a_{23} & 0 & 0 & 0 \\ 0 & -a_{32} & a_{33} & 0 & 0 & 0 \\ b_{14} & 0 & 0 & b_{11} & 0 & 0 \\ 0 & 0 & 0 & 0 & a_{33} & -a_{32} \\ 0 & 0 & 0 & 0 & -a_{23} & a_{22} \end{pmatrix}. \tag{5.3}$$

Here we have chosen an ordering of the coordinates of $\mathbf{R}^{12}$ given by ($x_1$, $x_2$, $x_3$, $x_4$, $x_5$, $x_6$, $y_1$, $y_2$, $y_3$, $y_4$, $y_5$, $y_6$), so, for example, $a_{11} = \frac{\partial g_1^r}{\partial x_1}$ and $a_{23} = \frac{\partial g_2^r}{\partial x_3}$. Moreover, by determining the null directions associated with the two-dimensional group orbit of $\text{Rh}_{h1,\alpha,\beta}$ solutions we find that $b_{11} = b_{14} = 0$. The computations for the other axial solution branches are similar. Note that rolls and simple hexagons lie in the six-dimensional fixed-point subspace $\text{Fix}(\mathbf{S}_{1,2,3})$ on which the hidden reflection $\tilde{\tau}_{x_1}$ acts (equation 2.28); for these solutions we take the hidden symmetry into account in determining the eigenvalues of $\mathbf{Dg}$.

In the case of super hexagons, we find the computation of the eigenvalues of $\mathbf{Dg}$ is simplified by forming the $\mathbf{D}_6$-isotypic decomposition of $\mathbf{C}^6$. It is

$$\mathbf{C}^6 = \mathbf{R}(1,1,1,1,1,1) \oplus \mathbf{R}(1,1,1,-1,-1,-1) \oplus \mathbf{R}(i,i,i,i,i,i) \oplus \mathbf{R}(i,i,i,-i,-i,-i)$$

$\oplus$ $\mathbf{R}\{(1,-1,0,0,-1,1),(-1,0,1,-1,1,0),(0,1,-1,-1,1,0),(1,-1,0,1,0,-1)\}$

$\oplus$ $\mathbf{R}\{(-i,0,i,i,0,-i),(0,-i,i,0,i,-i),(0,i,-i,-i,i,0),(-i,i,0,-i,0,i)\}$ . (5.4)

The $12 \times 12$ Jacobian matrix $\mathbf{Dg}$ is block diagonal with respect to this basis [12]. The four eigenvalues associated with the four one-dimensional isotypic components are thereby determined. In addition, each four-dimensional isotypic component is a direct sum of two isomorphic $\mathbf{D}_6$-absolutely irreducible subspaces of dimension two. Thus the Jacobian matrix, restricted to either of these subspaces, must have the form

$$\begin{pmatrix} a\mathbf{I}_2 & b\mathbf{I}_2 \\ c\mathbf{I}_2 & d\mathbf{I}_2 \end{pmatrix},$$ (5.5)

where $\mathbf{I}_2$ is the $2 \times 2$ identity matrix, and $a, b, c, d \in \mathbf{R}$. The eigenvalues, $\mu_1, \mu_2$, each have multiplicity two and satisfy $\mu_1 \mu_2 = ad - bc$, $\mu_1 + \mu_2 = a + d$. Moreover, one of the repeated eigenvalues associated with the last isotypic component in (5.4) must be zero due to the $\mathbf{T}^2$ symmetry; thus the nonzero eigenvalue (of multiplicity two) is determined by computing the trace of $\mathbf{Dg}$ restricted to this isotypic component.

## 5.2 Normal Form for $\mathbf{D}_6 \dotplus \mathbf{T}^2$.

In this section we determine the Taylor expansion of the equivariant bifurcation problem (2.16) to sufficient order to determine the signs of the eigenvalues of $\mathbf{Dg}$ given in Table 12. We focus, in particular, on the degenerate bifurcation problem $\epsilon = 0$, where $\epsilon$ is the coefficient of the quadratic term (5.1).

Our approach to determining the leading terms in the $\Gamma_h$-equivariant normal form is the same as that employed in Section 4.3 for the $\Gamma_s$-equivariant normal form. The general $\Gamma_h$-equivariant vector field that satisfies the equivariance condition (2.17) is

$$\begin{aligned} \dot{z}_1 &= g_1(z_1, z_2, z_3, z_4, z_5, z_6) \\ \dot{z}_2 &= g_1(z_2, z_3, z_1, z_5, z_6, z_4) \\ \dot{z}_3 &= g_1(z_3, z_1, z_2, z_6, z_4, z_5) \\ \dot{z}_4 &= g_1(z_4, z_6, z_5, z_1, z_3, z_2) \\ \dot{z}_5 &= g_1(z_5, z_4, z_6, z_2, z_1, z_3) \\ \dot{z}_6 &= g_1(z_6, z_5, z_4, z_3, z_2, z_1) \,, \end{aligned}$$ (5.6)

where

$$\overline{g_1(\mathbf{z})} = g_1(\bar{\mathbf{z}}) \,,$$ (5.7)

and

$$\Theta\left(\bar{z}_1 g_1(\mathbf{z})\right) = \bar{z}_1 g_1(\mathbf{z}) \,, \text{ for all } \Theta \in \mathbf{T}^2 \,.$$ (5.8)

Equivariance with respect to $\mathbf{D}_6 \subset \Gamma_h$ is guaranteed by conditions (5.6) and (5.7), while equivariance with respect to $\mathbf{T}^2 \subset \Gamma_h$, with action given by (2.32), is equivalent to condition (5.8). Finally, the hidden reflection (3.1) puts an additional restriction on the function $g_1(\mathbf{z})$, namely

$$g_1(z_1, z_2, z_3, 0, 0, 0) = g_1(z_1, z_3, z_2, 0, 0, 0)$$ (5.9)

Table 12: Eigenvalues for axial planforms associated with 12-dim. representations of $\Gamma_h$.

| Axial Planform | Eigenvalues |
|---|---|
| Rolls<br>$\mathbf{z} = x(1,0,0,0,0,0)$<br>$\Sigma = \mathbf{Z}_2^c \dot{+} \mathbf{S}^1$ | $\frac{\partial g_1^r}{\partial x_1}$, $\frac{\partial g_4^r}{\partial x_4}$ (mult. 2), $\frac{\partial g_5^r}{\partial x_5}$ (mult. 2), $\frac{\partial g_6^r}{\partial x_6}$ (mult. 2),<br><br>$\frac{\partial g_2^r}{\partial x_2} + \frac{\partial g_2^r}{\partial x_3}$ $^{(a)}$ (mult. 2), $\frac{\partial g_2^r}{\partial x_2} - \frac{\partial g_2^r}{\partial x_3}$ $^{(a)}$ (mult. 2), 0 |
| Simple Hexagons<br>$\mathbf{z} = x(1,1,1,0,0,0)$<br>$\Sigma = \mathbf{Z}_6 \dot{+} \mathbf{S}_{1,2,3}$ | $\frac{\partial g_1^r}{\partial x_1} + 2\frac{\partial g_1^r}{\partial x_2}$ $^{(a)}$, $\frac{\partial g_1^r}{\partial x_1} - \frac{\partial g_1^r}{\partial x_2}$ $^{(a)}$ (mult. 2),<br><br>$\frac{\partial g_4^r}{\partial x_4}$ (mult. 6), $3\frac{\partial g_1^i}{\partial y_1}$, 0 (mult. 2) |
| Rhombs ($\mathrm{Rh}_{h1,\alpha,\beta}$)<br>$\mathbf{z} = x(1,0,0,1,0,0)$<br>$\Sigma = \mathbf{D}_2^n \dot{+} \mathbf{S}_{1,4}$ | $\frac{\partial g_1^r}{\partial x_1} + \frac{\partial g_1^r}{\partial x_4}$, $\frac{\partial g_1^r}{\partial x_1} - \frac{\partial g_1^r}{\partial x_4}$, 0 (mult. 2)<br><br>$\mu_1, \mu_2$ (mult. 4); $\mu_1 + \mu_2 = \frac{\partial g_2^r}{\partial x_2} + \frac{\partial g_3^r}{\partial x_3}$, $\mu_1 \mu_2 = \frac{\partial g_2^r}{\partial x_2}\frac{\partial g_3^r}{\partial x_3} - \frac{\partial g_2^r}{\partial x_3}\frac{\partial g_3^r}{\partial x_2}$ |
| Rhombs ($\mathrm{Rh}_{h2,\alpha,\beta}$)<br>$\mathbf{z} = x(1,0,0,0,1,0)$<br>$\Sigma = \mathbf{D}_2^m \dot{+} \mathbf{S}_{1,5}$ | $\frac{\partial g_1^r}{\partial x_1} + \frac{\partial g_1^r}{\partial x_5}$, $\frac{\partial g_1^r}{\partial x_1} - \frac{\partial g_1^r}{\partial x_5}$, 0 (mult. 2)<br><br>$\mu_1, \mu_2$ (mult. 4); $\mu_1 + \mu_2 = \frac{\partial g_2^r}{\partial x_2} + \frac{\partial g_3^r}{\partial x_3}$, $\mu_1 \mu_2 = \frac{\partial g_2^r}{\partial x_2}\frac{\partial g_3^r}{\partial x_3} - \frac{\partial g_2^r}{\partial x_3}\frac{\partial g_3^r}{\partial x_2}$ |
| Rhombs ($\mathrm{Rh}_{h3,\alpha,\beta}$)<br>$\mathbf{z} = x(1,0,0,0,0,1)$<br>$\Sigma = \mathbf{D}_2^x \dot{+} \mathbf{S}_{1,6}$ | $\frac{\partial g_1^r}{\partial x_1} + \frac{\partial g_1^r}{\partial x_6}$, $\frac{\partial g_1^r}{\partial x_1} - \frac{\partial g_1^r}{\partial x_6}$, 0 (mult. 2)<br><br>$\mu_1, \mu_2$ (mult. 4); $\mu_1 + \mu_2 = \frac{\partial g_2^r}{\partial x_2} + \frac{\partial g_3^r}{\partial x_3}$, $\mu_1 \mu_2 = \frac{\partial g_2^r}{\partial x_2}\frac{\partial g_3^r}{\partial x_3} - \frac{\partial g_2^r}{\partial x_3}\frac{\partial g_3^r}{\partial x_2}$ |
| Super Hexagons<br>$\mathbf{z} = x(1,1,1,1,1,1)$<br>$\Sigma = \mathbf{D}_6$ | $\frac{\partial g_1^r}{\partial x_1} + \frac{\partial g_1^r}{\partial x_2} + \frac{\partial g_1^r}{\partial x_3} + \frac{\partial g_1^r}{\partial x_4} + \frac{\partial g_1^r}{\partial x_5} + \frac{\partial g_1^r}{\partial x_6}$, $\frac{\partial g_1^r}{\partial x_1} + \frac{\partial g_1^r}{\partial x_2} + \frac{\partial g_1^r}{\partial x_3} - \frac{\partial g_1^r}{\partial x_4} - \frac{\partial g_1^r}{\partial x_5} - \frac{\partial g_1^r}{\partial x_6}$,<br><br>$\frac{\partial g_1^i}{\partial y_1} + \frac{\partial g_1^i}{\partial y_2} + \frac{\partial g_1^i}{\partial y_3} + \frac{\partial g_1^i}{\partial y_4} + \frac{\partial g_1^i}{\partial y_5} + \frac{\partial g_1^i}{\partial y_6}$, $\frac{\partial g_1^i}{\partial y_1} + \frac{\partial g_1^i}{\partial y_2} + \frac{\partial g_1^i}{\partial y_3} - \frac{\partial g_1^i}{\partial y_4} - \frac{\partial g_1^i}{\partial y_5} - \frac{\partial g_1^i}{\partial y_6}$,<br><br>$2\frac{\partial g_1^i}{\partial y_1} - \frac{\partial g_1^i}{\partial y_2} - \frac{\partial g_1^i}{\partial y_3}$ (mult. 2), 0 (mult. 2), $\mu_1, \mu_2$ (mult. 2)<br><br>$\mu_1 \mu_2 = \frac{1}{2}\left\{\left(\frac{\partial g_1^r}{\partial x_1} - \frac{\partial g_1^r}{\partial x_2}\right)^2 + \left(\frac{\partial g_1^r}{\partial x_1} - \frac{\partial g_1^r}{\partial x_3}\right)^2 + \left(\frac{\partial g_1^r}{\partial x_2} - \frac{\partial g_1^r}{\partial x_3}\right)^2 - \left(\frac{\partial g_1^r}{\partial x_4} - \frac{\partial g_1^r}{\partial x_5}\right)^2\right.$<br>$\left. -\left(\frac{\partial g_1^r}{\partial x_4} - \frac{\partial g_1^r}{\partial x_6}\right)^2 - \left(\frac{\partial g_1^r}{\partial x_5} - \frac{\partial g_1^r}{\partial x_6}\right)^2\right\}$, $\mu_1 + \mu_2 = 2\frac{\partial g_1^r}{\partial x_1} - \frac{\partial g_1^r}{\partial x_2} - \frac{\partial g_1^r}{\partial x_3}$ |

$^{(a)}$ Here the effect on $\mathbf{Dg}$ of the hidden symmetry $\tilde{\tau}_{x_1}$ (3.1) is included.

The problem of determining the general form of the $\Gamma_h$-equivariant vector field then reduces to one of finding the function $g_1(\mathbf{z})$ that satisfies equations (5.7)-(5.9).

As in the square lattice case, we assume that local to the bifurcation point we can Taylor expand the function $g_1(\mathbf{z})$. We proceed by determining the $\mathbf{T}^2$-invariant monomials, which represent all terms present in the Taylor expansion of the $\mathbf{T}^2$-invariant function $\bar{z}_1 g_1(\mathbf{z})$. In the following, we assume that overall factors of $|z_j|^2$, $j = 1, ..., 6$, which are manifestly $\mathbf{T}^2$-invariant, have been removed from the monomials that we consider. We follow the convention that $z_j^n \equiv \bar{z}_j^{|n|}$ if $n < 0$, and focus on $\mathbf{T}^2$-invariants of the form

$$z_1^m z_2^n z_3^p z_4^q z_5^r z_6^s, \tag{5.10}$$

where $m, n, p, q, s \in \mathbf{Z}$ satisfy

$$\begin{aligned}(m - n + q - s)\alpha + (n - p - r + s)\beta &= 0, \\ -(n - p - q + r)\alpha + (m - p - q + s)\beta &= 0.\end{aligned} \tag{5.11}$$

Since $\alpha$ and $\beta$ are relatively prime, we have

$$\begin{aligned}m - n + q - s = j\beta, & \quad n - p - r + s = -j\alpha, \\ n - p - q + r = k\beta, & \quad m - p - q + s = k\alpha,\end{aligned} \tag{5.12}$$

where $j, k \in \mathbf{Z}$.

There are no nontrivial solutions of (5.12) with more than three of $m, n, p, q, r, s$ zero. In the case that $j = k = 0$ in (5.12), then $m = n = p$ and $q = r = s$, which yield the invariants

$$z_1 z_2 z_3 \quad \text{and} \quad z_4 z_5 z_6, \tag{5.13}$$

and their complex conjugates.

We claim that the lowest order $(\alpha, \beta)$-dependent invariants $(|j| + |k| \neq 0)$ are

$$z_1^\beta \bar{z}_2^{\alpha-\beta} \bar{z}_4^{\alpha-\beta} z_5^\beta, \qquad z_2^\beta \bar{z}_3^{\alpha-\beta} \bar{z}_5^{\alpha-\beta} z_6^\beta, \qquad z_3^\beta \bar{z}_1^{\alpha-\beta} \bar{z}_6^{\alpha-\beta} z_4^\beta, \tag{5.14}$$

and their complex conjugates, which are order $2\alpha$. Note that the set of the three invariants (5.14) together with their complex conjugates is invariant under the action of $\mathbf{D}_6$. We justify the above assertion by the following steps:

1. Show that we can assume that one of $m, n, p$ is zero and that one of $q, r, s$ is zero. We then focus on the specific case with $p = 0$ since the invariants with $m = 0$ and $n = 0$ can be transformed to $p = 0$ by the action of $R_{\pi/3} \in \mathbf{D}_6$ on the invariant.

2. Consider the cases where $p = 0$ and exactly two of $m, n, q, r, s$ are zero.

3. Consider the cases where $p = 0$ and only one of $q, r, s$ is zero.

Step 1. Consider $m, n, p \geq 0$. If $m, n, p$ are all nonzero, then we can construct a lower degree $\mathbf{T}^2$-invariant monomial from (5.10) by factoring out the invariant $z_1 z_2 z_3$. Thus the lowest degree monomials cannot have $m, n, p > 0$. Similarly, if $m, n, p < 0$, then we can lower the degree by factoring out the invariant $\bar{z}_1 \bar{z}_2 \bar{z}_3$.

Suppose now that $m, n, p$ do not all have the same sign; for example, consider the case $m \geq n \geq 0 \geq p$. Then the order of the monomial is $m + n + |p| + |q| + |r| + |s|$. However, if $z_1^m z_2^n \bar{z}_3^{|p|} z_4^q z_5^r z_6^s$ is invariant, then so is $z_1^{m-n} \bar{z}_3^{|p|+n} z_4^q z_5^r z_6^s$, which is of lower order $m + |p| + |q| + |r| + |s|$ unless $n = 0$, in which case it has the same order. Since we only aim to find the lowest order $(\alpha, \beta)$-dependent monomials, we can assume $n = 0$ in (5.10). The argument is the same for the other orderings of $m, n, p$, and 0; in each case we can find a lower degree invariant monomial unless one of $m, n, p$ is zero.

Hence, the lowest order $(\alpha, \beta)$-dependent monomial has at least one of $m, n, p$ equal to zero, and by a similar argument we can assume that one of $q, r, s$ is zero. In the following steps, we assume that $p = 0$. The invariants with $m = 0$ or $n = 0$ and $p \neq 0$ are obtained from the invariants with $p = 0$ by applying $R_{\pi/3} \in \mathbf{D}_6$ to the monomials.

Step 2. Let $p = 0$ and exactly two of $m, n, q, r, s$ be zero. From step 1, we can assume that at least one of $q, r, s$ is zero. There are nine combinations to consider; in each case we obtain an invariant of degree greater than $2\alpha$. As an example, consider $p = r = s = 0$. Then equations (5.12) can be solved provided we choose $j$ and $k$ such that

$$(k - j)\alpha = (j + 2k)\beta . \tag{5.15}$$

Thus

$$k - j = l\beta, \qquad j + 2k = l\alpha , \tag{5.16}$$

where $l \in \mathbf{Z}$. Solving for $j$ and $k$ gives

$$j = \frac{1}{3}l(\alpha - 2\beta), \qquad k = \frac{1}{3}l(\alpha + \beta) , \tag{5.17}$$

and, since $(3, \alpha + \beta) = 1$, $l$ must be divisible by 3. Let $l = 3l', l' \in \mathbf{Z}$, then we find

$$\begin{aligned} m &= l'\beta(2\alpha - \beta) , \\ n &= l'\alpha(2\beta - \alpha) , \\ q &= -l'(\alpha^2 - \alpha\beta + \beta^2) . \end{aligned} \tag{5.18}$$

The order of the invariant $z_1^m z_2^n z_4^q$ is $3l'\alpha\beta$, which is greater than the order of the invariants (5.14). The other eight combinations, with three non-zero exponents, also give invariants of order either $3\alpha\beta$, $\alpha^2 + \alpha\beta$, or $2\alpha^2 + \alpha\beta - \beta^2$, all of which are greater than $2\alpha$.

Step 3. If $p = 0$ then it follows from (5.12) that

$$m = j\beta - \frac{1}{3}(j - k)(\alpha + \beta) ,$$

$$n = k\beta - j\alpha + \frac{1}{3}(j-k)(\alpha + \beta), \tag{5.19}$$
$$q - s = j\beta - k\alpha - \frac{1}{3}(j-k)(\alpha + \beta),$$
$$r - s = k\beta + \frac{1}{3}(j-k)(\alpha + \beta).$$

We set $(k - j) = 3l$, $l \in \mathbf{Z}$, because $(3, \alpha + \beta) = 1$. Hence,

$$\begin{aligned} m &= j\beta + l(\alpha + \beta), \\ n &= -j(\alpha - \beta) + l(2\beta - \alpha), \\ q - s &= -j(\alpha - \beta) - l(2\alpha - \beta), \\ r - s &= j\beta + l(2\beta - \alpha), \end{aligned} \tag{5.20}$$

where $j, l \in \mathbf{Z}$.

We consider the three cases $q = 0$, $r = 0$, and $s = 0$ separately. The invariant $z_1^\beta \bar{z}_2^{\alpha-\beta} \bar{z}_4^{\alpha-\beta} z_5^\beta$ in (5.14) is obtained in the case $p = s = 0$ for $l = 0$, $j = 1$ in (5.20). The cases $p = q = 0$ and $p = r = 0$ lead to nontrivial invariant monomials of degree greater than $2\alpha$.

For $p = q = 0$, the degree of the monomial is $|m| + |n| + |r| + |s|$. It follows from the restriction $\alpha > \beta > \alpha/2$ in Table 2 that

$$|m| + |n| + |r| + |s| > |m + n + r| = |l|(\alpha + 4\beta) > 2|l|\alpha. \tag{5.21}$$

Hence, the degree of the monomial is greater than $2\alpha$ unless $l = 0$. However, if $l = 0$, then

$$|m| + |n| + |r| + |s| = |j|(3\alpha - \beta) > 2|j|\alpha. \tag{5.22}$$

This proves that the monomials associated with $p = q = 0$, $mnrs \neq 0$, are all of degree greater than $2\alpha$. The argument in the case $p = r = 0$ is similar.

From the invariant function $\bar{z}_1 g_1$, we can compute the general form of the equivariant vector field through $\mathcal{O}(2\alpha - 1)$. Specifically,

$$\begin{aligned} \dot{z}_1 &= z_1 \, f_1(u_1, u_2, u_3, u_4, u_5, u_6, q_1, \bar{q}_1, q_4, \bar{q}_4) + \bar{z}_2 \bar{z}_3 \, f_2(u_1, u_2, u_3, u_4, u_5, u_6, q_1, \bar{q}_1, q_4, \bar{q}_4) \\ &\quad + e_1 \, \bar{z}_1^{\alpha-\beta-1} z_3^\beta z_4^\beta \bar{z}_6^{\alpha-\beta} + e_2 \, \bar{z}_1^{\beta-1} z_2^{\alpha-\beta} z_4^{\alpha-\beta} \bar{z}_5^\beta + \mathcal{O}(2\alpha), \end{aligned} \tag{5.23}$$

where

$$u_j \equiv |z_j|^2, \quad q_1 \equiv z_1 z_2 z_3, \quad q_4 \equiv z_4 z_5 z_6, \tag{5.24}$$

and $e_1, e_2 \in \mathbf{R}$ are constants. It follows from (5.7) and (5.9), respectively that

$$\begin{aligned} \overline{f_j(u_1, u_2, u_3, u_4, u_5, u_6, q_1, \bar{q}_1, q_4, \bar{q}_4)} &= f_j(u_1, u_2, u_3, u_4, u_5, u_6, \bar{q}_1, q_1, \bar{q}_4, q_4), \\ f_j(u_1, u_2, u_3, 0, 0, 0, q_1, \bar{q}_1, 0, 0) &= f_j(u_1, u_3, u_2, 0, 0, 0, q_1, \bar{q}_1, 0, 0), \quad j = 1, 2. \end{aligned} \tag{5.25}$$

We use the $\mathbf{D}_6$ symmetry to determine the other components of the normal form from (5.23), as indicated by (5.6).

Table 13: Stability results for the hexagonal lattice bifurcation problem in the degenerate case $\epsilon = 0$, from Table 12, and equations 5.6, 5.23 and 5.26. The branching equations are given in Table 14.

| Axial Planform | Signs of Nonzero Eigenvalues |
|---|---|
| Rolls $\mathbf{z} = x(1,0,0,0,0,0)$ | $sgn(a_1)$, $sgn(a_4 - a_1)$, $sgn(a_5 - a_1)$, $sgn(a_6 - a_1)$, $sgn(a_2 - a_1)$ |
| Simple Hexagons $\mathbf{z} = x(1,1,1,0,0,0)$ | $sgn(a_1 + 2a_2)$, $sgn(a_1 - a_2)$, $sgn(a_4 + a_5 + a_6 - a_1 - 2a_2)$, $sgn[x(c_1 - b_1 - 2b_2)]$ |
| Rhombs ($\text{Rh}_{h1,\alpha,\beta}$) $\mathbf{z} = x(1,0,0,1,0,0)$ | $sgn(a_1 + a_4)$, $sgn(a_1 - a_4)$, $sgn(a_2 + a_5 - a_1 - a_4)$, $sgn(a_2 + a_6 - a_1 - a_4)$ |
| Rhombs ($\text{Rh}_{h2,\alpha,\beta}$) $\mathbf{z} = x(1,0,0,0,1,0)$ | $sgn(a_1 + a_5)$, $sgn(a_1 - a_5)$, $sgn(a_2 + a_4 - a_1 - a_5)$, $sgn(a_2 + a_6 - a_1 - a_5)$ |
| Rhombs ($\text{Rh}_{h3,\alpha,\beta}$) $\mathbf{z} = x(1,0,0,0,0,1)$ | $sgn(a_1 + a_6)$, $sgn(a_1 - a_6)$, $sgn(a_2 + a_4 - a_1 - a_6)$, $sgn(a_2 + a_5 - a_1 - a_6)$ |
| Super Hexagons $\mathbf{z} = x(1,1,1,1,1,1)$ | $sgn(a_1 + 2a_2 + a_4 + a_5 + a_6)$, $sgn(a_1 + 2a_2 - a_4 - a_5 - a_6)$ $sgn[x(-b_1 - 2b_2 - b_4 - b_5 - b_6 + c_1 + c_2 - c_3)]$, $sgn[x(-b_1 - 2b_2 - b_4 - b_5 - b_6 + c_1 - c_2 + c_3)]$, $sgn[-(2\alpha - \beta)e_1 - (\alpha + \beta)e_2]$, $sgn(\mu_1 + \mu_2) = sgn(a_1 - a_2)$, $sgn(\mu_1 \mu_2) = sgn[2(a_1 - a_2)^2 - (a_4 - a_5)^2 - (a_4 - a_6)^2 - (a_5 - a_6)^2]$ |

## 5.3 Stability for degenerate bifurcation problem.

Expanding (5.23) through quartic order, we obtain

$$\begin{aligned}
\dot{z}_1 &= \lambda z_1 + \epsilon \bar{z}_2 \bar{z}_3 + z_1(a_1|z_1|^2 + a_2|z_2|^2 + a_2|z_3|^2 + a_4|z_4|^2 + a_5|z_5|^2 + a_6|z_6|^2) \\
&+ \bar{z}_2 \bar{z}_3 (b_1|z_1|^2 + b_2|z_2|^2 + b_2|z_3|^2 + b_4|z_4|^2 + b_5|z_5|^2 + b_6|z_6|^2) \\
&+ z_1(c_1 z_1 z_2 z_3 + c_2 z_4 z_5 z_6 + c_3 \bar{z}_4 \bar{z}_5 \bar{z}_6) + \mathcal{O}(|\mathbf{z}|^5) \ .
\end{aligned} \quad (5.26)$$

The remaining components of $\dot{\mathbf{z}} = \mathbf{g}(\mathbf{z})$ are determined from (5.26) using (5.6). Again we assume time has been scaled so that the linear term in $\mathbf{g}(\mathbf{z})$ is $\lambda \mathbf{z}$. Table 13 gives the signs of the eigenvalues in Table 12 for the degenerate case $\epsilon = 0$. The quartic truncation is sufficient to determine the sign of all eigenvalues except

$$\frac{\partial g_1^i}{\partial y_1} - \frac{1}{2}\Big(\frac{\partial g_1^i}{\partial y_2} + \frac{\partial g_1^i}{\partial y_3}\Big) \ , \quad (5.27)$$

for super hexagons. The sign of this eigenvalue is determined by retaining the leading order $(\alpha, \beta)$-dependent terms in the normal form (5.23).

We assume that $\epsilon = 0$ and that the following nondegeneracy conditions are satisfied:

$$\begin{aligned}
a_1 &\neq 0, \ a_2, \ \pm a_4, \ \pm a_5, \ \pm a_6 \ , \\
(a_1 + 2a_2) &\neq 0, \ \pm(a_4 + a_5 + a_6) \ ,
\end{aligned}$$

$$\begin{aligned}
a_1 - a_2 &\neq \pm(a_4 - a_5),\ \pm(a_4 - a_6),\ \pm(a_5 - a_6)\,, \quad &(5.28)\\
2(a_1 - a_2)^2 &\neq (a_4 - a_5)^2 + (a_4 - a_6)^2 + (a_5 - a_6)^2\,, \\
c_1 - b_1 - 2b_2 &\neq 0,\ b_4 + b_5 + b_6 \pm (c_2 - c_3)\,, \\
\frac{e_1}{e_2} &\neq -\frac{\alpha + \beta}{2\alpha - \beta}\,.
\end{aligned}$$

In this case we can draw a number of conclusions from Table 13.

1. While all axial solution branches bifurcate unstably when $\epsilon \neq 0$, we find, in the degenerate case $\epsilon = 0$, that any one of the axial solution branches can bifurcate supercritically to produce a stable solution.

2. There are two distinct branches of simple and super hexagons, denoted SiH$^\pm$ and SuH$^\pm$, respectively, associated with $x > 0$ and $x < 0$. If simple hexagons are neutrally stable at cubic order, then one and only one of the two branches SiH$^\pm$ is stable. If super hexagons are neutrally stable at cubic order, then one and only one of the two branches will be stable if $(2\alpha - \beta)e_1 + (\alpha + \beta)e_2 > 0$, while they are both unstable if $(2\alpha - \beta)e_1 + (\alpha + \beta)e_2 < 0$.

3. If $(2\alpha - \beta)e_1 + (\alpha + \beta)e_2) < 0$ then it is possible for all of the axial planforms to bifurcate supercritically, but none be stable. On the other hand, if $(2\alpha - \beta)e_1 + (\alpha + \beta)e_2 > 0$ and all axial planforms bifurcate supercritically, then at least one of them must be stable.

4. If *any* axial solution branch bifurcates subcritically, then rolls and super hexagons are unstable.

5. If rolls or super hexagons bifurcate subcritically, then *all* axial planforms are unstable at bifurcation.

6. If simple hexagons is the only axial solution branch to bifurcate subcritically, then it is still possible that one, but not more, of the rhombs solutions is stable. Similarly, if rhombs is the only axial solution branch to bifurcate subcritically, then it is possible for simple hexagons to be stable, or for one or more of the remaining rhombs solutions to be stable. However, if simple hexagons *and* one of the rhombs bifurcate subcritically, then all axial solution branches are unstable.

7. If two of the rhombs solution branches bifurcate subcritically, then it is possible that the remaining rhombs solution or simple hexagons is stable, but not both. However, if all three rhombs solution branches are subcritical, then all axial planforms are unstable.

8. The only solution branches that can co-exist stably are simple hexagons SiH and the rhombs Rh$_{h1,\alpha,\beta}$, Rh$_{h2,\alpha,\beta}$, Rh$_{h3,\alpha,\beta}$. Any combination of two of these states can bifurcate stably. It is also possible for all three types of rhombs to be stable simultaneously. However, if two or more of the rhombs are stable, then simple hexagons are unstable.

## 5.4 Secondary Bifurcations.

In this section we address briefly the unfolding of the degenerate bifurcation problem $\epsilon = 0$ analyzed in the previous section. Specifically, we indicate how the stability of the axial solutions change along the solution branch in the case that $|\epsilon| \ll 1$. While a complete analysis of the unfolding is beyond the scope of the present paper, we do present an example in which part of a bifurcation diagram is computed. This example indicates the wealth of secondary transitions that occur close to $\lambda = 0$ when $|\epsilon| \ll 1$. When $\epsilon \neq 0$, certain eigenvalues given in Table 13 are modified to those given in Table 14. Note that, as discussed above, the presence of the quadratic term in the bifurcation problem ensures that at least one of the eigenvalues for each axial planform is positive for $(\lambda, \mathbf{z})$ sufficiently close to the origin.

As a specific example, we consider the bifurcation problem

$$\begin{aligned}
\dot{z}_1 &= \lambda\, z_1 + \epsilon\, \bar{z}_2 \bar{z}_3 + z_1\, (a_1|z_1|^2 + a_2|z_2|^2 + a_2|z_3|^2 + a_4|z_4|^2 + a_5|z_5|^2 + a_6|z_6|^2) \\
&\quad + b_2\, \bar{z}_2 \bar{z}_3\, (|z_2|^2 + |z_3|^2) + e_1\, \bar{z}_1^{\alpha-\beta-1} z_3^\beta z_4^\beta \bar{z}_6^{\alpha-\beta} + e_2\, \bar{z}_1^{\beta-1} z_2^{\alpha-\beta} z_4^{\alpha-\beta} \bar{z}_5^\beta,
\end{aligned} \quad (5.29)$$

where

$$a_1 = -1.5,\ a_2 = -3.5,\ a_4 = 0.5,\ a_5 = 0.6,\ a_6 = 0.7,\ b_2 = 0.6,\ e_1 = 1.0,\ e_2 = 0.5. \quad (5.30)$$

This choice satisfies the non-degeneracy conditions (5.28). It follows from Table 13 that all three rhomb states are stable when $\epsilon = 0$. We show two bifurcation diagrams for $0 < \epsilon \ll 1$. The bifurcation diagrams indicate, schematically, the amplitude $|\mathbf{z}|$ as a function of the bifurcation parameter $\lambda$ for each axial planform. Solutions on the same group orbit are identified and bifurcation points are indicated by solid circles. We follow the convention that solid lines indicate stable solutions and dotted lines indicate unstable solutions. Figure 8 is a well-known bifurcation diagram that applies to the six-dimensional representation of $\Gamma_h$; here it is obtained by restricting our analysis to the six-dimensional subspace where $\mathbf{z} = (z_1, z_2, z_3, 0, 0, 0)$. Figure 9 gives the bifurcation diagram that applies, for the same coefficient values (5.30), in the full twelve-dimensional space.

In the six-dimensional subspace, where $z_4 = z_5 = z_6 = 0$, only two axial planforms exist, rolls and simple hexagons. In this subspace, and for the choice of coefficients (5.30), Figure 8 indicates that as $\lambda$ increases through 0, the trivial solution becomes unstable and there is a transition to stable simple hexagons. On further increase of $\lambda$ the hexagons become unstable and there is a transition to rolls. Both the transition to hexagons and that to rolls exhibit hysteresis. The bifurcation scenario of Figure 8 has been investigated in a wide variety of hydrodynamic systems [3, 10, 18, 23], in solidification problems [28, 2, 21], and in chemical reaction-diffusion systems [9, 20].

Figure 9 indicates how the familiar bifurcation diagram in Figure 8 is modified when we consider stability within the full twelve-dimensional space. In this case, rolls are always unstable to rhombs, and the range of stability of simple hexagons is greatly decreased. Indeed simple hexagons are stable only in a subcritical regime where super hexagons are also stable. In this case, on increasing $\lambda$, there is first a jump at $\lambda = 0$ to stable super hexagons and then a transition to one of the three stable rhombs states. All of the transitions exhibit hysteresis. While the bifurcations to simple and super hexagons are transcritical, all other primary bifurcations are pitchforks. All of the secondary bifurcation points indicated in the diagram approach $\lambda = |\mathbf{z}| = 0$ as $\epsilon \to 0$. The paths of the secondary branches have not been computed.

Table 14: Stability results for the hexagonal lattice bifurcation problem in the case $|\epsilon| \ll 1$, from Table 12, and equations 5.6, 5.23 and 5.26. Also see Table 13; only the eigenvalues that depend on $\epsilon$ are given here.

| Planform | $\epsilon$-Dependent Eigenvalues | Branching Equation |
|---|---|---|
| Rolls | $\epsilon x + (a_2 - a_1)x^2 + \cdots$ <br> $-\epsilon x + (a_2 - a_1)x^2 + \cdots$ | $0 = \lambda x + a_1 x^3$ <br> $+\mathcal{O}(x^5)$ |
| $\text{SiH}^{\pm}_{\alpha,\beta}$ | $\epsilon x + 2(a_1 + 2a_2)x^2 + \cdots$ <br> $-2\epsilon x + 2(a_1 - a_2)x^2 + \cdots$ <br> $-\epsilon x + (a_4 + a_5 + a_6 - a_1 - 2a_2)x^2 + \cdots$ <br> $-3\epsilon x + 3(c_1 - b_1 - 2b_2)x^3 + \cdots$ | $0 = \lambda x + \epsilon x^2$ <br> $+(a_1 + 2a_2)x^3$ <br> $+\mathcal{O}(x^4)$ |
| $\text{Rh}_{h1,\alpha,\beta}$ [a] | $\mu_1, \mu_2;\ \mu_1 + \mu_2 = (-2a_1 - 2a_4 + 2a_2 + a_5 + a_6)x^2 + \cdots$ <br> $\mu_1 \mu_2 = -\epsilon^2 x^2$ <br> $+(a_1 + a_4 - a_2 - a_5)(a_1 + a_4 - a_2 - a_6)x^4 + \cdots$ | $0 = \lambda x + (a_1 + a_4)x^3$ <br> $+\mathcal{O}(x^5)$ |
| $\text{SuH}^{\pm}_{\alpha,\beta}$ | $\epsilon x + 2(a_1 + 2a_2 + a_4 + a_5 + a_6)x^2 + \cdots$ <br> $\epsilon x + 2(a_1 + 2a_2 - a_4 - a_5 - a_6)x^2 + \cdots$ <br> $-3[\epsilon x + (b_1 + 2b_2 + b_4 + b_5 + b_6 - c_1 - c_2 + c_3)x^3] + \cdots$ <br> $-3[\epsilon x + (b_1 + 2b_2 + b_4 + b_5 + b_6 - c_1 + c_2 - c_3)x^3] + \cdots$ <br> $\mu_1, \mu_2;\ \mu_1 + \mu_2 = -4\epsilon x + 4(a_1 - a_2)x^2 + \cdots$ <br> $\mu_1 \mu_2 = 4\epsilon^2 x^2 - 8(a_1 - a_2)\epsilon x^3 + 4(a_1 - a_2)^2 x^4$ <br> $-2[(a_4 - a_5)^2 + (a_4 - a_6)^2 + (a_5 - a_6)^2]x^4 + \cdots$ | $0 = \lambda x + \epsilon x^2$ <br> $+(a_1 + 2a_2)x^3$ <br> $+(a_4 + a_5 + a_6)x^3$ <br> $+\mathcal{O}(x^4)$ |

[a] The results for $\text{Rh}_{h2,\alpha,\beta}$ ($\text{Rh}_{h3,\alpha,\beta}$) are obtained from those for $\text{Rh}_{h1,\alpha,\beta}$ by interchanging the 4 and 5 (4 and 6) subscripts.

## 5.5 Stability results: $\Gamma = \Gamma_h + \mathbf{Z}_2$.

In this section we consider the consequences of the additional $\mathbf{Z}_2$ symmetry, $\kappa(\mathbf{z}) = -\mathbf{z}$, for the generic bifurcation problem on the hexagonal lattice, $\dot{\mathbf{z}} = \mathbf{g}(\mathbf{z}, \lambda)$, $\mathbf{g} : \mathbf{C}^6 \times \mathbf{R} \to \mathbf{C}^6$. Specifically we consider the branching and stability assignments for the axial planforms listed in Table 8.

The $\mathbf{Z}_2$ symmetry places some additional restrictions on the eigenvalues of rolls and rhombs listed in Table 12; specifically, it ensures that $\frac{\partial g_2^r}{\partial x_3} = \frac{\partial g_3^r}{\partial x_2} = 0$ on these solution branches. The eigenvalues for simple and super hexagons, listed in Table 12, are unchanged. The eigenvalues of $\mathbf{Dg}$ for the remaining axial planforms are listed in Table 15.

We note that the $\mathbf{D}_6$-isotypic decomposition of $\mathbf{C}^6$ is the same for the super hexagons, anti-hexagons, super-triangles and anti-triangles planforms; it is given by (5.4). Indeed, the only difference between the eigenvalue structure for the triangle states and the hexagon states is that the null vectors lie in different isotypic components in the two cases. For example, the null vectors, associated with translations of the super hexagons and anti-hexagons, lie in the four-dimensional

isotypic component

$$\mathbf{R}\{(-i,0,i,i,0,-i),(0,-i,i,0,i,-i),(0,i,-i,-i,i,0),(-i,i,0,-i,0,i)\}, \tag{5.31}$$

while the null vectors for the super and anti-triangles lie in

$$\mathbf{R}\{(1,-1,0,0,-1,1),(-1,0,1,-1,1,0),(0,1,-1,-1,1,0),(1,-1,0,1,0,-1)\}. \tag{5.32}$$

The additional $\mathbf{Z}_2$ symmetry forces the coefficients of all even order terms in the Taylor expansion of the $\Gamma_h$-equivariant normal form (5.23) to be zero. Hence there are no quadratic terms; the differences between the degenerate bifurcation problem with $\Gamma_h$-symmetry ($\epsilon = 0$) and the generic bifurcation problem with $\Gamma_h + \mathbf{Z}_2$-symmetry arise at $\mathcal{O}(|\mathbf{z}|^4)$. Thus the eigenvalues for the rolls and the rhombs in Table 13, which are determined by a cubic truncation of the normal form, are unchanged by the extra $\mathbf{Z}_2$ symmetry. Note that certain eigenvalues of simple and super hexagons for the degenerate $\Gamma_h$ bifurcation problem depend on the coefficients of quartic terms (see Table 13). These eigenvalues are now determined at quintic order.

The stability results for the axial planforms are summarized in Table 16. The quintic truncation of the normal form is

$$\begin{aligned}
\dot{z}_1 &= \lambda z_1 + z_1(a_1|z_1|^2 + a_2|z_2|^2 + a_2|z_3|^2 + a_4|z_4|^2 + a_5|z_5|^2 + a_6|z_6|^2) \\
&\quad + z_1(f_{11}|z_1|^4 + f_{12}|z_1|^2|z_2|^2 + \cdots + f_{56}|z_5|^2|z_6|^2 + f_{66}|z_6|^2) \\
&\quad + \bar{z}_2\bar{z}_3(d_1 z_1 z_2 z_3 + d_2 \bar{z}_1 \bar{z}_2 \bar{z}_3 + d_3 z_4 z_5 z_6 + d_4 \bar{z}_4 \bar{z}_5 \bar{z}_6) \\
&\quad + e_1 \bar{z}_1^{\alpha-\beta-1} z_3^\beta z_4^\beta \bar{z}_6^{\alpha-\beta} + e_2 \bar{z}_1^{\beta-1} z_2^{\alpha-\beta} z_4^{\alpha-\beta} \bar{z}_5^\beta + \mathcal{O}(|\mathbf{z}|^7).
\end{aligned} \tag{5.33}$$

Note that the leading order $\alpha$, $\beta$ dependent terms are $\mathcal{O}(2\alpha - 1)$, where $2\alpha - 1 \geq 5$, with $2\alpha - 1 = 5$ only in the case of $(\alpha, \beta) = (3, 2)$.

We assume that the following nondegeneracy conditions are satisfied:

$$\begin{aligned}
a_1 &\neq 0, \pm a_2, \pm a_4, \pm a_5, \pm a_6, \\
(a_1 + a_2) &\neq (a_4 + a_5), (a_4 + a_6), (a_5 + a_6), \\
(a_1 + 2a_2) &\neq 0, \pm(a_4 + a_5 + a_6), \\
(a_1 - a_2) &\neq \pm(a_4 - a_5), \pm(a_4 - a_6), \pm(a_5 - a_6), \\
2(a_1 - a_2)^2 &\neq (a_4 - a_5)^2 + (a_4 - a_6)^2 + (a_5 - a_6)^2, \\
\frac{e_1}{e_2} &\neq -\frac{\alpha + \beta}{2\alpha - \beta}, \\
d_2 &\neq 0, \pm d_3, \\
d_2 \pm d_4 &\neq \begin{cases} 0 & \text{if } (\alpha,\beta) \neq (3,2) \\ \pm\left(\frac{e_1 - e_2}{3}\right) & \text{if } (\alpha,\beta) = (3,2) \end{cases}
\end{aligned} \tag{5.34}$$

In this case we can draw a number of conclusions from Tables 13 and 16.

1. The rhombs $\mathrm{Rh}_{h0}$ always bifurcate unstably [13].

Table 15: Eigenvalues for axial planforms associated with 12-dim. representations of $\Gamma_h + \mathbf{Z}_2$. Also see Table 12.

| Axial Planform | Eigenvalues |
|---|---|
| Rhombs ($\mathrm{Rh}_{h0}$) $x(1,1,0,0,0,0)$ | $\frac{\partial g_1^r}{\partial x_1} + \frac{\partial g_1^r}{\partial x_2}$ (a), $\quad \frac{\partial g_1^r}{\partial x_1} - \frac{\partial g_1^r}{\partial x_2}$ (a), $\quad \frac{\partial g_3^r}{\partial x_3}$, $\quad \frac{\partial g_3^i}{\partial y_3^i}$, $\frac{\partial g_4^r}{\partial x_4}$ (mult. 2), $\frac{\partial g_5^r}{\partial x_5}$ (mult. 2), $\frac{\partial g_6^r}{\partial x_6}$ (mult. 2), $\quad 0$ (mult. 2) |
| Simple Triangles $y(i,i,i,0,0,0)$ | $\frac{\partial g_1^i}{\partial y_1} + 2\frac{\partial g_1^i}{\partial y_2}$ (a), $\quad \frac{\partial g_1^i}{\partial y_1} - \frac{\partial g_1^i}{\partial y_2}$ (a) (mult. 2), $\frac{\partial g_4^r}{\partial x_4}$ (mult. 6), $\quad 3\frac{\partial g_1^r}{\partial x_1}$, $\quad 0$ (mult. 2) |
| Anti-Hexagons $x(1,1,1,-1,-1,-1)$ | Same as Super Hexagons. See Table 12 |
| Super Triangles $y(i,i,i,i,i,i)$ | $\frac{\partial g_1^i}{\partial y_1} + \frac{\partial g_1^i}{\partial y_2} + \frac{\partial g_1^i}{\partial y_3} + \frac{\partial g_1^i}{\partial y_4} + \frac{\partial g_1^i}{\partial y_5} + \frac{\partial g_1^i}{\partial y_6}$, $\frac{\partial g_1^i}{\partial y_1} + \frac{\partial g_1^i}{\partial y_2} + \frac{\partial g_1^i}{\partial y_3} - \frac{\partial g_1^i}{\partial y_4} - \frac{\partial g_1^i}{\partial y_5} - \frac{\partial g_1^i}{\partial y_6}$, $\frac{\partial g_1^r}{\partial x_1} + \frac{\partial g_1^r}{\partial x_2} + \frac{\partial g_1^r}{\partial x_3} + \frac{\partial g_1^r}{\partial x_4} + \frac{\partial g_1^r}{\partial x_5} + \frac{\partial g_1^r}{\partial x_6}$, $\frac{\partial g_1^r}{\partial x_1} + \frac{\partial g_1^r}{\partial x_2} + \frac{\partial g_1^r}{\partial x_3} - \frac{\partial g_1^r}{\partial x_4} - \frac{\partial g_1^r}{\partial x_5} - \frac{\partial g_1^r}{\partial x_6}$, $2\frac{\partial g_1^r}{\partial x_1} - \frac{\partial g_1^r}{\partial x_2} - \frac{\partial g_1^r}{\partial x_3}$ (mult. 2), $0$ (mult. 2), $\mu_1$, $\mu_2$ (mult. 2) $\mu_1 \mu_2 = \frac{1}{2}\left\{ \left(\frac{\partial g_1^i}{\partial y_1} - \frac{\partial g_1^i}{\partial y_2}\right)^2 + \left(\frac{\partial g_1^i}{\partial y_1} - \frac{\partial g_1^i}{\partial y_3}\right)^2 + \left(\frac{\partial g_1^i}{\partial y_2} - \frac{\partial g_1^i}{\partial y_3}\right)^2 - \left(\frac{\partial g_1^i}{\partial y_4} - \frac{\partial g_1^i}{\partial y_5}\right)^2 \right.$ $\left. - \left(\frac{\partial g_1^i}{\partial y_4} - \frac{\partial g_1^i}{\partial y_6}\right)^2 - \left(\frac{\partial g_1^i}{\partial y_5} - \frac{\partial g_1^i}{\partial y_6}\right)^2 \right\}$, $\mu_1 + \mu_2 = 2\frac{\partial g_1^i}{\partial y_1} - \frac{\partial g_1^i}{\partial y_2} - \frac{\partial g_1^i}{\partial y_3}$ |
| Anti-Triangles $y(i,i,i,-i,-i,-i)$ | Same as Super Triangles. |

(a) Here the effect on $\mathbf{Dg}$ of the hidden symmetry $\tilde{\tau}_{x_1}$ (3.1) is included.

Table 16: Stability results for the hexagonal lattice bifurcation problem with $\Gamma = \Gamma_h + \mathbf{Z}_2$, from Tables 12 and 15, and equations 5.6 and 5.33. For rolls and other rhombs, see Table 13.

| Axial Planform | Signs of Nonzero Eigenvalues | Branching Equation ($x \in \mathbf{R}$) and $\mathbf{z} \in \text{Fix}(\Sigma)$ |
|---|---|---|
| $\text{Rh}_{h0}$ | $sgn(a_1 + a_2)$, $sgn(a_1 - a_2)$, $-sgn(a_1 - a_2)$, $sgn(a_4 + a_6 - a_1 - a_2)$, $sgn(a_4 + a_5 - a_1 - a_2)$, $sgn(a_5 + a_6 - a_1 - a_2)$ | $\lambda x + (a_1 + a_2)x^3 + \mathcal{O}(x^5) = 0$, and $\mathbf{z} = (x, x, 0, 0, 0, 0)$ |
| SiH | $sgn(a_1 + 2a_2)$, $sgn(a_1 - a_2)$, $sgn(a_4 + a_5 + a_6 - a_1 - 2a_2)$, $-sgn(d_2)$ | $\lambda x + (a_1 + 2a_2)x^2 + \mathcal{O}(x^5) = 0$, and $\mathbf{z} = (x, x, x, 0, 0, 0)$ |
| SiT | $sgn(a_1 + 2a_2)$, $sgn(a_1 - a_2)$, $sgn(a_4 + a_5 + a_6 - a_1 - 2a_2)$, $sgn(d_2)$ | $\lambda x + (a_1 + 2a_2)x^3 + \mathcal{O}(x^5) = 0$, and $\mathbf{z} = (ix, ix, ix, 0, 0, 0)$ |
| Common to $\text{SuH}_{\alpha,\beta}$, $\text{AH}_{\alpha,\beta}$, $\text{SuT}_{\alpha,\beta}$, $\text{AT}_{\alpha,\beta}$ | $sgn(a_1 + 2a_2 + a_4 + a_5 + a_6)$, $sgn(a_1 + 2a_2 - a_4 - a_5 - a_6)$, $sgn(\mu_1 + \mu_2) = sgn(a_1 - a_2)$, $sgn(\mu_1 \mu_2) = sgn[2(a_1 - a_2)^2 - (a_4 - a_5)^2 - (a_4 - a_6)^2 - (a_5 - a_6)^2]$, and signs of remaining eigenvalues given below. | $\lambda x + (a_1 + 2a_2 + a_4 + a_5 + a_6)x^3 + \mathcal{O}(x^5) = 0$, and $\mathbf{z}$ given below. |
| $\text{SuH}_{\alpha,\beta}$ | $-sgn[(2\alpha - \beta)e_1 + (\alpha + \beta)e_2]$, $-sgn(d_2 + d_3)$, $-sgn[3(d_2 + d_4) - (2\beta - \alpha)(e_1 - e_2)x^{2(\alpha-3)}]$ [a] | $\mathbf{z} = (x, x, x, x, x, x)$ |
| $\text{AH}_{\alpha,\beta}$ | $(-1)^{\alpha+1} sgn[(2\alpha - \beta)e_1 + (\alpha + \beta)e_2]$, $-sgn(d_2 - d_3)$, $-sgn[3(d_2 - d_4) - (-1)^\alpha(2\beta - \alpha)(e_1 - e_2)x^{2(\alpha-3)}]$ [a] | $\mathbf{z} = (x, x, x, -x, -x, -x)$ |
| $\text{SuT}_{\alpha,\beta}$ | $(-1)^{\alpha+1} sgn[(2\alpha - \beta)e_1 + (\alpha + \beta)e_2]$, $sgn(d_2 - d_3)$, $sgn[3(d_2 + d_4) + (-1)^\alpha(2\beta - \alpha)(e_1 - e_2)x^{2(\alpha-3)}]$ [a] | $\mathbf{z} = (ix, ix, ix, ix, ix, ix)$ |
| $\text{AT}_{\alpha,\beta}$ | $-sgn[(2\alpha - \beta)e_1 + (\alpha + \beta)e_2]$, $sgn(d_2 + d_3)$, $sgn[3(d_2 - d_4) + (2\beta - \alpha)(e_1 - e_2)x^{2(\alpha-3)}]$ [a] | $\mathbf{z} = (ix, ix, ix, -ix, -ix, -ix)$ |

[a] Note that the $(\alpha, \beta)$-dependent terms can be neglected here for all cases except $(\alpha, \beta) = (3, 2)$.

2. If simple triangles and simple hexagons are neutrally stable at cubic order, then one and only one of the two branches is stable. The relative stability properties of these two solutions is determined at quintic order.

3. It is possible for super hexagons, anti-hexagons, super triangles and anti-triangles to be unstable, even if they are all neutrally stable at cubic order.

4. It is possible for all of the axial planforms to bifurcate supercritically, but none be stable.

5. If *any* axial solution branch bifurcates subcritically, then rolls, super hexagons, super triangles, anti-hexagons and anti-triangles are all unstable.

6. If rolls or super hexagons bifurcate subcritically, then *all* axial planforms are unstable at bifurcation.

7. If simple hexagons and simple triangles are the only axial solution branches to bifurcate subcritically, then it is still possible that one, but not more, of the rhombs solutions is stable. Similarly, if rhombs $Rh_{hj,\alpha,\beta}$ ($j = 1, 2$, or $3$) is the only axial solution branch to bifurcate subcritically, then it is possible for simple hexagons (or simple triangles) to be stable, or for one or more of the remaining rhombs solutions to be stable.

8. If simple hexagons and the rhombs $Rh_{h0}$ bifurcate subcritically, then it is possible of one, but not more, of the other rhombs to be stable. However, if simple hexagons and one of the rhombs other than $Rh_{h0}$ bifurcate subcritically, then all axial solution branches are unstable.

9. If two of the rhombs $Rh_{hj,\alpha,\beta}$ ($j = 1, 2, 3$) solution branches bifurcate subcritically, then it is possible that one, and only one, of the following solutions is stable: the remaining rhombs $Rh_{hj,\alpha,\beta}$, simple hexagons or simple triangles. However, if $Rh_{h0}$ and one of the other rhombs bifurcate subcritically, or if all three of the rhombs $Rh_{hj,\alpha,\beta}$ solution branches are subcritical, then all of the axial solutions are unstable.

10. If $\alpha$ is odd, then the only solution branches that can co-exist stably are simple hexagons (or simple triangles) and the rhombs $Rh_{h1,\alpha,\beta}$, $Rh_{h2,\alpha,\beta}$, $Rh_{h3,\alpha,\beta}$. Any combination of two of these states can bifurcate stably. It is also possible for all three types of rhombs to be stable simultaneously. However, if two or more of the rhombs are stable, then simple hexagons and simple triangles are unstable. If $\alpha$ is even, then it is also possible for super hexagons and anti-hexagons to be stable simultaneously, or for super triangles and anti-triangles to both be stable.

# 6 Conclusions.

We have investigated steady, spatially-periodic planforms which bifurcate from a spatially-uniform time-independent solution of $E(2)$-equivariant and $E(2) + \mathbf{Z}_2$-equivariant PDEs. We have done this within the framework of finite-dimensional equivariant normal forms of steady state bifurcation problems. Our analysis applies on a center manifold associated with a doubly-periodic solution space of the PDEs. We considered separately the cases where the solutions are doubly-periodic on a square lattice and on a hexagonal lattice.

In the case of generic bifurcation problems with $\mathbf{D}_6\dot{+}\mathbf{T}^2$-symmetry, a result of Ihrig and Golubitsky [15] ensures that all of the axial solutions on the hexagonal lattice bifurcate unstably due to the presence of a quadratic term in the normal form. This means that not only rolls and simple hexagons are unstable at bifurcation, but also a countable set of rhombs are unstable. In order to capture stable solutions within a local bifurcation analysis, we considered a degenerate bifurcation problem in which the coefficient of the quadratic term is zero. We also considered the case in which an extra reflection symmetry ensures that no even terms are present in the normal form.

For both lattices we determined the stability of the planforms which are guaranteed to bifurcate from the trivial solution by the equivariant branching lemma [12, 27]. In order to do this, we derived the normal form of each bifurcation problem. An order $2(\alpha+\beta)-1$ truncation of the normal form is required to completely determine the signs of the eigenvalues for all axial planforms periodic on a square lattice. In the case of the hexagonal lattice, an order $(2\alpha-1)$ truncation is necessary. However, an important practical consideration is that much is already determined at cubic order.

Previous studies focused on the "small box" limit for which the size of the periodic domain coincides with the wavelength of the instability; this leads to a bifurcation problem on $\mathbf{C}^2$ for the square lattice and a bifurcation problem on $\mathbf{C}^3$ for the hexagonal lattice. We have used $\mathbf{C}^4$ and $\mathbf{C}^6$ representations respectively for the symmetry groups associated with the square and hexagonal lattices. These apply when the periodicity of the lattice is much greater than the wavelength of the instability. This analysis extends the results of the earlier $\mathbf{C}^2$ and $\mathbf{C}^3$ bifurcation studies, both enlarging the number of planforms which are supported by the lattice and allowing for a wider class of disturbances in the stability analysis. For example, by considering all of the irreducible representations the stability of rolls, simple squares, simple hexagons, and simple triangles, to an infinite number of perturbations, can be determined. In particular, our bifurcation analysis provides a framework for addressing the relative stability of simple hexagons (or simple squares) and a countably-infinite set of rhombs. This is of particular interest in light of recent laboratory experiments on chemical Turing patterns in which a transition from a hexagonal pattern to a rhombic one is observed [14]. Our unfolding of the degenerate bifurcation problem on the hexagonal lattice provides a mathematical setting for investigating such a transition.

There is no general bifurcation theoretic framework for computing the relative stability of squares and hexagons since no lattice supports them both. However, we are able to compute the stability of hexagons relative to rhombs, which are "almost square", *i.e.*, which are composed of rectangles with aspect ratio that is close to 1 (see Table 9). For example, for the representation of $\mathbf{D}_6\dot{+}\mathbf{T}^2$ with $(\alpha,\beta)=(4,3)$, the rhombs $\mathrm{Rh}_{h3,4,3}$ are made up of rectangles with aspect ratio approximately 0.96; the angle between the wave vectors $\mathbf{K}_1$ and $\mathbf{K}_6$ in this case is about $92°$.

By not requiring the periodicity $\ell$ of the lattice to coincide with the wavelength of the instability $1/k_c$, we were able to investigate axial solution branches with periodicity $1/k_c$ and simultaneously solution branches that have fundamental periodicity $\ell \gg 1/k_c$. We called the latter states super squares, super hexagons, super triangles, and anti-squares, anti-hexagons, anti-triangles. For $E(2)$-equivariant PDEs, there is an infinite family of these solution branches that is parameterized by an integer pair $(\alpha, \beta)$. For the square lattice, the periodicity of these axial planforms is $\ell = \sqrt{\alpha^2 + \beta^2}/k_c$, while for the hexagonal lattice $\ell = \sqrt{\alpha^2 + \beta^2 - \alpha\beta}/k_c$. The wavelength of the instability $1/k_c$ determines the scale of the internal structure of these super and anti-states, while $\ell$ determines their periodicity. By increasing $\alpha$ and $\beta$ we obtain axial planforms that are periodic on larger and larger scales, all of which bifurcate from the trivial solution at $\lambda = 0$. This is perhaps interesting in light of recent hydrodynamic experiments on quasi-patterns [10]. We emphasize, however, that the existence of a center manifold in the quasi-periodic case has not been established. Thus our analysis, which assumes that there is a center manifold, breaks down in the limit that $\ell$ goes to infinity.

## Acknowledgements

We would like to thank I. Melbourne for helpful discussions. This work was begun during a visit by the authors to the Fields Institute, and we are grateful to the Fields Institute for support. The research of BD was supported by the Natural Sciences and Engineering Research Council of Canada. The research of MS was supported by NSF grants DMS-9410115 and DMS-9404266, and by an NSF CAREER award DMS-9502266. The research of ACS was supported by the EPSRC under grant GR/K41311.

## 7 Appendix.

In this appendix we give some of the details of our calculations of the isotropy subgroups that are presented in section 3. The computations are divided into two parts. In the first part, we compute the isotropy subgroups $\Sigma$ of $\Gamma$ which are *translation free*, i.e., for which $\Sigma \cap \mathbf{T}^2 = \{Id\}$, where $Id$ is the identity element in $\Gamma$. In the second part, we consider the isotropy subgroups of $\Gamma$ which are not translation free.

The computation of the translation free subgroups of $\Gamma$ is simplified by the following observation. Let

$$\Pi_{\mathbf{H}} : \Gamma \to \mathbf{H} \tag{7.1}$$

be the projection of $\Gamma$ into the holohedry $\mathbf{H}$ defined by

$$(h, \Theta) \mapsto h . \tag{7.2}$$

(In the case that $\Gamma = \Gamma_h + \mathbf{Z}_2$, let $\Pi_{\mathbf{H}} : \Gamma \to \mathbf{D}_6 + \mathbf{Z}_2$.) The projection $\Pi_{\mathbf{H}}$ is a group homomorphism. If $\Sigma \subset \Gamma$ is translation free then $\Sigma$ is $\Pi_{\mathbf{H}}$-isomorphic to a subgroup of $\mathbf{H}$ because $\ker(\Pi_{\mathbf{H}}) \subset \Sigma \cap \mathbf{T}^2 = \{Id\}$. Moreover, if $\Sigma_1$ and $\Sigma_2$ are conjugate subgroups of $\Gamma$ then $\Pi_{\mathbf{H}}(\Sigma_1)$ and $\Pi_{\mathbf{H}}(\Sigma_2)$ are conjugate subgroups of $\mathbf{H}$.

To find the translation free isotropy subgroups of $\Gamma$, we proceed as follows:

1. We list all the subgroups $\mathbf{G}$ of $\mathbf{H}$ up to conjugacy.

2. For each subgroup **G** we then compute, up to conjugacy, all subgroups **K** of Γ that are isomorphic by $\Pi_\mathbf{H}$ to **G**.

3. We determine which of the subgroups **K** are *isotropy* subgroups.

For the isotropy subgroups $\Sigma = \Sigma_\mathbf{z}$ that are not translation free, we first show that certain components of **z** must be zero for it to possess a nontrivial translation symmetry. We then classify the isotropy subgroups associated with these points **z**.

## 7.1 Square lattice case.

In this section, we compute the isotropy subgroups for the eight-dimensional representations of $\Gamma_s = \mathbf{D}_4 \dot{+} \mathbf{T}^2$.

**Translation free isotropy subgroups.**

**Proposition 7.1** *Up to conjugacy, the subgroups of $\mathbf{D}_4[R_{\pi/2}, \tau_{x_1}]$ are* **1**, $\mathbf{Z}_2^c[R_\pi]$, $\mathbf{Z}_2^x[\tau_{x_1}]$, $\mathbf{Z}_2^d[\tau_d]$, $\mathbf{Z}_4[R_{\pi/2}]$, $\mathbf{D}_2^x[R_\pi, \tau_{x_1}]$, $\mathbf{D}_2^d[R_\pi, \tau_d]$ *and* $\mathbf{D}_4[R_{\pi/2}, \tau_{x_1}]$.

The proof follows from elementary group theory.

**Proposition 7.2** *Up to conjugacy, the translation free subgroups of $\Gamma_s$ are*

$$\begin{array}{llll}
\mathbf{1}, & \mathbf{Z}_2^c[R_\pi], & \mathbf{Z}_2^x[\tau_{x_1}], & \mathbf{Z}_2^d[\tau_d], \\
\widehat{\mathbf{Z}}_2^x[(\tau_{x_1}, (1/2, 0))], & \mathbf{Z}_4[R_{\pi/2}], & \mathbf{D}_2^x[R_\pi, \tau_{x_1}], & \widehat{\mathbf{D}}_2^x[R_\pi, (\tau_{x_1}, (1/2, 0))], \\
\widetilde{\mathbf{D}}_2^x[R_\pi, (\tau_{x_1}, (1/2, 1/2))], & \mathbf{D}_2^d[R_\pi, \tau_d], & \widetilde{\mathbf{D}}_4[R_{\pi/2}, (\tau_{x_1}, (1/2, 1/2))], & \mathbf{D}_4[R_{\pi/2}, \tau_{x_1}].
\end{array}$$

The proof can be found in [8], where the translation free subgroups are called "shifted subgroups".

**Proposition 7.3** *The subgroups of Proposition 7.2 are all isotropy subgroups.*

**Proof:** To show that a subgroup is an isotropy subgroup, it is sufficient to find a point $\mathbf{z} \in \mathbf{C}^4$ with symmetry $\Sigma$, *i.e.*, to show that $\Sigma = \Sigma_\mathbf{z} \equiv \{\gamma \in \Gamma_s : \gamma(\mathbf{z}) = \mathbf{z}\}$ for some $\mathbf{z} \in \mathbf{C}^4$. The conclusion of the proposition follows from

$$\begin{array}{rclrcl}
\mathbf{D}_4 & = & \Sigma_{(1,1,1,1)} & \widetilde{\mathbf{D}}_4 & = & \Sigma_{(1,1,-1,-1)} \\
\mathbf{D}_2^d & = & \Sigma_{(1,2,1,2)} & \widetilde{\mathbf{D}}_2^x & = & \Sigma_{(1,2,-1,-2)} \\
\mathbf{D}_2^x & = & \Sigma_{(1,2,2,1)} & \widehat{\mathbf{D}}_2^x & = & \left\{ \begin{array}{ll} \Sigma_{(1,2,2,-1)} & \text{if } \alpha \text{ is odd} \\ \Sigma_{(1,2,-2,1)} & \text{if } \beta \text{ is odd} \end{array} \right. \\
\mathbf{Z}_4 & = & \Sigma_{(1,1,3,3)} & \mathbf{Z}_2^d & = & \Sigma_{(i,2i,i,-2i)} \\
\mathbf{Z}_2^x & = & \Sigma_{(i,2i,-2i,-i)} & \widehat{\mathbf{Z}}_2^x & = & \left\{ \begin{array}{ll} \Sigma_{(i,2i,-2i,i)} & \text{if } \alpha \text{ is odd} \\ \Sigma_{(i,2i,2i,-i)} & \text{if } \beta \text{ is odd} \end{array} \right. \\
\mathbf{Z}_2^c & = & \Sigma_{(1,2,3,4)} & \mathbf{1} & = & \Sigma_{(i,2i,3i,4i)}
\end{array}$$

The computations are simplified by the following observations:

1. The action of $\Gamma_s$ on $\mathbf{C}^4$ is translation free.

2. The action of $\mathbf{D}_4$ on $\mathbf{C}^4$ permutes and/or conjugates the coordinates of $\mathbf{z}$.

Suppose that $(h, \Theta)(\mathbf{z}) = \mathbf{z}$ for one of the $\mathbf{z}$ above. Because of 2, the only possible nontrivial action of $\Theta$ on $\mathbf{z} \in \mathbf{C}^4$ is to multiply some of the coordinates by $-1$. Hence $2\Theta$ acts trivially on $\mathbf{C}^4$. By 1, this implies that $2\Theta = \mathbf{0}$ in $\mathbf{T}^2$. Therefore we only need to consider $(h, \Theta)$ where $\Theta = \mathbf{0}, \frac{1}{2}\ell_1, \frac{1}{2}\ell_2$ or $\frac{1}{2}\ell_1 + \frac{1}{2}\ell_2$ and $h \in \mathbf{D}_4$. ∎

### Non-translation-free isotropy subgroups.

We now find the isotropy subgroups $\Sigma \subset \Gamma_s$ such that $\Sigma \cap \mathbf{T}^2 \neq \{Id\}$. Our approach relies on first showing that if there is a nontrivial translation that acts trivially on some $\mathbf{z} \in \mathbf{C}^4$, then at least two of the coordinates of $\mathbf{z}$ must be zero. This follows from the next lemma.

**Lemma 7.4** *If $\Theta \in \mathbf{T}^2$ acts trivially on three coordinates of $\mathbf{z} \in \mathbf{C}^4$, then $\Theta = \mathbf{0} \in \mathbf{T}^2$.*

**Proof:** It is sufficient to show that if $\Theta$ acts trivially on $z_j$ for $j = 1, 2$ and 3, then $\Theta$ also acts trivially on $z_4$. Because the action of $\Gamma_s$ on $\mathbf{C}^4$ is translation free, it then follows that $\Theta = \mathbf{0} \in \mathbf{T}^2$. We now show that
$$\mathbf{K}_4 = a\mathbf{K}_1 + b\mathbf{K}_2 + c\mathbf{K}_3 , \tag{7.3}$$
where $a$, $b$ and $c$ are integers. The conclusion of the lemma follows immediately from (7.3), since if $\mathbf{K}_1 \cdot \Theta, \mathbf{K}_2 \cdot \Theta, \mathbf{K}_3 \cdot \Theta \in \mathbf{Z}$, then $\mathbf{K}_4 \cdot \Theta \in \mathbf{Z}$. If we substitute the expressions for the $\mathbf{K}_j$'s given in Table 1 into (7.3), we find that
$$\begin{aligned}(a+1)\alpha - (b-c)\beta &= 0 , \\ (a-1)\alpha + (b+c)\beta &= 0 .\end{aligned} \tag{7.4}$$

The following statements are equivalent:

(i) There exist three integers $a$, $b$ and $c$ such that (7.4) is satisfied.

(ii) There exist integers $a$, $b$, $c$, $r$ and $s$ such that
$$\begin{aligned}a+1 &= r\beta , & a-1 &= s\alpha , \\ b-c &= r\alpha , & b+c &= -s\beta .\end{aligned}$$

(iii) There exist integers $a$, $b$, $c$, $r$ and $s$ such that
$$\begin{aligned}2a &= r\beta + s\alpha , & 2b &= r\alpha - s\beta , \\ 2c &= -r\alpha - s\beta , & 2 &= r\beta - s\alpha .\end{aligned}$$

Statements (i) and (ii) are equivalent because $\alpha$ and $\beta$ are relatively prime. Statements (ii) and (iii) are equivalent because the two systems of equations are equivalent.

It follows from a well known result of number theory that if $(\alpha, \beta) = 1$, then there exist integers $r'$ and $s'$ such that $1 = r'\beta - s'\alpha$. Setting $r = 2r'$ and $s = 2s'$ proves (iii). ∎

Lemma 7.4 ensures that the non-translation-free isotropy subgroups are, up to conjugacy, $\Sigma = \Sigma_{\mathbf{z}}$, where $\mathbf{z}$ satisfies one of the following conditions:

1. $\mathbf{z} = \mathbf{0}$

2. $\mathbf{z} = (z_1, 0, 0, 0)$, $z_1 \in \mathbf{R}$, $z_1 > 0$

3. $\mathbf{z} = (z_1, z_2, 0, 0)$, $z_1, z_2 \in \mathbf{R}$, $z_1, z_2 > 0$

4. $\mathbf{z} = (z_1, 0, z_3, 0)$, $z_1, z_3 \in \mathbf{R}$, $z_1, z_3 > 0$

5. $\mathbf{z} = (z_1, 0, 0, z_4)$, $z_1, z_4 \in \mathbf{R}$, $z_1, z_4 > 0$

Note that we have exploited the $\mathbf{D}_4$ symmetry to assume, without loss of generality, that $z_1 \neq 0$ whenever $\mathbf{z} \neq \mathbf{0}$. Moreover, the $\mathbf{T}^2$ symmetry allows us to assume that the coordinates of $\mathbf{z}$ are real and nonnegative for the non-translation-free isotropy subgroups $\Sigma = \Sigma_{\mathbf{z}}$.

It is now a simple computation to determine, up to conjugacy, the non-translation-free isotropy subgroups listed in table 3. Presented below are the details of the computations for case (3). The others are similar.

Let $z_1$ and $z_2$ be the only nonzero coordinates of $\mathbf{z}$. The largest subgroup of $\mathbf{T}^2$ acting trivially on $z_1$ and $z_2$ is $\mathbf{S}_{1,2}$ given in Table 3. We obtain $\mathbf{S}_{1,2}$ by solving

$$\mathbf{K}_1 \cdot \Theta = n_1 \in \mathbf{Z} \quad \text{and} \quad \mathbf{K}_2 \cdot \Theta = n_2 \in \mathbf{Z}$$

for $\Theta \in \mathbf{T}^2$, where $\mathbf{K}_1$ and $\mathbf{K}_2$ are given in Table 1. We find that

$$\Theta = n_1 \left( \frac{\alpha}{\alpha^2 + \beta^2}, \frac{\beta}{\alpha^2 + \beta^2} \right) + n_2 \left( \frac{-\beta}{\alpha^2 + \beta^2}, \frac{\alpha}{\alpha^2 + \beta^2} \right) .$$

The group $\mathbf{S}_{1,2}$ does not act trivially on any of the other coordinates because of Lemma 7.4 and the fact that $\mathbf{S}_{1,2} \neq \{Id\}$.

We used the action of $\mathbf{T}^2$ to assume that $\Sigma = \Sigma_{\mathbf{z}}$ where $z_1$ and $z_2$ are positive real numbers. Up to conjugacy, we obtain the following isotropy subgroups:

$$\Sigma = \Sigma_{\mathbf{z}} = \mathbf{Z}_2^c \dot{+} \mathbf{S}_{1,2} \quad \text{if } z_1 \neq z_2$$

and

$$\Sigma = \Sigma_{\mathbf{z}} = \mathbf{Z}_4 \dot{+} \mathbf{S}_{1,2} \quad \text{if } z_1 = z_2 .$$

Note that $\mathbf{Z}_4$ is the largest subgroup of $\mathbf{D}_4$ that preserves the set $\{\pm\mathbf{K}_1, \pm\mathbf{K}_2\}$.

## 7.2 Hexagonal lattice case: $\Gamma = \Gamma_h$.

In this section, we compute the isotropy subgroups for the twelve-dimensional representations of $\Gamma_h = \mathbf{D}_6 \dot{+} \mathbf{T}^2$.

**Translation free isotropy subgroups.**

As for the square lattice case, if $\Sigma \subset \Gamma_h$ is translation free then $\Sigma$ is $\Pi_{\mathbf{D}_6}$-isomorphic to a subgroup of $\mathbf{D}_6$. Thus, to find all translation free isotropy subgroups of $\Gamma_h = \mathbf{D}_6 \dot{+} \mathbf{T}^2$, we first identify those subgroups of $\Gamma_h$ that are isomorphic by $\Pi_{\mathbf{D}_6}$ to a subgroup of $\mathbf{D}_6$.

**Proposition 7.5** *Up to conjugacy, the subgroups of $\mathbf{D}_6[R_{\pi/3}, \tau_{x_1}]$ are $\mathbf{1}$, $\mathbf{Z}_2^c[R_\pi]$, $\mathbf{Z}_2^x[\tau_{x_1}]$, $\mathbf{Z}_2^n[\tau_n]$, $\mathbf{Z}_3[R_{\pi/3}^2]$, $\mathbf{Z}_6[R_{\pi/3}]$, $\mathbf{D}_2^x[R_\pi, \tau_{x_1}]$, $\mathbf{D}_3[R_{\pi/3}^2, \tau_{x_1}]$, $\mathbf{D}_3^n[R_{\pi/3}^2, \tau_n]$ and $\mathbf{D}_6[R_{\pi/3}, \tau_{x_1}]$.*

The proof follows from elementary group theory.

**Proposition 7.6** *Up to conjugacy, the translation free subgroups of $\Gamma_h$ are the subgroups of $\mathbf{D}_6$ given in Proposition 7.5.*

The proof can be found in [8].

**Proposition 7.7** *The subgroups of Proposition 7.5 are all isotropy subgroups.*

**Proof:** The conclusion of the proposition follows from

$$\begin{array}{rclrcl}
\mathbf{D}_6 & = & \Sigma_{(1,1,1,1,1,1)} & \mathbf{D}_3^n & = & \Sigma_{(i,i,i,-i,-i,-i)} \\
\mathbf{D}_3 & = & \Sigma_{(i,i,i,i,i,i)} & \mathbf{Z}_6 & = & \Sigma_{(1,1,1,2,2,2)} \\
\mathbf{D}_2^x & = & \Sigma_{(1,2,3,3,2,1)} & \mathbf{Z}_3 & = & \Sigma_{(i,i,i,2i,2i,2i)} \\
\mathbf{Z}_2^n & = & \Sigma_{(i,2i,3i,-i,-3i,-2)} & \mathbf{Z}_2^x & = & \Sigma_{(i,2i,3i,3i,2i,i)} \\
\mathbf{Z}_2^c & = & \Sigma_{(1,2,3,4,5,6)} & \mathbf{1} & = & \Sigma_{(i,2i,3i,4i,5i,6i)}
\end{array}$$

The computations are similar to those for Proposition 7.3. In particular, if $(h, \Theta)(\mathbf{z}) = \mathbf{z}$ for one of the $\mathbf{z}$ above, then $2\Theta = \mathbf{0}$ in $\mathbf{T}^2$. Therefore we need only consider $(h, \Theta)$ where $\Theta = \mathbf{0}, \frac{1}{2}\ell_1$, $\frac{1}{2}\ell_2$ or $\frac{1}{2}\ell_1 + \frac{1}{2}\ell_2$ and $h \in \mathbf{D}_6$. ∎

**Non-translation-free isotropy subgroups.**

We now find the isotropy subgroups $\Sigma \subset \Gamma_h$ such that $\Sigma \cap \mathbf{T}^2 \neq \{Id\}$. We use the following lemma.

**Lemma 7.8** *If $\Theta \in \mathbf{T}^2$ acts trivially on four of the coordinates of $\mathbf{z} \in \mathbf{C}^6$, then $\Theta = \mathbf{0} \in \mathbf{T}^2$. When $\Theta$ acts trivially on only three of the coordinates, then either (1) $\Theta = \mathbf{0} \in \mathbf{T}^2$, or (2) the three coordinates are $z_1, z_2, z_3$, or they are $z_4, z_5, z_6$.*

**Proof:** If $\Theta$ acts trivially on two of the coordinates $z_j$, where j = 1, 2 or 3, then $\Theta$ acts trivially on all three, because $\mathbf{K}_1 + \mathbf{K}_2 + \mathbf{K}_3 = \mathbf{0}$. Similarly, if $\Theta$ acts trivially on two of the coordinates $z_4, z_5, z_6$ then it necessarily acts trivially on all three because $\mathbf{K}_4 + \mathbf{K}_5 + \mathbf{K}_6 = \mathbf{0}$.

We complete the proof by showing that if $\Theta$ acts trivially on $z_j$ for j = 1, 2 and 4, then $\Theta$ also acts trivially on $z_5$; thus it acts trivially on all $\mathbf{z} \in \mathbf{C}^6$. Hence $\Theta = \mathbf{0} \in \mathbf{T}^2$ since the action of $\Gamma_h$ on $\mathbf{C}^6$ is translation free.

Assume that $\Theta$ acts trivially on $z_1, z_2, z_4$. It follows from the observation that if there exist integers $a$, $b$ and $c$ such that
$$\mathbf{K}_5 = a\mathbf{K}_1 + b\mathbf{K}_2 + c\mathbf{K}_4 \tag{7.5}$$
then $\Theta$ also acts trivially on $z_5$. Substituting the expressions for the $\mathbf{K}_j$'s given in Table 2 into (7.5), we obtain the following conditions on $a$, $b$, and $c$:
$$\begin{aligned}(1 - b + c)\alpha + (a - c)\beta &= 0, \\ (a - b + c)\alpha + (b + 1)\beta &= 0.\end{aligned} \tag{7.6}$$

The following statements are equivalent:

**(i)** There exist three integers $a$, $b$ and $c$ such that (7.6) is satisfied.

**(ii)** There exist integers $a$, $b$, $c$, $r$ and $s$ such that
$$\begin{aligned} 1 - b + c &= r\beta, & a - c &= -r\alpha, \\ a - b + c &= s\beta, & b + 1 &= -s\alpha.\end{aligned}$$

**(iii)** There exist integers $a$, $b$, $c$, $r$ and $s$ such that
$$\begin{aligned} a &= r(\beta - \alpha) - s\alpha - 2, & b &= -s\alpha - 1, \\ c &= r\beta - s\alpha - 2, & 3 &= r(2\beta - \alpha) - s(\alpha + \beta).\end{aligned}$$

Statements (i) and (ii) are equivalent because $\alpha$ and $\beta$ are relatively prime. Statements (ii) and (iii) are equivalent because the two systems of equations are equivalent.

The proof proceeds by showing that $(2\beta - \alpha)$ and $(\alpha + \beta)$ are relatively prime. To see this, suppose that $d$ divides both $2\beta - \alpha$ and $\alpha + \beta$, then $d$ divides their sum $3\beta$. Since $(3, \alpha + \beta) = 1$, we know that $d \neq 3$, so $d$ must divide $\beta$ as well as $\alpha + \beta$. Putting this all together we have $d$ divides both $\alpha$ and $\beta$, which are relatively prime so $d = 1$. Hence $(2\beta - \alpha, \alpha + \beta) = 1$ as claimed. It follows that there exist integers $r'$ and $s'$ such that $1 = r'(2\beta - \alpha) - s'(\beta + \alpha)$. Setting $r = 3r'$ and $s = 3s'$ proves (iii). ∎

Lemma 7.8 ensures that the non-translation-free isotropy subgroups are, up to conjugacy, $\Sigma = \Sigma_{\mathbf{z}}$, where $\mathbf{z}$ satisfies one of the following conditions:

1. $\mathbf{z} = \mathbf{0}$

2. $\mathbf{z} = (z_1, 0, 0, 0, 0, 0)$, $z_1 \in \mathbf{R}$, $z_1 > 0$

3. $\mathbf{z} = (z_1, z_2, z_3, 0, 0, 0)$, $z_1, z_2 \in \mathbf{R}$, $z_1, z_2 > 0$, $z_3 \neq 0$

4. $\mathbf{z} = (z_1, z_2, 0, 0, 0, 0)$, $z_1, z_2 \in \mathbf{R}$, $z_1, z_2 > 0$

5. $\mathbf{z} = (z_1, 0, 0, z_4, 0, 0)$, $z_1, z_4 \in \mathbf{R}$, $z_1, z_4 > 0$

6. $\mathbf{z} = (z_1, 0, 0, 0, z_5, 0)$, $z_1, z_5 \in \mathbf{R}$, $z_1, z_5 > 0$

7. $\mathbf{z} = (z_1, 0, 0, 0, 0, z_6)$, $z_1, z_6 \in \mathbf{R}$, $z_1, z_6 > 0$

It is now a straightforward computation to determine, up to conjugacy, the non-translation-free isotropy subgroups in table 4. We present the details of the computations for cases 3 and 4, only.

Assume that $z_4 = z_5 = z_6 = 0$ and that $z_1$ and $z_2$ are nonzero coordinates of $\mathbf{z}$. The largest subgroup of $\mathbf{T}^2$ acting trivially on $z_1$ and $z_2$ is $\mathbf{S}_{1,2,3}$ given in Table 4. We obtain $\mathbf{S}_{1,2,3}$ by solving

$$\mathbf{K}_1 \cdot \Theta = n_1 \in \mathbf{Z} \quad \text{and} \quad \mathbf{K}_2 \cdot \Theta = n_2 \in \mathbf{Z}$$

for $\Theta \in \mathbf{T}^2$, where $\mathbf{K}_1$ and $\mathbf{K}_2$ are given in Table 2. It follows from $\mathbf{K}_1 + \mathbf{K}_2 + \mathbf{K}_3 = \mathbf{0}$ that $\mathbf{K}_3 \cdot \Theta$ is also in $\mathbf{Z}$. The group $\mathbf{S}_{1,2,3}$ does not act trivially on any of the other coordinates because of Lemma 7.8 and the fact that $\mathbf{S}_{1,2,3} \neq \{Id\}$.

Note that $\mathbf{Z}_6$ is the largest subgroup of $\mathbf{D}_6$ that preserves the set $\{\pm \mathbf{K}_1, \pm \mathbf{K}_2, \pm \mathbf{K}_3\}$. Hence, isotropy subgroups $\Sigma = \Sigma_\mathbf{z}$ for cases 3 and 4 are determined by considering, in turn, the action of $(R_\pi, \Theta)$, $(R_{2\pi/3}, \Theta)$, and $(R_{\pi/3}, \Theta)$, $\Theta \in \mathbf{T}^2$, on $\mathbf{z} = (r_1, r_2, r_3 e^{2\pi i \varphi}, 0, 0, 0)$, where $r_1, r_2, r_3 \in \mathbf{R}$ with $r_1, r_2 > 0$, $r_3 \geq 0$ and $\varphi \in [0, 1)$. Here we have used the translation symmetry $\mathbf{T}^2$ to assume that $z_1, z_2$ are real and positive for some $\Sigma_\mathbf{z}$ in each conjugacy class. Moreover, we analyze cases 3 and 4 simultaneously by allowing the possibility that $z_3 = 0$ in the following.

1. $(R_\pi, \Theta)\mathbf{z} = \mathbf{z}$. Substituting $-\mathbf{K}_1 - \mathbf{K}_2$ for $\mathbf{K}_3$, we have

$$(R_\pi, \Theta)\mathbf{z} = (e^{-2\pi i (\mathbf{K}_1 \cdot \Theta)} r_1, e^{-2\pi i (\mathbf{K}_2 \cdot \Theta)} r_2, e^{-2\pi i \varphi} e^{2\pi i (\mathbf{K}_1 + \mathbf{K}_2) \cdot \Theta} r_3, 0, 0, 0). \qquad (7.7)$$

Hence $(R_\pi, \Theta)\mathbf{z} = \mathbf{z}$ only if $(Id, \Theta) \in \mathbf{S}_{1,2,3}$ and either $r_3 = 0$ or $\varphi = 0, \frac{1}{2}$, i.e., $z_3 \in \mathbf{R}$.

2. $(R_{2\pi/3}, \Theta)\mathbf{z} = \mathbf{z}$ only if $r_1 = r_2 = r_3 = r$. In this case,

$$(R_{2\pi/3}, \Theta)\mathbf{z} = (e^{2\pi i \varphi} e^{-2\pi i (\mathbf{K}_1 \cdot \Theta)} r, e^{-2\pi i (\mathbf{K}_2 \cdot \Theta)} r, e^{2\pi i (\mathbf{K}_1 + \mathbf{K}_2) \cdot \Theta} r, 0, 0, 0), \qquad (7.8)$$

and $(R_{2\pi/3}, \Theta)\mathbf{z} = \mathbf{z}$ provided we choose $\Theta \in \mathbf{T}^2$ such that $\mathbf{K}_2 \cdot \Theta \in \mathbf{Z}$ and $\mathbf{K}_1 \cdot \Theta - \varphi \in \mathbf{Z}$, which we can always do.

3. Finally, we consider the equation $(R_{\pi/3}, \Theta)\mathbf{z} = \mathbf{z}$, where, again, $r_1 = r_2 = r_3 = r$ and

$$(R_{\pi/3}, \Theta)\mathbf{z} = (e^{-2\pi i (\mathbf{K}_1 \cdot \Theta)} r, e^{-2\pi i \varphi} e^{-2\pi i (\mathbf{K}_2 \cdot \Theta)} r, e^{2\pi i (\mathbf{K}_1 + \mathbf{K}_2) \cdot \Theta} r, 0, 0, 0). \qquad (7.9)$$

Thus $\Theta \in \mathbf{T}^2$ and $\varphi \in [0, 1)$ must satisfy the following equations:

$$\mathbf{K}_1 \cdot \Theta \in \mathbf{Z}, \quad \mathbf{K}_2 \cdot \Theta \pm \varphi \in \mathbf{Z}. \qquad (7.10)$$

Hence $(R_{\pi/3}, \Theta)\mathbf{z} = \mathbf{z}$ for some $\Theta \in \mathbf{T}^2$ only if $\varphi = 0, \frac{1}{2}$.

From the above considerations we obtain, up to conjugacy, the following isotropy subgroups associated with the subspace $\{\mathbf{z} \in \mathbf{C}^6 : z_4 = z_5 = z_6 = 0\}$:

1. $\Sigma = \Sigma_{\mathbf{z}} = \mathbf{S}_{1,2,3}$, where $z_1$, $z_2$ and $z_3$ are not all of the same norm, $z_1$ and $z_2$ are positive real numbers, and $z_3 \in \mathbf{C} \setminus \mathbf{R}$.

2. $\Sigma = \Sigma_{\mathbf{z}} = \mathbf{Z}_2^c \dotplus \mathbf{S}_{1,2,3}$, where $z_1$, $z_2$ and $z_3$ are not all of the same norm, $z_1$ and $z_2$ are positive real numbers, and $z_3 \in \mathbf{R}$. (Note that this gives the isotropy associated with case 4 for which $z_3 = 0$.)

3. $\Sigma = \Sigma_{\mathbf{z}} = \mathbf{Z}_3 \dotplus \mathbf{S}_{1,2,3}$, where $z_1 = z_2 = z_3 \in \mathbf{C} \setminus \mathbf{R}$. Here we have used the observation that $\mathbf{z}$ is on the group orbit of $(r, r, re^{2\pi i \varphi}, 0, 0, 0)$, where $r > 0$ is real and $\varphi \in (0, 1)$, $\varphi \neq \frac{1}{2}$.

4. $\Sigma = \Sigma_{\mathbf{z}} = \mathbf{Z}_6 \dotplus \mathbf{S}_{1,2,3}$, where $z_1 = z_2 = z_3 \in \mathbf{R}$.

## 7.3 Hexagonal lattice case: $\Gamma = \Gamma_h + \mathbf{Z}_2$.

In this section we determine the additional isotropy subgroups that result from enlarging $\Gamma_h$ to $\Gamma_h + \mathbf{Z}_2$.

**Translation free isotropy subgroups.**

Our approach is the same as for the previous two cases. We begin by finding, up to conjugacy, all subgroups of $\Gamma_h + \mathbf{Z}_2$ that are isomorphic by $\Pi_{\mathbf{D}_6 + \mathbf{Z}_2}$ to a subgroup of $\mathbf{D}_6 + \mathbf{Z}_2$.

**Proposition 7.9** *Up to conjugacy, the subgroups of $\mathbf{D}_6[R_{\pi/3}, \tau_{x_1}] + \mathbf{Z}_2[\kappa]$ are*

**first class:** $1$, $\mathbf{Z}_2^c[R_\pi]$, $\mathbf{Z}_2^x[\tau_{x_1}]$, $\mathbf{Z}_2^n[\tau_n]$, $\mathbf{Z}_3[R_{\pi/3}^2]$, $\mathbf{Z}_6[R_{\pi/3}]$, $\mathbf{D}_2^x[R_\pi, \tau_{x_1}]$, $\mathbf{D}_3[R_{\pi/3}^2, \tau_{x_1}]$, $\mathbf{D}_3^n[R_{\pi/3}^2, \tau_n]$, *and* $\mathbf{D}_6[R_{\pi/3}, \tau_{x_1}]$.

**second class:** *Groups of the form $\mathbf{A} + \mathbf{Z}_2$ where $\mathbf{A}$ is one of the groups of the first class.*

**third class:** $\mathbf{Z}_2[((R_\pi, 0), \kappa)]$, $\mathbf{Z}_2[((\tau_{x_1}, 0), \kappa)]$, $\mathbf{Z}_2[((\tau_n, 0), \kappa)]$, $\mathbf{Z}_6[((R_{\pi/3}, 0), \kappa)]$,
$\mathbf{D}_2[((\tau_{x_1}, 0), 0), ((R_\pi, 0), \kappa)]$, $\mathbf{D}_2[((\tau_{x_1}, 0), \kappa), ((R_\pi, 0), 0)]$,
$\mathbf{D}_2[((\tau_{x_1}, 0), \kappa), ((R_\pi, 0), \kappa)]$, $\mathbf{D}_3[((R_{\pi/3}^2, 0), 0), ((\tau_{x_1}, 0), \kappa)]$,
$\mathbf{D}_3[((R_{\pi/3}^2, 0), 0), ((\tau_n, 0), \kappa)]$, $\mathbf{D}_6[((R_{\pi/3}, 0), 0), ((\tau_{x_1}, 0), \kappa)]$,
$\mathbf{D}_6[((R_{\pi/3}, 0), \kappa), ((\tau_{x_1}, 0), 0)]$, *and* $\mathbf{D}_6[((R_{\pi/3}, 0), \kappa), ((\tau_{x_1}, 0), \kappa)]$.

The proof follows from elementary group theory.

**Proposition 7.10** *Up to conjugacy, the translation free subgroups of $\Gamma_h + \mathbf{Z}_2$ are the subgroups given in Proposition 7.9.*

The proof is similar to the proof of Proposition 7.5 given in [8].

**Proposition 7.11** *Only the subgroups of first and third classes in Proposition 7.9 are isotropy subgroups.*

**Proof:** Because $\text{Fix}(\mathbf{Z}_2) = \{\mathbf{0}\}$, we have $\text{Fix}(\Sigma) = \{\mathbf{0}\}$ for any subgroup in the second class. However, the origin has full symmetry $\Gamma_h + \mathbf{Z}_2$, so no subgroup of the second class is an isotropy subgroup.

For the groups in the first class, the conclusion of the proposition follows from

$$\begin{array}{rclcrcl}
\mathbf{D}_6 & = & \Sigma_{(1,1,1,1,1,1)} & & \mathbf{D}_3^n & = & \Sigma_{(1+i,1+i,1+i,1-i,1-i,1-i)} \\
\mathbf{D}_3 & = & \Sigma_{(1+i,1+i,1+i,1+i,1+i,1+i)} & & \mathbf{Z}_6 & = & \Sigma_{(1,1,1,2,2,2)} \\
\mathbf{D}_2^x & = & \Sigma_{(1,2,3,3,2,1)} & & \mathbf{Z}_3 & = & \Sigma_{(1+i,1+i,1+i,1+2i,1+2i,1+2i)} \\
\mathbf{Z}_2^n & = & \Sigma_{(1+i,1+2i,1+3i,1-i,1-3i,1-2i)} & & \mathbf{Z}_2^x & = & \Sigma_{(1+i,1+2i,1+3i,1+3i,1+2i,1+i)} \\
\mathbf{Z}_2^c & = & \Sigma_{(1,2,3,4,5,6)} & & \mathbf{1} & = & \Sigma_{(1+i,1+2i,1+3i,1+4i,1+5i,1+6i)}
\end{array}$$

For the groups in the third class, the conclusion of the proposition follows from

$$\begin{array}{rcl}
\mathbf{D}_6[((R_{\pi/3},0),\kappa),((\tau_{x_1},0),\kappa)] & = & \Sigma_{(i,i,i,-i,-i,-i)} \\
\mathbf{D}_6[((R_{\pi/3},0),\kappa),((\tau_{x_1},0),0)] & = & \Sigma_{(i,i,i,i,i,i)} \\
\mathbf{D}_6[((R_{\pi/3},0),0),((\tau_{x_1},0),\kappa)] & = & \Sigma_{(1,1,1,-1,-1,-1)} \\
\mathbf{D}_3[((R_{\pi/3}^2,0),0),((\tau_{x_1},0),\kappa)] & = & \Sigma_{(1+i,1+i,1+i,-1-i,-1-i,-1-i)} \\
\mathbf{D}_3[((R_{\pi/3}^2,0),0),((\tau_n,0),\kappa)] & = & \Sigma_{(1+i,1+i,1+i,-1+i,-1+i,-1+i)} \\
\mathbf{D}_2[((\tau_{x_1},0),\kappa),((R_\pi,0),\kappa)] & = & \Sigma_{(i,2i,3i,-3i,-2i,-i)} \\
\mathbf{D}_2[((\tau_{x_1},0),\kappa),((R_\pi,0),0)] & = & \Sigma_{(1,2,3,-3,-2,-1)} \\
\mathbf{D}_2[((\tau_{x_1},0),0),((R_\pi,0),\kappa)] & = & \Sigma_{(i,2i,3i,3i,2i,i)} \\
\mathbf{Z}_6[((R_{\pi/3},0),\kappa)] & = & \Sigma_{(i,i,i,2i,2i,2i)} \\
\mathbf{Z}_2[((\tau_{x_1},0),\kappa)] & = & \Sigma_{(1+i,1+2i,1+3i,-1-3i,-1-2i,-1-i)} \\
\mathbf{Z}_2[((\tau_n,0),\kappa)] & = & \Sigma_{(1+i,1+2i,1+3i,-1+i,-1+3i,1+2i)} \\
\mathbf{Z}_2[\kappa] & = & \Sigma_{(i,2i,3i,4i,5i,6i)}
\end{array}$$

The computations are almost identical to those for the proof of Proposition 7.7. ∎

### Non-translation-free isotropy subgroups.

Lemma 7.8 still applies in the present context. We proceed, as in the case of $\Gamma = \Gamma_h$, by determining the isotropy of the following points in $\mathbf{C}^6$:

1. $\mathbf{z} = \mathbf{0}$

2. $\mathbf{z} = (z_1, 0, 0, 0, 0, 0)$, $z_1 \in \mathbf{R}$, $z_1 > 0$

3. $\mathbf{z} = (z_1, z_2, z_3, 0, 0, 0)$, $z_1, z_2 \in \mathbf{R}$, $z_1, z_2 > 0$, $z_3 \neq 0$,

4. $\mathbf{z} = (z_1, z_2, 0, 0, 0, 0)$, $z_1, z_2 \in \mathbf{R}$, $z_1, z_2 > 0$

5. $\mathbf{z} = (z_1, 0, 0, z_4, 0, 0)$, $z_1, z_4 \in \mathbf{R}$, $z_1, z_4 > 0$

6. $\mathbf{z} = (z_1, 0, 0, 0, z_5, 0)$, $z_1, z_5 \in \mathbf{R}$, $z_1, z_5 > 0$

7. $\mathbf{z} = (z_1, 0, 0, 0, 0, z_6)$, $z_1, z_6 \in \mathbf{R}$, $z_1, z_6 > 0$

A straightforward computation determines the non-translation-free isotropy subgroups in table 6. We present the details for case 4, only.

Using the action of $\mathbf{T}^2$, we can assume that $z_1$ and $z_2$ are real. The group $\mathbf{Z}_2^c[R_\pi] + \mathbf{Z}_2[\kappa]$ is the largest subgroup of $\mathbf{D}_6 + \mathbf{Z}_2$ that preserves the set $\{\pm \mathbf{K}_1, \pm \mathbf{K}_2\}$. The element $R_\pi$ acts trivially on $\mathbf{z}$ in this case, and $\kappa(\mathbf{z}) = -\mathbf{z}$. We determine the generators of the new $\mathbf{S}_{1,2}$ by solving

$$\mathbf{K}_1 \cdot \Theta = \frac{1}{2}, \quad \text{and} \quad \mathbf{K}_2 \cdot \Theta = \pm \frac{1}{2},$$

for $\Theta \in \mathbf{T}^2$. We find that $\mathbf{S}_{1,2}$ is generated by $((Id, \frac{\alpha+\beta}{2(\alpha^2-\alpha\beta+\beta^2)}\ell_1 - \frac{2\alpha-\beta}{2(\alpha^2-\alpha\beta+\beta^2)}\ell_2), \kappa)$ and $((Id, \frac{\alpha-\beta}{2(\alpha^2-\alpha\beta+\beta^2)}\ell_1 + \frac{\beta}{2(\alpha^2-\alpha\beta+\beta^2)}\ell_2), \kappa)$.

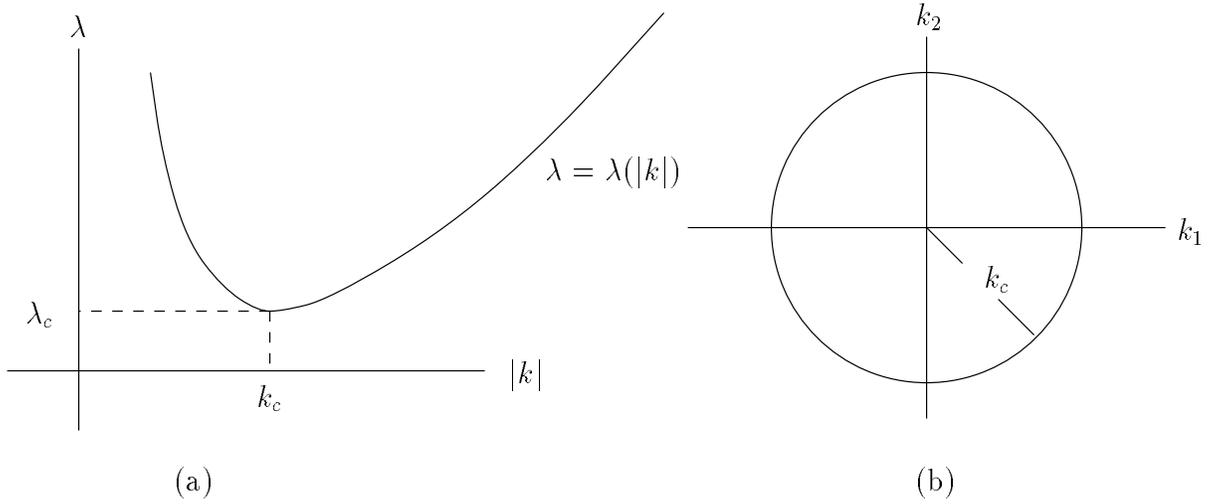

Figure 1: (a) Typical neutral stability curve. (b) Circle of critical wave vectors in **k**-space.

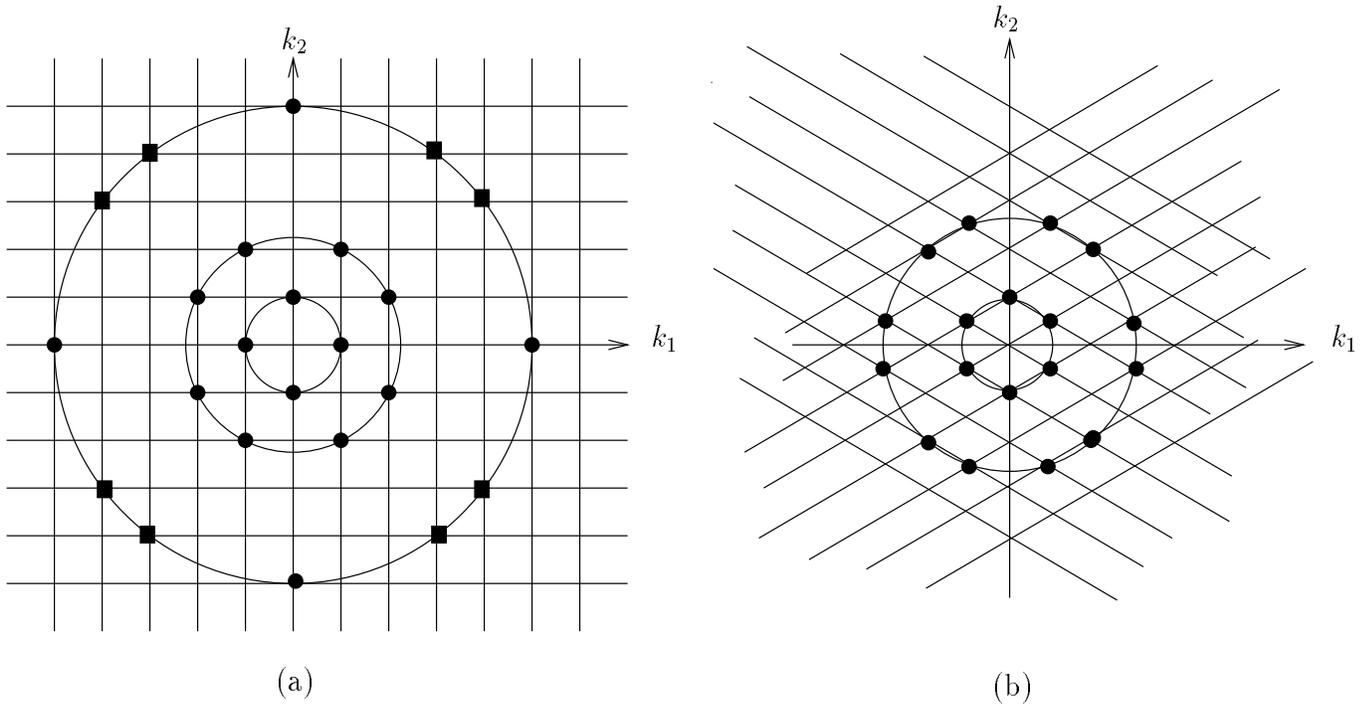

Figure 2: (a) Critical circles for the square lattice when $\mathbf{k}_c = 1$, $\sqrt{3}$ and 5. (b) Critical circles for the hexagonal lattice when $\mathbf{k}_c = 1$ and $\sqrt{7}$.

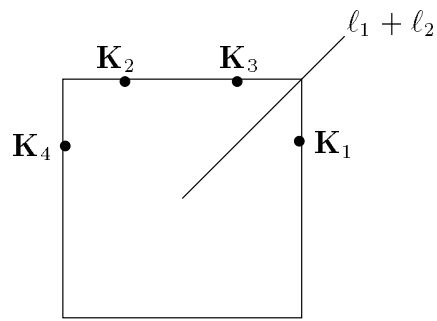

Figure 3: Square lattice wave vectors $\mathbf{K}_j$ for the eight-dimensional representations of $\Gamma_s$ in Table 1.

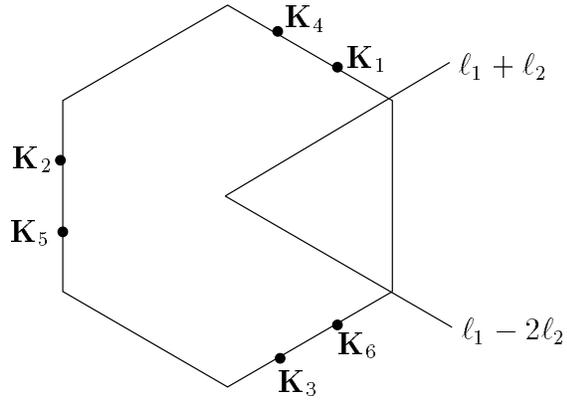

Figure 4: Hexagonal lattice wave vectors $\mathbf{K}_j$ for the twelve-dimensional representations of $\Gamma_h$ in Table 2.

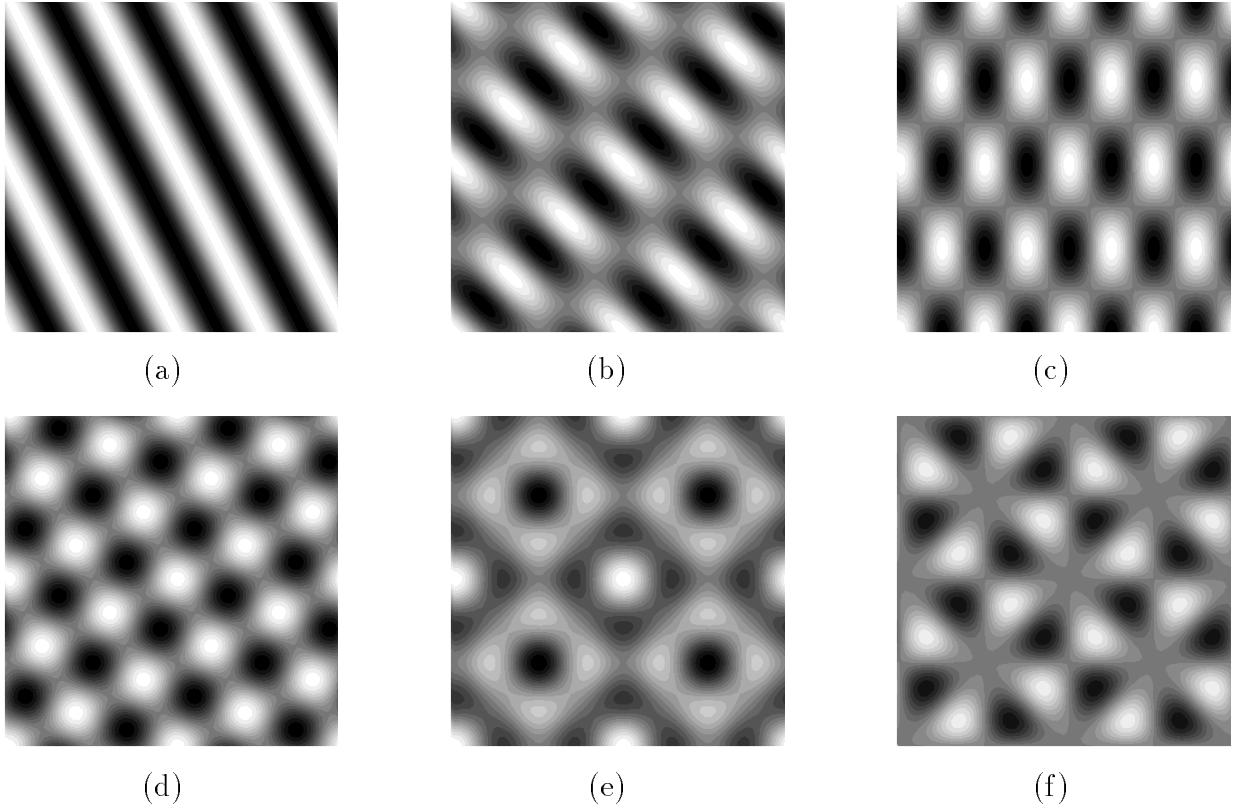

Figure 5: Axial planforms associated with 8-dimensional representation of $\Gamma_s$ with $(\alpha, \beta) = (2, 1)$ and $x_1, x_2 \in [-1, 1]$ (i.e., four copies of the fundamental domain are shown); (a) rolls, (b) rhombs ($\text{Rh}_{s1,2,1}$), (c) rhombs ($\text{Rh}_{s2,2,1}$), (d) simple squares, (e) super squares, and (f) anti-squares.

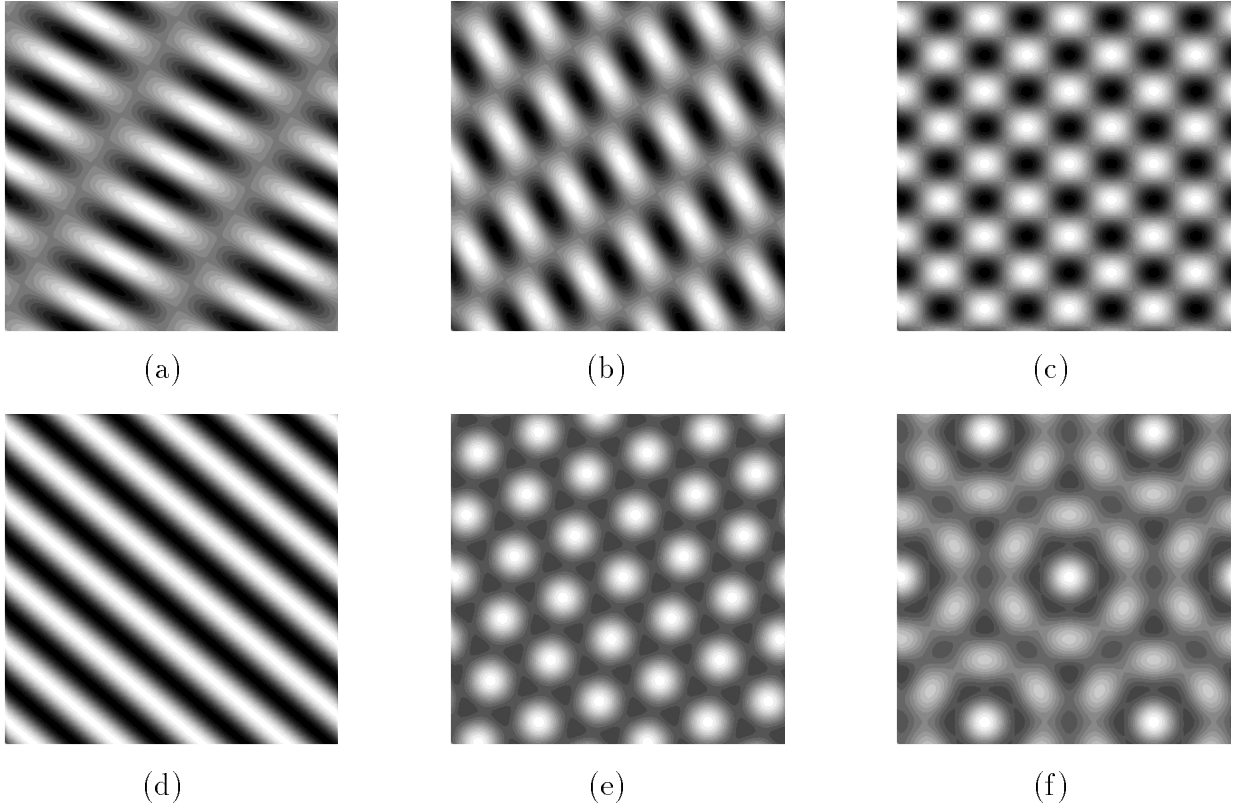

Figure 6: Axial planforms associated with 12-dimensional representation of $\Gamma_h$ with $(\alpha, \beta) = (3, 2)$, $x_1, x_2 \in [-\frac{2}{\sqrt{3}}, \frac{2}{\sqrt{3}}]$; (a) rhombs ($\text{Rh}_{h1,3,2}$), (b) rhombs ($\text{Rh}_{h2,3,2}$), (c) rhombs ($\text{Rh}_{h3,3,2}$), (d) rolls, (e) simple hexagons ($\text{SiH}^+$), and (f) super hexagons ($\text{SuH}^+$).

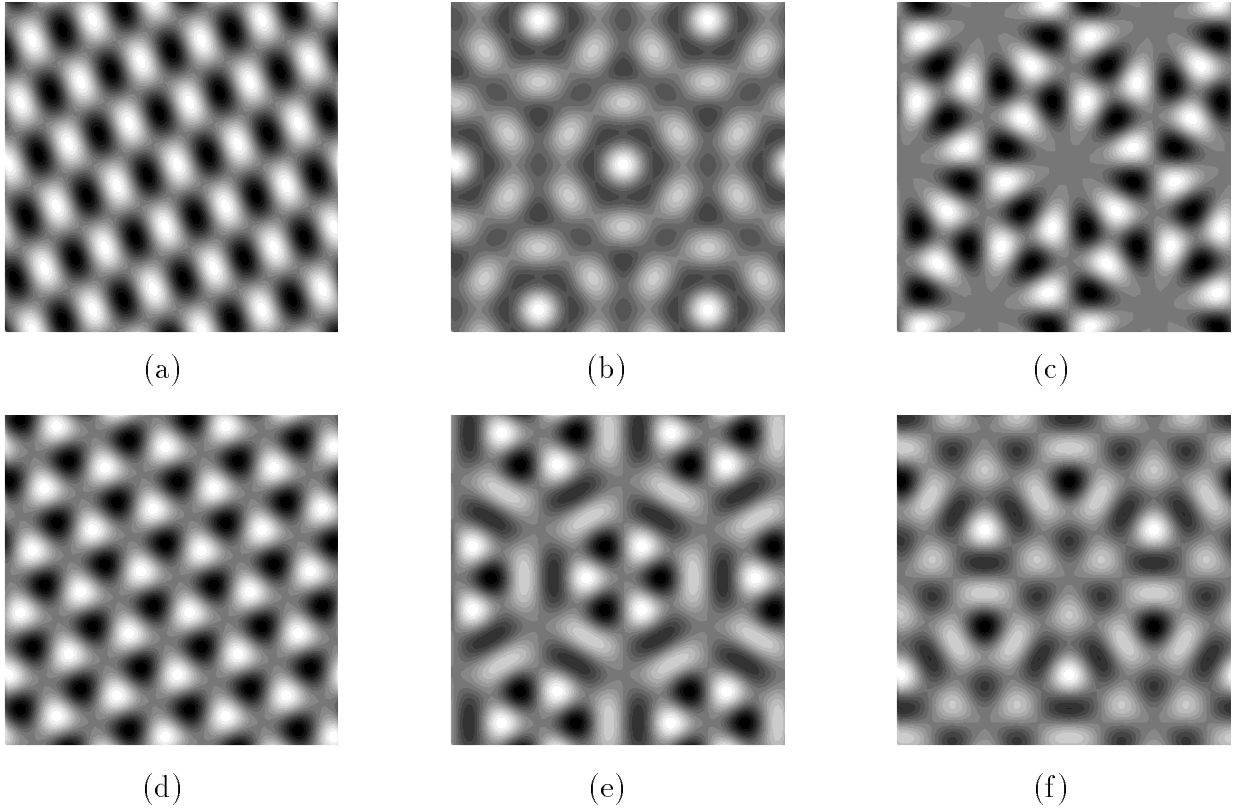

Figure 7: Axial planforms associated with 12-dimensional representation of $\Gamma_h + \mathbf{Z}_2$ with $(\alpha, \beta) = (3, 2)$, $x_1, x_2 \in [-\frac{2}{\sqrt{3}}, \frac{2}{\sqrt{3}}]$; (a) rhombs ($\mathrm{Rh}_{h0}$), (b) super hexagons, (c) anti-hexagons, (d) simple triangles, (e) super triangles, and (f) anti-triangles. (See Figure 6 for the additional axial planforms: rolls, simple hexagons, $\mathrm{Rh}_{h1,3,2}$, $\mathrm{Rh}_{h2,3,2}$, and $\mathrm{Rh}_{h3,3,2}$.)

Figure 8: Example of an hexagonal lattice bifurcation diagram for solutions in the six-dimensional subspace, $\mathbf{z} = (z_1, z_2, z_3, 0, 0, 0)$. Solid (dotted) lines indicate stable (unstable) solutions. The secondary solution branch has the form $\mathbf{z} = (x_1, x_2, x_2, 0, 0, 0)$, where $x_1, x_2 \in \mathbf{R}$.

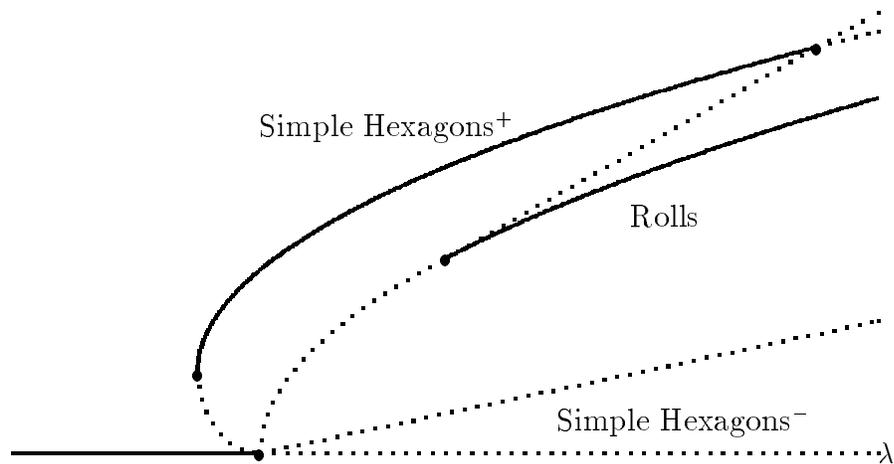

Figure 9: Example of an hexagonal lattice bifurcation diagram for the twelve-dimensional representations of $\Gamma_h$. Here $0 < \epsilon \ll 1$ in equation 5.29; see equation 5.30 for the other coefficients. Secondary bifurcation points are indicated by a solid circle; no secondary solution branches are shown.

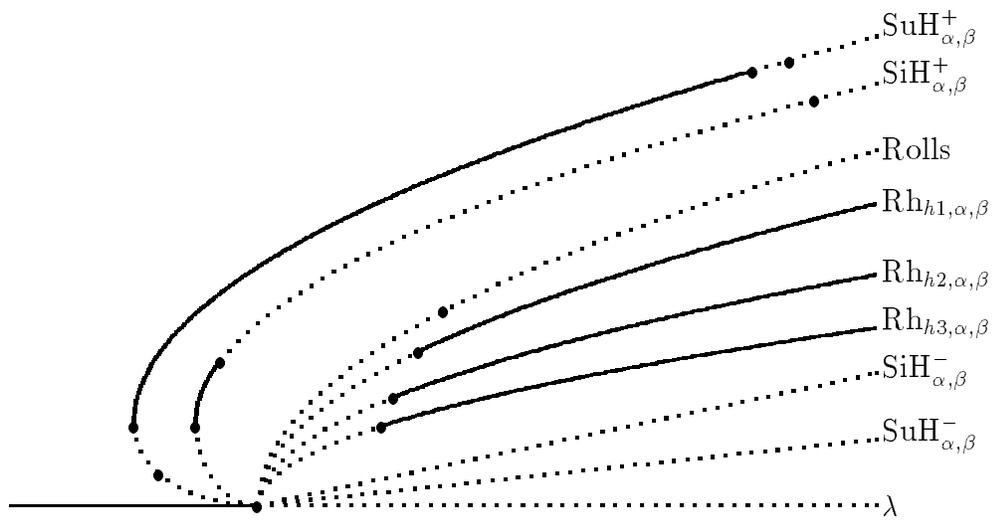